\RequirePackage{lineno} 

\documentclass[twocolumn,showpacs,preprintnumbers,amsmath,amssymb,nofootinbib]{revtex4}

\usepackage{graphicx}
\usepackage{dcolumn}
\usepackage{bm}
 
\usepackage{epstopdf}
\def\bea{\begin{eqnarray}}
\def\eea{\end{eqnarray}}

\def\pp{\mbox{$p$-$p$}}

\def\pa{\mbox{$p$-A}}

\def\auau{\mbox{Au-Au}}

\def\pbpb{\mbox{Pb-Pb}}
\def\ppb{\mbox{$p$-Pb}}

\def\pn{\mbox{$p$-N}}
\def\aa{\mbox{A-A}}
\def\nn{\mbox{N-N}}

\def\ee{\mbox{$e^+$-$e^-$}}

\def\pt{$p_t$}
\def\mt{$m_t$}
\def\yt{$y_t$}

\def\nch{$n_{ch}$}
\def\mmpt{$\bar p_t$}

\begin{document} 

\setlength{\pdfpagewidth}{8.5in}
\setlength{\pdfpageheight}{11in}

\setpagewiselinenumbers
\modulolinenumbers[5]

\preprint{version 1.9}

\title{Differential comparison of identified-hadron $\bf p_t$ spectra from high-energy A-B \\ nuclear collisions based on a  two-component model of hadron production
}

\author{Thomas A.\ Trainor}\affiliation{University of Washington, Seattle, WA 98195}


\date{\today}

\begin{abstract}
	
Identified-hadron (PID) spectra from 2.76 TeV Pb-Pb and $p$-$p$ collisions are analyzed via a two-component (soft + hard) model (TCM) of hadron production in high-energy nuclear collisions. The PID TCM is adapted with minor changes from a recent analysis of PID hadron spectra from 5 TeV $p$-Pb collisions. Results from LHC data are compared with a PID TCM for 200 GeV Au-Au pion and proton spectra. 2.76 TeV proton spectra exhibit strong inefficiencies above 1 GeV/c estimated by comparing the $p$-$p$ spectrum with the corresponding TCM. After inefficiency correction Pb-Pb proton spectra are very similar to \auau\ proton spectra. PID A-A spectra are generally inconsistent with radial flow. Jet-related Pb-Pb and Au-Au spectrum hard components exhibit strong suppression at higher $p_t$ in more-central collisions corresponding to results from spectrum ratio $R_{AA}$ but also, for pions and kaons, exhibit dramatic enhancements below $p_t = 1$ GeV/c that are concealed by $R_{AA}$. In contrast, enhancements of proton hard components appear only above 1 GeV/c suggesting  that the baryon/meson ``puzzle'' is a jet phenomenon. Modification of spectrum hard components in more-central A-A collisions is consistent with increased gluon splitting during jet formation but with approximate conservation of  leading-parton energy within a jet via the lower-$p_t$ enhancements. 

\end{abstract}


\maketitle

 \section{Introduction}
 
 This article reports analysis of identified-hadron (PID) \pt\ spectra from six centrality classes of 2.76 TeV \pbpb\ collisions~\cite{alicepbpbpidspec} via the  two-component (soft + hard) model (TCM) of hadron production in high-energy nuclear collisions~\cite{ppprd,ppquad,hardspec,alicetomspec,anomalous,tommpt}. Data from 2.76 TeV \pp\ collisions~\cite{alicespec2} and more-limited peripheral data from \pbpb\ collisions~\cite{alicepbpbspecx} are also included.  This analysis is intended to explore the centrality systematics of jet modification (jet quenching) in \pbpb\ collisions and test the hypothesis that quark-gluon plasma (QGP), a dense flowing QCD medium, is formed in \aa\ collisions. In particular, spectra are examined for evidence that transverse or radial flow plays a significant role in \aa\ collision dynamics.
 
 Conventional approaches to analysis of \pt\ spectra from high-energy nuclear collisions have been motivated by an adopted narrative in which partial equilibration of flowing QCD matter at the parton level occurs within the collision space-time volume, and most detected hadrons emerge from a low-viscosity flowing medium or ``perfect liquid''~\cite{perfect} via a subsequent bulk hadronization process.%
 \footnote{For example, see the introduction to Ref.~\cite{alicepbpbpidspec}.}  
 The \pt\ spectra of those hadrons should then reveal the velocity profile of the flowing medium and possible modification of jet formation due to passage of energetic scattered partons through the conjectured dense medium. Flows are assumed to dominate the low-\pt\ part of a spectrum (e.g.\ below 2-3 GeV/c) while jet fragments (a small fraction of all hadrons) are assumed to dominate at higher \pt\ (e.g.\ above 6-10 GeV/c)~\cite{alicepbpbpidspec}. In the former case a blast-wave model function is fitted to spectra over a limited low-\pt\ interval, with parameters representing a ``kinetic freezeout'' temperature and a radial-flow velocity~\cite{blastwave}. In the latter case the spectrum ratio $R_{AA}$ is interpreted to reveal the effects of jet quenching as reductions from unity (jet suppression) at higher \pt~\cite{raastar}.
 
 An alternative approach involving no {\em a priori} assumptions addresses efficient and accurate representation of any information carried by spectrum data. An example is given by Ref.~\cite{ppprd} describing a process in which a two-component spectrum model is inferred from the charge multiplicity (\nch) dependence of \pt\ spectra from 200 GeV \pp\ collisions. The physical origins of the two spectrum components have been inferred by comparisons with QCD theory~\cite{fragevo} and correlation data~\cite{porter2,porter3}. 
 
 The TCM has been applied subsequently to 200 GeV \auau\ collisions~\cite{hardspec}, a large volume of \pp\ data over several collision energies~\cite{alicetomspec}, \ppb\ PID spectrum data~\cite{ppbpid} and two-particle angular correlations~\cite{anomalous,ppquad}.  As a {\em composite}  production model the TCM is apparently required by spectrum~\cite{alicetomspec} and correlation~\cite{ppquad} data for all A-B collision systems, and the two components overlap strongly on \pt.  The present study is an extension of that program to determine the properties of \pbpb\ PID \pt\ spectra.
 
Establishment of a PID spectrum TCM for 2.76 TeV \pbpb\ collisions in this study is based on a previous analysis of 5 TeV \ppb\ PID spectra~\cite{ppbpid}.  The \ppb\ PID spectrum TCM is adopted with almost no change (exceptions noted below) to provide a reference for \pbpb\ data. PID spectra from 2.76 TeV \pp\ collisions and spectra with more-limited \pt\ acceptance from 80-90\% central \pbpb\ collisions are also analyzed to provide a null reference for possible jet modification and evidence for radial flow.

Some results of the present analysis include: (a) PID spectra from peripheral 2.76 TeV \pbpb\ and \pp\ collisions  are consistent with 5 TeV \ppb\ results reported in Ref.~\cite{ppbpid} including no significant evidence for radial flow or jet modification for any \ppb\ \nch\ class spanning a thirty-fold variation in dijet production. (b) For more-central \pbpb\ collisions jet-related spectrum hard components deviate strongly from \pp\ and \ppb\ references, including suppression at higher \pt\ and complementary {\em large enhancement} at lower \pt. (c) A {\em sharp transition} in the jet modification trend is similar to the transition observed in jet properties from 200 GeV \auau\ collisions~\cite{anomalous}. (d) For all \pbpb\ data, spectrum soft components are consistent with a fixed shape as for \pp\ and \ppb\ reference cases, suggesting no evidence for radial flow (i.e.\ no significant boost of the soft component on transverse rapidity \yt). (e) Overall spectrum shape evolution is dominated by effects that scale with  the number of \nn\ binary collisions as expected for dijet production.

This article is arranged as follows:
Section~\ref{alicedata} presents 2.76 TeV  \pbpb\ PID spectrum data from Ref.~\cite{alicepbpbpidspec}.
Section~\ref{spectrumtcm} describes  \pp\ and A-B PID spectrum TCMs.
Section~\ref{auau} reviews a previous TCM analysis of 200 GeV \auau\ PID spectra.
Section~\ref{pidspec} reports a 2.76 TeV  \pbpb\ PID spectrum TCM resulting from the present study.
Section~\ref{ratcompare} compares spectrum ratios in various formats.
Section~\ref{fractions}  compares PID particle fractions from data and TCM.
Section~\ref{sys}  reviews systematic uncertainties.
Sections~\ref{disc} and~\ref{summ} present discussion and summary.

\section{2.76 $\bf TeV$ $\bf Pb$-$\bf Pb$ PID Spectrum data} \label{alicedata}

Identified-hadron spectrum data for the present analysis, obtained from Ref.~\cite{alicepbpbpidspec}, were produced by the ALICE collaboration at the LHC. The azimuth acceptance is $2\pi$ and pseudorapidity acceptance is $|\eta|< 0.8$ or $\Delta \eta = 1.6$.  \pbpb\ Collisions ($40\times 10^6$ events) are divided into six centrality classes: 0-5\%, 5-10\%, 10-20\%, 20-40\%, 40-60\% and 60-80\%. Corresponding estimated centrality parameters from Ref.~\cite{pbpbcent} are presented in Table \ref{ppbparams1}.
Hadron species include charged pions $\pi^\pm$, charged kaons $K^\pm$ and protons $p,~\bar p$. The \pp\ spectra ($11\times 10^6$ events) are reported in Ref.~\cite{alicespec2}. Peripheral 80-90\% \pbpb\ data with more-limited \pt\ acceptance are reported in Ref.~\cite{alicepbpbspecx}.

\subsection{Pb-Pb PID  spectrum data} \label{piddataa}

\begin{figure}[h]
	\includegraphics[width=1.65in]{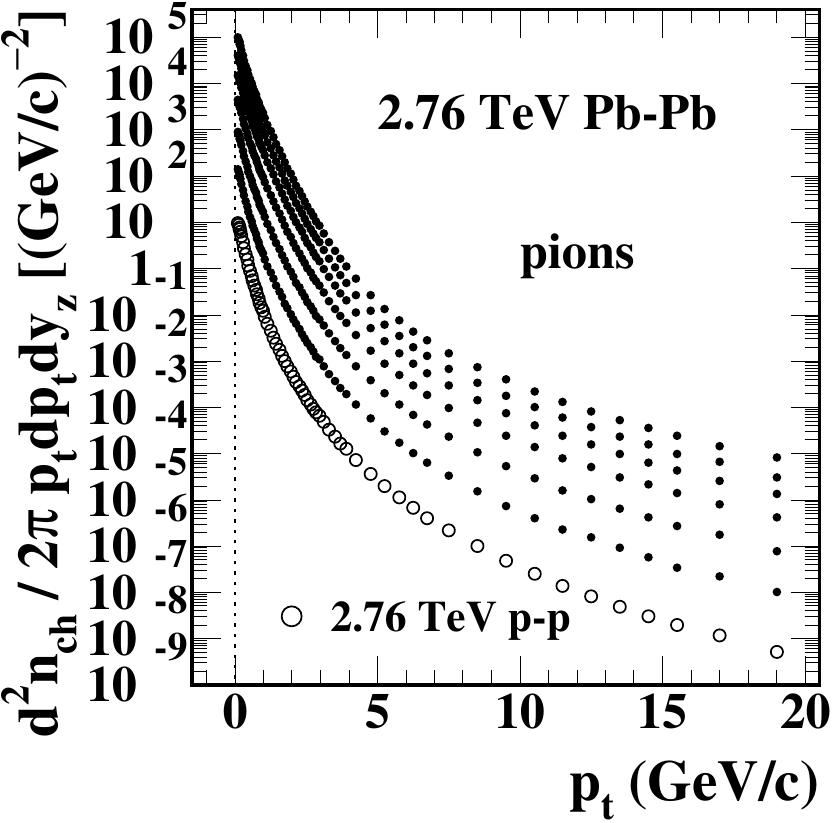}
	\includegraphics[width=1.65in]{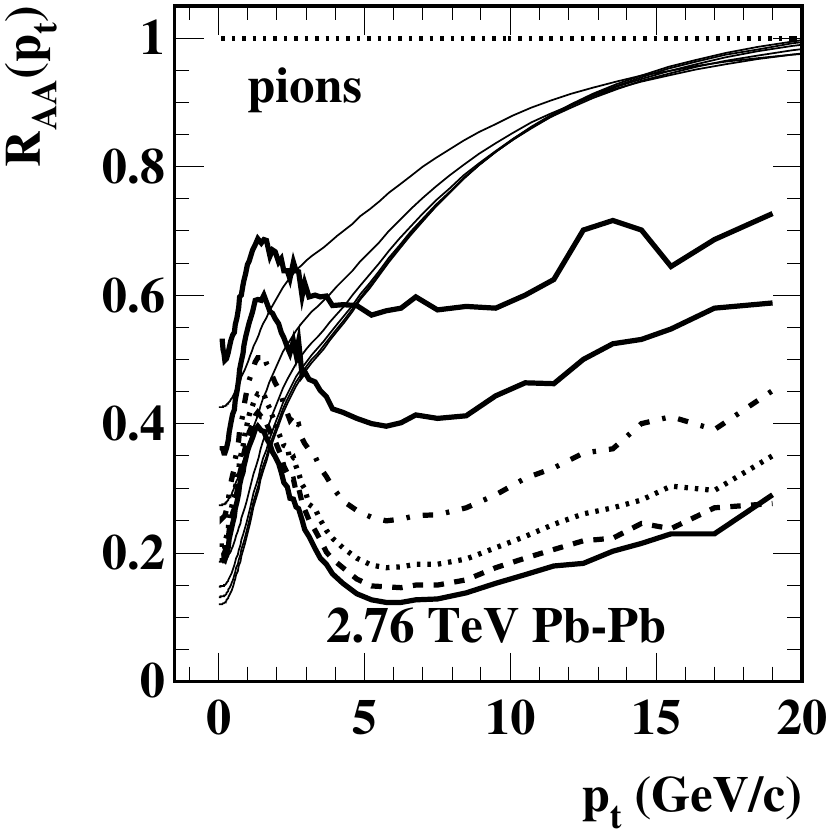}
\put(-142,85) {\bf (a)}
\put(-21,98) {\bf (d)}\\
	\includegraphics[width=1.65in]{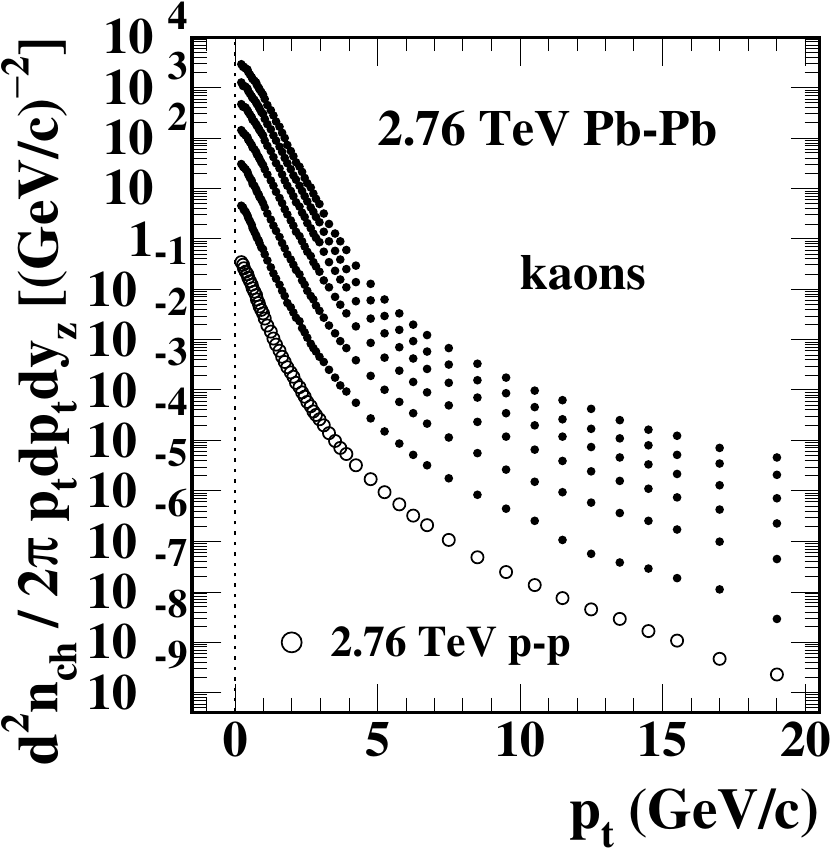}
\includegraphics[width=1.65in]{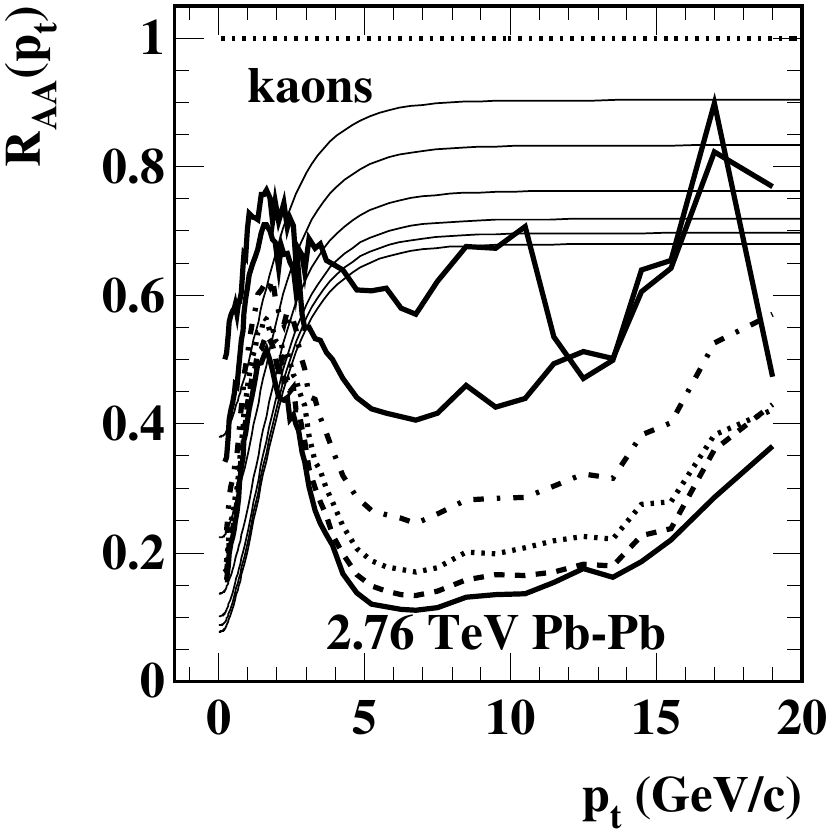}
\put(-142,85) {\bf (b)}
\put(-21,32) {\bf (e)}\\
	\includegraphics[width=1.65in]{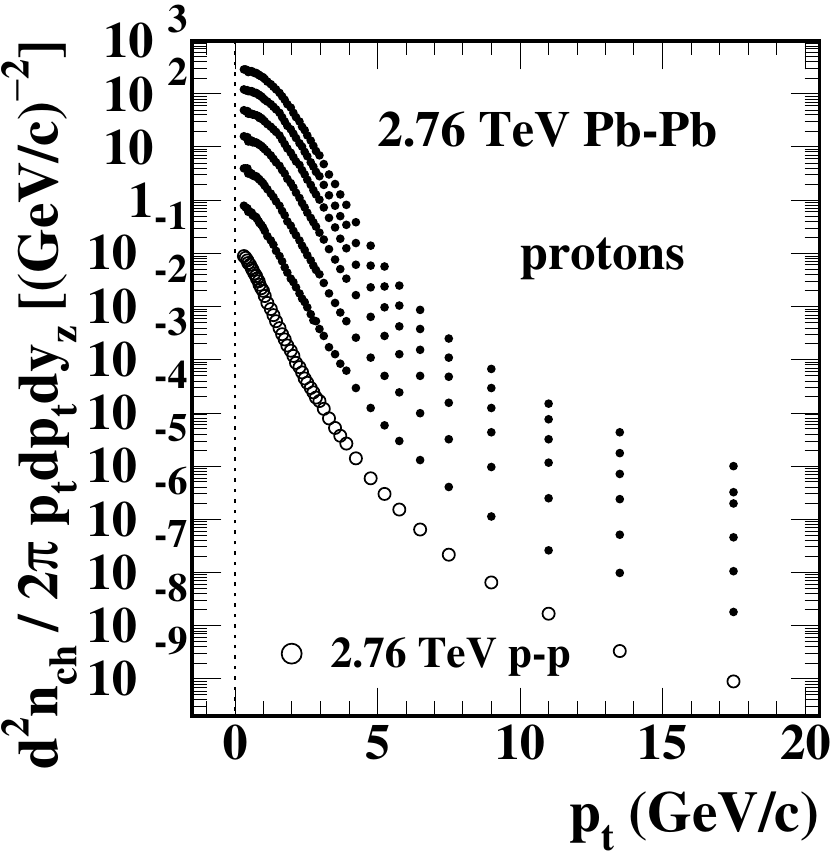}
\includegraphics[width=1.65in]{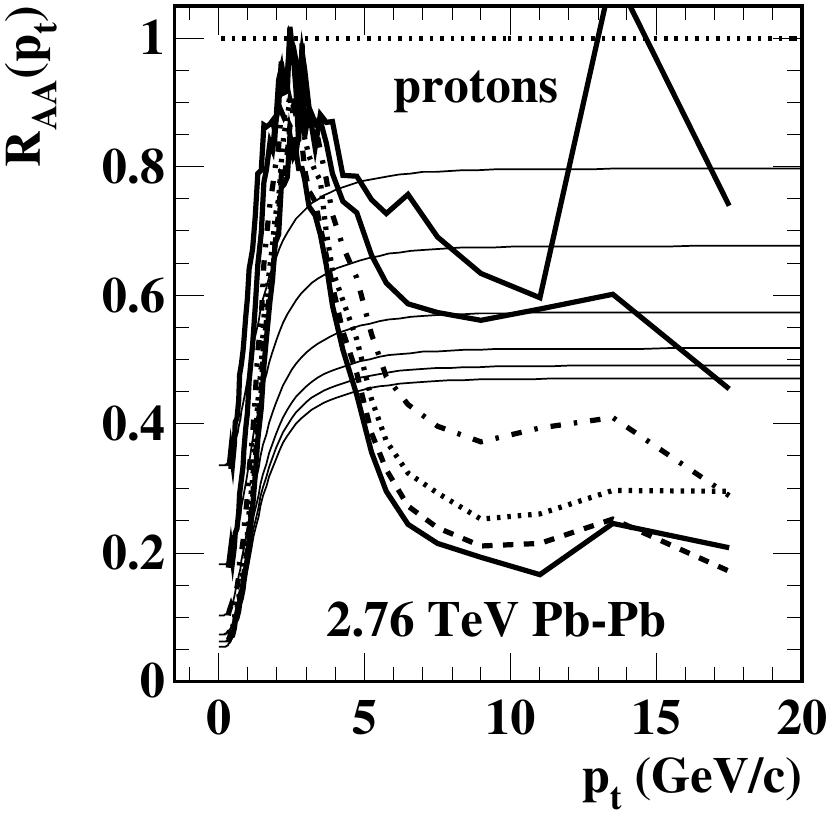}
\put(-142,85) {\bf (c)}
\put(-21,102) {\bf (f)}\\
	\caption{\label{piddata}
Left:
\pt\ spectra for three species of identified hadrons from six centrality classes of 2.76 TeV \pbpb\ collisions (solid) and from \pp\ collisions (open)~\cite{aliceppbpid}.
Right: Spectrum ratios (nuclear modification factors) $R_\text{AA}(p_t)$ derived from data in the left panels (bold curves of various line styles). The thin solid reference curves are described in Sec.~\ref{piddiff}.
} 
\end{figure}

Figure~\ref{piddata} (a) - (c) shows PID spectra $\bar \rho_\text{0PbPbi}(p_t,b)$ and $\bar \rho_{0ppi}(p_t)$ from Ref.~\cite{alicepbpbpidspec} (points) in a conventional  semilog plotting format vs linear hadron \pt. The symbol $i$ represents a specific hadron species. The spectra are scaled by powers of 2 according to $2^{m-1}$ where $m \in [1,6]$ and $m = 6$ is most central (not to be confused with centrality index $n \in [1,7]$ for \ppb\ collisions in Ref.~\cite{ppbpid}). The selected plot format limits visual access to {\em differential} spectrum features as they vary with hadron species and \pbpb\ centrality and especially at lower \pt\ were most jet fragments appear~\cite{ppprd,fragevo}. Compare with corresponding figures presented in a TCM format in Sec.~\ref{pidspec}.

Figure~\ref{piddata} (d) - (f) show spectrum ratios  (nuclear modification factors) of \pbpb\ spectra vs \pp\ spectra normalized by the number of binary collisions $N_{bin}$ estimated by a Glauber Monte Carlo as in Table~\ref{ppbparams1}~\cite{pbpbcent}
\bea \label{raa}
R_\text{AAi}(p_t;n_{ch}) &=& \frac{1}{N_{bin}}\frac{\bar \rho_\text{0PbPbi}(p_t;n_{ch})}{\bar \rho_{0ppi}(p_t)}.
\eea
The independent variable in these plots is transverse momentum \pt. The majority of the plot space on the right is then occupied by slowly-varying data while rapid changes are compressed into a small interval on the left.
Transverse rapidity $y_t \equiv \ln[(p_t + m_{t\pi})/m_\pi]$ is an alternative independent variable assuming the pion mass for all hadron species. Spectra can be transformed from \pt\ to \yt\ via Jacobian $p_t m_{t\pi}/y_{t\pi}$. As such, \yt\ is simply a logarithmic representation of transverse momentum \pt\ with a well-defined zero. The logarithmic momentum scale gives much-improved visual access to lower-\pt\ spectrum structure {\em where the great majority of jet fragments resides}~\cite{fragevo,alicetomspec,ppquad}. The thin solid curves are TCM reference trends defined in Sec.~\ref{piddiff} assuming that \pbpb\ collisions are linear superpositions of \nn\ collisions. For data in panel (f) the TCM reference accurately describes the particularly low values of proton $R_{AA}(y_t)$ at low \pt.

\subsection{$\bf Pb$-Pb Glauber-model centrality parameters}

Table~\ref{ppbparams1} shows centrality parameters for 2.76 TeV \pbpb\ collisions from Table~1 of Ref.~\cite{pbpbcent} corresponding to the spectra in Fig.~\ref{piddata}. The centrality parameters are inferred from a Glauber Monte Carlo. To interpret PID \pt\ spectra from Ref.~\cite{alicepbpbpidspec} properly centralities and geometry parameters should be estimated accurately.

\begin{table}[h]
	\caption{Glauber parameters for 2.76 TeV \pbpb\ collisions from Table 1 of Ref.~\cite{pbpbcent}.  $b_\text{max}$ is the maximum impact parameter for each centrality interval. The lead (Pb) nucleus diameter is 14.2 fm, equivalent to eight tangent nucleons. $\nu \equiv 2 N_{bin} / N_{part}$ is the mean number of binary nucleon-nucleon (N-N) collisions per participant-nucleon pair.
	}
	\label{ppbparams1}
	\begin{center}
		\begin{tabular}{|c|c|c|c|c|} \hline
			centrality (\%)  & $b_\text{max}$ (fm) & $N_{part}$ & $N_{bin}$ & $\nu$  \\ \hline
			0 - 5   &  3.5   & 383 & 1685 & 8.8  \\ \hline
			~\,5 - 10   & 4.9  & 329& 1316  & 8.0 \\ \hline
			10 - 20   &7.0  & 260 & 921  &  7.1  \\ \hline
			20 - 40   & 9.9  & 157  & 438 &  5.6  \\ \hline
			40 - 60     & 12.1  & 68.6 & 128  & 3.7  \\ \hline
			60 - 80    & 14.0  &22.5 & 26.7  & 2.4  \\ \hline
		\end{tabular}
	\end{center}
\end{table}

As a reference the Pb nucleus is approximately eight tangent nucleons in diameter, and $\nu  \equiv 2 N_{bin} / N_{part}$ is the mean number of \nn\ binary collisions per \nn\ pair averaged over the \pbpb\ overlap region. For central \pbpb\ collisions $\nu \approx 9$ implies that on average a participant nucleon must collide each time {\em simultaneously} with 2-3 participant nucleons. A study of \ppb\ geometry in Ref.~\cite{tomglauber} and follow-up study of Glauber-model inconsistencies in Ref.~\cite{tomexclude} suggest that at least in the asymmetric \ppb\ context multiple simultaneous \nn\ collisions are unlikely.

\subsection{Pb-Pb PID spectrum data interpretations} \label{interpret}

Reference~\cite{alicepbpbpidspec} interprets the \pbpb\ spectrum data in the context of assumed formation of a dense and flowing QCD medium or QGP. Two \pt\ intervals are  assumed to represent distinct physical mechanisms. For $p_t < 2$ GeV/c spectrum features should reflect ``bulk production,'' whereas for $p_t > 10$ GeV/c jet formation and in-medium jet modification (``jet quenching'') should be the relevant issues. In the lower-\pt\ interval ``hardening of the spectra [with increasing centrality] is mass dependent and is characteristic of hydrodynamic flow....'' In the higher-\pt\ interval ``the spectra follow a power-law shape as expected from perturbative QCD (pQCD) calculations.'' Trends in the intermediate-\pt\ region (e.g.\ the  ``baryon/meson puzzle'' \cite{rudy,duke,tamu}) are still not understood.

Concerning jet quenching and PID $R_\text{AA}$ data, Ref.~\cite{alicepbpbpidspec} concludes that for $p_t > 10$ GeV/c ($y_t > 5$) three hadron species follow {\em statistically similar trends}, suggesting that while ``jet quenching'' represents in some sense parton energy loss to a dense medium the hadrochemistry (``species composition'') of the jet ``core'' is not significantly influenced by the medium. That summary is related to broad issues of hadron-species-dependent fragmentation functions (FFs) in various collision contexts~\cite{eeprd,fragevo}. See the responding comment at the end of Sec.~\ref{compare}.

\section{$\bf p$-$\bf p$ and A-B Spectrum $\bf TCMs$} \label{spectrumtcm}

High-energy nuclear data exhibit a  basic property:  certain data features are {\em composite} and require a two-component (soft + hard) model of hadron production, as demonstrated initially for \pt\ spectra from \pp\ collisions~\cite{ppprd,ppquad}. The two components overlap strongly on \pt. PID spectra from 2.76 TeV \pp\ and \pbpb\ collisions are analyzed in this study in the context of previous analysis of \pp~\cite{ppprd,ppquad}, \auau~\cite{hardspec} and \ppb~\cite{tomglauber,ppbpid} data which provide reference trends. The TCM for \pp, \auau\ and \ppb\  collisions has been the product of phenomenological analysis of data from a variety of collision systems and data formats~\cite{ppprd,hardspec,ppquad,alicetomspec,tommpt}. Physical interpretations of TCM soft and hard components have been derived {\em a posteriori} by comparing inferred TCM characteristics with other relevant measurements~\cite{hardspec,fragevo}, in particular measured MB dijet properties~\cite{eeprd,jetspec2,mbdijets}. The TCM does not result from fits to individual spectra which would require many parameter values. The few TCM parameters are required to have simple $\log(\sqrt{s})$ trends on collision energy and simple extrapolations from measured \pp\ trends.

\subsection{A-B spectrum TCM for unidentified hadrons} \label{unidspec}

The \pt\ or \yt\ spectrum TCM is by definition the sum of soft and hard components with details inferred from data (e.g.\ Ref.~\cite{ppprd}). For \pp\ collisions
\bea  \label{rhotcm}
\bar \rho_{0}(y_t;n_{ch}) &=& S_{pp}(y_t;n_{ch}) + H_{pp}(y_t;n_{ch})
\\ \nonumber
&\approx& \bar \rho_{s}(n_{ch}) \hat S_{0}(y_t) + \bar \rho_{h}(n_{ch}) \hat H_{0}(y_t),
\eea
where hadron charge \nch\ is an event-class index, and factorization of the dependences on \yt\ and \nch\ is the main feature of the spectrum TCM. The motivation for transverse rapidity $y_{ti} \equiv \ln[(p_t + m_{ti})/m_i]$ (applied to hadron species $i$) is described in Sec.~\ref{tcmmodel}. The \yt\ integral of Eq.~(\ref{rhotcm}) is $\bar \rho_0 \equiv n_{ch} / \Delta \eta = \bar \rho_s + \bar \rho_h$, a sum of soft and hard charge densities. $\hat S_{0}(y_t)$ and $\hat H_{0}(y_t)$ are unit-normal model functions approximately independent of \nch, and the centrally-important relation $\bar \rho_{h} \approx \alpha \bar \rho_{s}^2$ with $\alpha \approx O(0.01)$ is inferred from \pp\ spectrum data~\cite{ppprd,ppquad,alicetomspec}. 

For composite A-B collisions the spectrum TCM can be generalized to
\bea \label{ppspectcm}
\bar \rho_{0}(y_t;n_{ch}) &=& \frac{N_{part}}{2}S_{AB}(y_t;n_{ch}) +N_{bin}  H_{AB}(y_t;n_{ch})~~~
\\ \nonumber
&\approx& \frac{N_{part}}{2}  \bar \rho_{sAB} \hat S_{0}(y_t) + N_{bin}  \bar \rho_{hAB} \hat H_{0}(y_t;n_{ch}),
\eea
which includes a further factorization of charge densities $\bar \rho_x = n_x / \Delta \eta$ into A-B Glauber geometry parameters $N_{part}$ (number of nucleon participants N) and $N_{bin}$ (\nn\ binary collisions) and mean charge densities $\bar \rho_{xAB}$  per \nn\ pair averaged over all \nn\ interactions within the A-B system.  For A-B collisions $\bar \rho_{s}(n_s) = [N_{part}(n_s)/2]\bar \rho_{sAB}(n_s)$ is a factorized soft-component density and $\bar \rho_h(n_s) = N_{bin}(n_s) \bar \rho_{hAB}(n_s)$ is a factorized hard-component density.   
Integrating Eq.~(\ref{ppspectcm}) over  \yt\ the mean charge density is
\bea \label{nchppb}
\bar \rho_0 &=& \frac{N_{part}}{2} \bar \rho_{sAB}(n_s) + N_{bin} \bar \rho_{hAB}(n_s)
\\ \nonumber
\frac{\bar \rho_0}{\bar \rho_{s}} &=& \frac{\bar n_{ch}}{\bar n_s} ~=~ 1 + x(n_s)\nu(n_s),
\eea
where the hard/soft ratio is $x(n_s) \equiv \bar \rho_{hAB}/\bar \rho_{sAB}$ and the mean number of binary collisions per participant pair is $\nu(n_s) \equiv 2 N_{bin} / N_{part}$. For \aa\ collisions $\bar \rho_{hAA}$ and $\hat H_{0}(y_t;n_{ch})$ may vary strongly with centrality whereas centrality variation of $ \bar \rho_{sAA}$ and $\hat S_{0}(y_t)$ may be negligible.

To obtain details of model functions and other aspects of the TCM the measured hadron spectra are normalized by charge-density soft component $\bar \rho_s$. Normalized data spectra then have the form
\bea  \label{norm}
\frac{\bar \rho_{0}(y_t;n_{ch})}{\bar \rho_{s}}  &\equiv& X(y_t;n_{ch})
\\ \nonumber
&=&  \hat S_{0}(y_t) +   x(n_s) \, \nu(n_s) \hat H_{0}(y_t;n_{ch}),
\eea
where $n_s$ is the soft component of event-class index \nch\ integrated within some $\eta$ acceptance $\Delta \eta$. Because $x(n_s) \sim n_{ch}$ (for \pp\ and \pa) the {\em data} soft component is defined as the limiting form of $X(y_t;n_{ch})$ as $n_{ch} \rightarrow 0$. 

For \pp\ collisions $x(n_s) \equiv \bar \rho_{h}/\bar \rho_{s} \approx \alpha \bar \rho_{s}$ is inferred from data, with $\alpha \approx O(0.01)$ over a range of \pp\ collision energies~\cite{alicetomspec}.  For \pa\ collisions $x(n_s)  \approx \alpha \bar \rho_{sNN}$ is assumed by analogy with \pp\ collisions. Other \ppb\ TCM and geometry elements  are then defined in terms of $x(n_s)$~\cite{tommpt,tomglauber,ppbpid}. For \aa\ collisions the {\em product} $x(n_s)\nu(n_s)$ may be inferred from yield data as follows: Given $\bar \rho_0$ data and values of $N_{part}/2$ for each centrality from a Monte Carlo Glauber, and assuming mean $\bar \rho_{sNN}$ is the same for all \aa\ collisions as for NSD \pp\ collisions, product $x(n_s)\nu(n_s)$ can be inferred from the second line of Eq.~(\ref{nchppb}). For unidentified hadrons the normalization factor in Eq.~(\ref{norm}) is
\bea \label{overrhos}
{\bar \rho_s} &=&  \frac{\bar \rho_0(n_s)}{1 + x(n_s) \nu(n_s)}  = {\bar \rho_{sNN} N_{part}/2}.
\eea
In the present study $N_{part}$ from Table~\ref{ppbparams1} is accepted as a reasonable estimate. The value $\bar \rho_{sNN} \approx 4.55$ for 2.76 TeV \pp\ collisions is obtained from Ref.~\cite{alicetomspec}. 

Normalized spectra $X(y_t;n_{ch})$ can be compare directly with a model function $\hat S_{0}(y_t)$ to extract data hard components in the form
\bea \label{yhard}
Y(y_t;n_{ch}) &\equiv& \frac{1}{ x(n_s) \nu(n_s) } \left[ X(y_t;n_{ch}) -  \hat S_{0}(y_t) \right].
\eea
Equations~(\ref{ppspectcm}) - (\ref{yhard}) with model parameters as in Tables~\ref{pidparams} and \ref{otherparams} describe PID \pp\ and \ppb\ spectra within data uncertainties as reported in Refs~\cite{ppprd,ppquad,ppbpid} and serve as a {\em reference} for 2.76 TeV \pbpb\ PID spectrum data as in the present study.
For the full \aa\ collision system $x(n_s)$ is required in Eq.~(\ref{overrhos}), but in Eq.~(\ref{yhard}) the alternative $x(n_s) \rightarrow x_{pp} = \alpha \bar \rho_{spp}$ may be preferred. Then \aa\ data in the form of quantity $Y(y_t)$ may be compared directly with the \pp\ $\hat H_0(y_t)$ model function to evaluate jet modification (as an alternative to spectrum ratio $R_{AA}$), as described in Sec.~\ref{piddiff} in connection with Eq.~(\ref{raax}).

\subsection{Spectrum TCM model functions} \label{tcmmodel}

Given normalized spectrum data as in Eq.~(\ref{norm}) and the trend $x(n_s) \sim n_s \sim n_{ch}$ spectrum soft component $\hat S_0(y_t)$ is defined as the asymptotic limit of normalized {\em data} spectra as $n_{ch} \rightarrow 0$. Hard components of data spectra are then defined as complementary to soft components. Thus, data soft and hard components are {\em inferred from data alone}, not imposed via assumed model functions.

The data soft component for specific hadron species $i$, once isolated as an asymptotic limit, is typically well described by a L\'evy distribution on $m_{ti}  = \sqrt{p_t^2 + m_i^2}$. The unit-integral soft-component model is 
\bea \label{s00}
\hat S_{0i}(m_{ti}) &=& \frac{A}{[1 + (m_{ti} - m_i) / n T]^n},
\eea
where $m_{ti}$ is the transverse mass-energy for hadrons $i$ of mass $m_i$, $n$ is the L\'evy exponent, $T$ is the slope parameter and coefficient $A$ is determined by the unit-integral condition. Reference parameter values for unidentified hadrons from 5 TeV \pp\ collisions reported in Ref.~\cite{alicetomspec} are $(T,n) \approx (145~ \text{MeV},8.3)$. Model parameters $(T,n)$ for each species of identified hadrons as in Table~\ref{pidparams} are determined from \ppb\ spectrum data as described below.

The unit-integral hard-component model  is a Gaussian on $y_{t\pi} \equiv \ln((p_t + m_{t\pi})/m_\pi)$ (as explained below) with exponential (on $y_t$) or power-law (on $p_t$) tail for larger \yt\
\bea \label{h00}
\hat H_{0}(y_t) &\approx & A \exp\left\{ - \frac{(y_t - \bar y_t)^2}{2 \sigma^2_{y_t}}\right\}~~~\text{near mode $\bar y_t$}
\\ \nonumber
&\propto &  \exp(- q y_t)~~~\text{for larger $y_t$ -- the tail},
\eea
where the transition from Gaussian to exponential on \yt\ is determined by slope matching~\cite{fragevo}. The $\hat H_0$ tail density varies on \pt\ approximately as power law $1/p_t^{q + 2}$. Coefficient $A$ is determined by the unit-integral condition. Model parameters $(\bar y_t,\sigma_{y_t},q)$ for identified hadrons as in Table~\ref{pidparams} are also derived from \ppb\ spectrum data.

All data spectra are plotted vs pion rapidity $y_{t\pi}$ with pion mass assumed.  Spectra on \pt\ are transformed to densities on $y_{t\pi}$ via Jacobian factor $m_{t\pi} p_t / y_{t\pi}$ where $m_{t\pi}^2 = p_t^2 + m_\pi^2$ and $y_{t\pi} = \ln[(m_{t\pi} + p_t)/m_\pi]$. The motivation is comparison of spectrum hard components apparently arising from a common underlying jet spectrum on \pt~\cite{fragevo}, in which case $y_{t\pi}$ serves simply as a logarithmic measure of hadron \pt\ with well-defined zero.  The $\hat S_{0i}(m_{ti})$  as defined in Eq.~(\ref{s00}) are converted to $\hat S_{0i}(y_{t\pi})$ via the Jacobian factor $m_{t\pi} p_t / y_{t\pi}$, and the $\hat H_{0i}(y_{t})$ in Eq.~(\ref{h00}) are always defined on $y_{t\pi}$ as noted. The accuracy of TCM-data comparisons requires precise normalization of all model functions $\hat S_{0i}(m_{ti})$ and $\hat H_{0i}(y_{t})$.

\subsection{p-Pb geometry inferred from non-PID data} \label{geom}

The analysis of PID \ppb\ spectrum data reported in Ref.~\cite{ppbpid} provides a reference for \pp\ and \pbpb\ PID spectrum analysis in the present study. The \ppb\ geometry parameters from the earlier study are here reviewed.
Table~\ref{rppbdata} presents geometry parameters for 5 TeV \ppb\ collisions inferred from a Glauber-model analysis in Ref.~\cite{aliceglauber} (primed values) and TCM values from the analysis in Ref.~\cite{tomglauber}  (unprimed values). The large differences between Glauber and TCM values are explained in Ref.~\cite{tomglauber}. Primed quantities based on a Glauber Monte Carlo study result from assumptions inconsistent with \mmpt\ data~\cite{tommpt}. Charge densities $\bar \rho_0$ are derived directly from data and correctly characterize the seven centrality classes.

\begin{table}[h]
	\caption{Glauber (primed~\cite{aliceglauber}) and  TCM (unprimed~\cite{tomglauber}) fractional cross sections and  geometry parameters, midrapidity charge density $\bar \rho_0$, \nn\ soft component $\bar \rho_{sNN}$ and TCM hard/soft ratio $x(n_s)$ used for 5 TeV \ppb\ PID spectrum data.
	}
	\label{rppbdata}
	\begin{center}
		\begin{tabular}{|c|c|c|c|c|c|c|c|c|c|} \hline
		$n$&	$\sigma' / \sigma_0$ &   $\sigma / \sigma_0$    & $N_{bin}'$ & $N_{bin}$ & $\nu'$ & $\nu$ & $\bar \rho_0$ & $\bar \rho_{sNN}$ & $x(n_s)$ \\ \hline
	1	& 	0.025         & 0.15 & 14.7  & 3.20 & 1.87  & 1.52 & 44.6 & 16.6  & 0.188 \\ \hline
	2	& 	0.075  & 0.24 &  13.0   & 2.59  &  1.86 & 1.43 & 35.9 &15.9  & 0.180 \\ \hline
	3	& 	0.15  & 0.37 &  11.7 & 2.16 & 1.84 &  1.37 & 30.0  & 15.2  & 0.172 \\ \hline
	4	& 	0.30 & 0.58 &  9.4 & 1.70 & 1.80  & 1.26  & 23.0  & 14.1  & 0.159  \\ \hline
	5	& 	0.50   &0.80  & 6.42  & 1.31 & 1.73  & 1.13 & 15.8 &   12.1 & 0.137  \\ \hline
	6	& 	0.70  & 0.95 & 3.81  & 1.07 & 1.58  & 1.03  & 9.7  &  8.7 & 0.098 \\ \hline
	7	& 	0.90 & 0.99 &  1.94 & 1.00 & 1.32  & 1.00  &  4.4  & 4.2 &0.047  \\ \hline
		\end{tabular}
	\end{center}
\end{table}

TCM (unprimed) geometry parameters in Table~\ref{rppbdata}, derived from \ppb\ \pt\ spectrum and \mmpt\ data for unidentified hadrons~\cite{tommpt,tomglauber}, are assumed to be valid also for each identified-hadron species and were used unchanged to process \ppb\ PID spectrum data in Ref.~\cite{ppbpid}. 

\subsection{A-B spectrum TCM for identified hadrons} \label{pidspecc}

To establish a TCM for A-B PID \pt\ spectra it is assumed that (a) \nn\ parameters $\alpha$, $\bar \rho_{sNN}$ and $\bar \rho_{hNN}$ have been inferred from unidentified-hadron data and (b) geometry parameters $N_{part}$, $N_{bin}$, $\nu$ and $x$ are a common property (relating to centrality) of a specific A-B collision system independent of identified-hadron species.

Given the A-B spectrum TCM for unidentified-hadron spectra in Eq.~(\ref{ppspectcm}) a corresponding TCM for identified hadrons can be generated by assuming that each hadron species $i$ comprises certain fractions of soft and hard TCM components denoted by $z_{si}$ and $z_{hi}$  (both $\leq 1$) and assumed {\em independent of \yt} (but not centrality). The PID spectrum TCM can then be expressed as
\bea \label{pidspectcm}
\bar \rho_{0i}(y_t)
&\approx& \frac{N_{part}}{2} z_{si} \bar \rho_{sNN} \hat S_{0i}(y_t) + N_{bin}  z_{hi} \bar \rho_{hNN} \hat H_{0i}(y_t) 
\nonumber \\
\frac{\bar \rho_{0i}(y_t)}{ \bar \rho_{si}} &\equiv&   X_i(y_t)
\\ \nonumber
&=& \hat S_{0i}(y_t) +  (z_{hi}/z_{si})x(n_s)\nu(n_s) \hat H_{0i}(y_t,b),
\eea
where unit-integral model functions $\hat S_{0i}(y_t)$ and $\hat H_{0i}(y_t)$ may depend on hadron species $i$.
For identified hadrons of species $i$ the normalization factor $ \bar \rho_{si}$ in the second line follows the form of Eq.~(\ref{overrhos}) but can be re-expressed in terms of $ \bar \rho_{s}$ for unidentified hadrons already inferred
\bea \label{rhosi}
{\bar \rho_{si}} &=& (N_{part}/2) \bar \rho_{sNNi}(y_t)
\\ \nonumber
 &=& \frac{\bar \rho_{0i} \equiv z_{0i} \bar \rho_0}{1 + (z_{hi}/z_{si}) x(n_s) \nu(n_s)}
\\ \nonumber
&\approx& \left[\frac{1 + x(n_s) \nu(n_s)}{1 + (z_{hi}/z_{si}) x(n_s) \nu(n_s)} \right]  {z_{0i}} \cdot {\bar \rho_{s}} 
\\ \nonumber
&\equiv& q_i(n_s) {z_{0i}} \cdot {\bar \rho_{s}},
\eea
defining $q_i(n_s)$ as representing the ratio in square brackets.
For \ppb\ data, ratio $z_{hi} / z_{si}$ for each hadron species $i$ is first adjusted to achieve coincidence of all seven normalized spectra as $y_t \rightarrow 0$. Parameter $z_{0i}$ is then adjusted to match rescaled spectra to unit-normal $\hat S_{0i}(y_t)$, also as $y_t \rightarrow 0$. Those expressions describe \ppb\ PID spectra within their data uncertainties as reported in Ref~\cite{ppbpid} and serve as fixed references for 2.76 TeV \pbpb\ and \pp\ PID spectrum data as reported in the present study.

Unit-normal model functions $\hat S_{0i}(y_t)$ and $\hat H_{0i}(y_t)$ for \pbpb\ and \pp\ data are retained from the \ppb\ analysis in Ref.~\cite{ppbpid}. As noted in Sec.~\ref{tcmmodel} $\hat S_{0i}(y_t)$ is first defined on proper $m_{ti}$ for a given hadron species $i$ and then transformed to $y_{t\pi}$. $\hat H_{0i}(y_t)$ is defined on $y_{t\pi}$ in all cases.

Normalized data spectra $X_i(y_t)$ can be combined with model function $\hat S_{0i}(y_t)$ per Eq.~(\ref{pidspectcm}) to extract data spectrum hard components in the form
\bea \label{pidhard}
Y_i(y_t) &\equiv& \frac{1}{(z_{hi} / z_{si}) x(n_s) \nu(n_s) } \left[ X_i(y_t) -  \hat S_{0i}(y_t) \right].~~~
\eea
As defined in Eq.~(\ref{pidhard}) $Y_i(y_t)$ for \aa\ collisions is, in effect, a unit-normal  function $\hat H_{0i,AA}(y_t)$ that is not directly comparable with $R_{AA}(y_t)$ at higher \pt. Details are discussed in Sec.~\ref{ratcompare}. As noted in connection with Eq.~(\ref{yhard}) it is preferable to replace $x(n_s)$ with $x_{pp} \equiv \alpha \bar \rho_{spp}$ so that $Y_i(y_t)$ can be compared directly with \pp\ model function $\hat H_{0i,pp}(y_t)$ as described in relation to Eq.~(\ref{raax}).

 \subsection{$\bf p$-$\bf Pb$ TCM PID spectrum parameters} \label{pidfracdata}
 
The TCM for \ppb\ data does not represent imposition of {\em a priori} physical models but does assume approximate {\em linear superposition} of \pn\ collisions within \ppb\ collisions consistent with no significant jet modification. That assumption is confirmed by the results of analysis.
 
Table~\ref{pidparams} shows TCM model parameters for hard component $\hat H_0(y_t)$ (first three) and soft component $\hat S_0(y_t)$ (last two). Hard-component model parameters vary slowly but significantly with hadron species. Centroids $\bar y_t$ shift to larger \yt\ with increasing hadron mass. Widths $\sigma_{y_t}$ are substantially larger for mesons than for baryons. Only $K_s^0$ and $\Lambda$ data extend to sufficiently high \pt\ to determine exponent $q$ which is substantially larger for baryons than for mesons. The combined centroid, width and exponent trends result in near coincidence among the several species for larger \yt. Evolution of TCM hard-component parameters with hadron species is consistent with measured PID FFs (refer to Fig.~7 of Ref.~\cite{ppbpid}) and with a common underlying parton (jet) spectrum for all TCM hard components.
  
 \begin{table}[h]
 	\caption{TCM model parameters for unidentified hadrons $h$ from Ref.~\cite{alicetomspec} and for identified hadrons from 5 TeV \ppb\ collisions from Ref.~\cite{ppbpid}: hard-component parameters $(\bar y_t,\sigma_{y_t},q)$ and soft-component parameters $(T,n)$. Numbers without uncertainties are adopted from a comparable hadron species with greater accuracy (e.g.\ $p$ vs $\Lambda$). 
 	}
 	\label{pidparams}
 	\begin{center}
 		\begin{tabular}{|c|c|c|c|c|c|} \hline
 			& $\bar y_t$ & $\sigma_{y_t}$ & $q$ & $T$ (MeV) &  $n$  \\ \hline
 			$ h $     &  $2.64\pm0.03$ & $0.57\pm0.03$ & $3.9\pm0.2$ & $145\pm3$ & $8.3\pm0.3$ \\ \hline
 			$ \pi^\pm $     &  $2.52\pm0.03$ & $0.56\pm0.03$ & $4.0\pm1$ & $145\pm3$ & $8.5\pm0.5$ \\ \hline
 			$K^\pm$    & $2.65$  & $0.58$ & $4.0$ & $200$ & $14$ \\ \hline
 			$K_s^0$          &  $2.65\pm0.03$ & $0.58\pm0.02$ & $4.0\pm0.2$ & $200\pm5$ & $14\pm2$ \\ \hline
 			$p$        & $2.92\pm0.02$  & $0.47$ & $4.8$ & $210\pm10$ & $14\pm4$ \\ \hline
 			$\Lambda$       & $2.96\pm0.02$  & $0.47\pm0.03$ & $4.8\pm0.5$  & $210$ & $14$ \\ \hline	
 		\end{tabular}
 	\end{center}
 \end{table}
  
 Soft-component model parameter $T \approx 145$ MeV for pions is the same as that for unidentified hadrons found to be universal over all A-B collision systems and collision energies~\cite{alicetomspec}. The values for higher-mass hadrons are substantially larger. L\'evy exponent $n \approx 8.5$ for pions is also consistent with that for unidentified hadrons at 5 TeV and has a  $\log(\sqrt{s}/\text{10 GeV})$ energy dependence~\cite{alicetomspec}. Soft-component exponent $n$ values for more-massive hadrons are not well-defined because the hard-component contribution is much larger than for pions. Varying $n$ then has little impact on the overall spectra.
  
 \begin{table}[h]
 	\caption{TCM model parameters for identified hadrons from 5 TeV \ppb\ collisions~\cite{ppbpid}. Numbers without uncertainties are adopted from a comparable hadron species with greater accuracy. Parameters $ \bar p_{tsi}$ and $\bar p_{th0i}$ are determined by model functions $\hat S_{0i}(y_t)$ and $\hat H_{0i}(y_t)$ with parameters from Table~\ref{pidparams}.  $h$ represents results for unidentified hadrons.
 	}
 	\label{otherparams}
 	\begin{center}
 		\begin{tabular}{|c|c|c|c|c|} \hline
 			&   $z_0$    &  $z_h / z_s$ &   $ \bar p_{ts}$ (GeV/c)  & $ \bar p_{th0}$ (GeV/c)  \\ \hline
 			$ h$        &  $\equiv 1$  & $\equiv 1$  & $ 0.40\pm0.02$ &    $1.30\pm0.03$  \\ \hline
 			$ \pi^\pm$        &   $0.70\pm0.02$  & $0.8\pm0.05$  & $0.40\pm0.02$ &    $1.15\pm0.03$  \\ \hline
 			$K^\pm $   &  $ 0.125\pm0.01$   &  $2.8\pm0.2$ &  $0.60$&  $1.34$   \\ \hline
 			$K_s^0$        &  $0.062\pm0.005$ &  $3.2\pm0.2$ &  $0.60\pm0.02$ &   $1.34\pm0.03$  \\ \hline
 			$p $        & $ 0.07\pm0.005$    &  $7.0\pm1$ &  $0.73\pm0.02$&   $1.57\pm0.03$   \\ \hline
 			$\Lambda $        &  $0.037\pm0.005$    & $7.0$ &   $0.76\pm0.02$ &    $1.65\pm0.03$ \\ \hline	
 		\end{tabular}
 	\end{center}
 \end{table}
  
Table~\ref{otherparams} shows PID parameters $z_0$ and $z_h / z_s$ for five hadron species that are determined from \ppb\ PID spectrum data as fixed values independent of centrality.  The choice to hold $z_0$ and $z_h / z_s$ fixed rather than $z_s$ and $z_h$ separately arises from  PID spectrum data structure as follows:
Given the TCM expression in Eq.~(\ref{pidspectcm}) the correct normalization $1/\bar \rho_{si}$ should result in \ppb\ data spectra coincident with $\hat S_0(y_t)$ as $y_t \rightarrow 0$ for all centralities. That condition is not achieved by fixing $z_{si}$. Empirically, the required TCM condition is met by holding $z_{hi} / z_{si}$ and $z_{0i}$ fixed as described in the previous subsection. 

The above remarks apply to \ppb\ data for which spectrum hard components are {\em observed} to make a negligible contribution to spectra at lower \pt\ (e.g.\ $< 0.2$ GeV/c)~\cite{ppbpid}. In contrast, as demonstrated below, hard components for \pbpb\ spectra may extend to substantially lower \pt\ due to jet modification in more-central collisions. The \ppb\ values are therefore retained unchanged to provide a reference for \pbpb\ spectrum data.

\subsection{Terminology and data-model comparisons}

Application of the TCM to spectrum data requires certain distinctions between data structures and model elements. Equation~(\ref{rhotcm}) represents both model functions and the inferred structure of spectrum data based on empirical analysis of spectrum evolution with \nch\ and collision energy~\cite{ppprd,alicetomspec}. That process is not based on {\em a priori} assumptions about spectrum structure. Implicit in the analysis is the empirical determination that for \pp\ collisions $\bar \rho_h \approx \alpha \bar \rho_s^2$ with percent accuracy~\cite{ppprd}. Given spectrum structure suggested by Eq.~(\ref{rhotcm}) normalized spectra $\bar \rho_0(y_t) / \bar \rho_s$ should have, in the limit $\bar \rho_{s} \rightarrow 0$, an asymptotic form represented by symbol $\hat S_0(y_t)$, a data ``soft component''  that can in turn be described by a model function also represented by $\hat S_0(y_t)$. Subtracting model function $\hat S_0(y_t)$ from data ratio $\bar \rho_0(y_t) / \bar \rho_s$ leads to a data ``hard component'' represented by the expression  $(\bar \rho_h / \bar \rho_s)\hat H_{0}(y_t)$ that can in turn be described by a hard-component model function also represented by $\hat H_0(y_t)$. One could introduce different symbols for data structures and model elements, but proliferation of symbols could also lead to confusion. Efforts are made in the text to maintain a clear distinction between data and models.

\section{200 $\bf GeV$ $\bf Au$-$\bf Au$ PID spectrum TCM}
\label{auau}

As an illustration, the PID TCM described in the previous section is applied to spectra from 200 GeV \auau\ collisions, the basis for the first \auau\ TCM analysis as reported in Ref.~\cite{hardspec}. The earlier analysis provided valuable experience and important results but can be substantially improved based on recent results from \ppb\ PID spectra as reported in Ref.~\cite{ppbpid}. The \auau\ analysis was based on the expression [compare with Eq.~(\ref{ppspectcm})]
\bea  \label{auaunorm}
\frac{2}{N_{part}} \bar \rho_{0}(y_t;n_{ch})  
&=& \bar \rho_{sNN}  \hat S_{0NN}(y_t) +   \nu \bar \rho_{hAA} \hat H_{0AA}(y_t)
\nonumber \\
&=& S_{NN}(y_t) + \nu H_{AA}(y_t)
\\ \nonumber
&\rightarrow &  S_{pp}(y_t) + H_{pp}(y_t)~~\text{for \pp}.
\eea
In Ref.~\cite{hardspec} $S_{NN}(y_t)$ is assumed to be $S_{pp}(y_t)$, a constant times the $\hat S_{0}(y_t)$ L\'evy distribution derived from 200 GeV \pp\ collisions~\cite{ppprd}. Subtracting a fixed $S_{NN}(y_t)$ model from spectrum data in the form of Eq.~(\ref{auaunorm}) should then reveal hard-component data in the form $\nu H_{AA}(y_t)$  that can be compared with a \pp\ reference in the form $ H_{pp}(y_t)$ to generate an alternative to spectrum ratio $R_{AA}(p_t)$. 

For \pp\ collisions $\bar \rho_s$ is uniquely determined from charge density $\bar \rho_0 = n_{ch} / \Delta \eta$ by the quadratic relation $\bar \rho_0 = \bar \rho_s + \bar \rho_h \approx \bar \rho_s + \alpha \bar \rho_s^2$ derived from spectrum data as first reported in Ref.~\cite{ppprd}. For \ppb\ collisions $\bar \rho_{sNN}$ and associated geometry parameters are inferred accurately, relative to charge density $\bar \rho_0$, from ensemble mean \mmpt\ data as described in Ref.~\cite{tommpt}. That analysis is based on the assumption that jet formation remains unmodified for any \ppb\ collision conditions, which has been verified via \ppb\ PID spectrum analysis as reported in Ref.~\cite{ppbpid}.

For the \auau\ analysis in Ref.~\cite{hardspec} the assumption was made that averaged over many \nn\ collisions $\bar \rho_{sNN}$ remains unchanged with increasing \auau\ centrality. However, PID spectra normalized by $N_{part}/2$ alone do not coincide in the limit $p_t \rightarrow 0$, which was interpreted as a spectrum normalization problem  noted in App.\ B of Ref.~\cite{hardspec}. The spectra were therefore rescaled by {\em ad hoc} factors $1+a(\nu-1)$ with different $a$ for pions and protons adjusted to achieve coincidence of spectra for  $p_t \rightarrow 0$. 

The PID analysis of \ppb\ PID spectra in Ref.~\cite{ppbpid} revealed that in addition to $N_{part}/2$ PID spectra must also be rescaled by factor $\bar \rho_{sNNi}$ as defined in Eq.~(\ref{rhosi}). The additional factor $q_i(n_s)$ is strongly centrality dependent and plays a major role in PID TCM analysis.  
For the definition of $\bar \rho_{si}$ in  Eq.~(\ref{rhosi}) $\nu$ (derived from $N_{part}$ and $N_{bin}$) is obtained for \auau\ collisions by a power-law approximation defined in Ref.~\cite{powerlaw} that accurately reproduces results from Monte Carlo Glauber simulations for 200 GeV \auau\ collision geometry. $x(\nu)$ as defined by
\bea \label{xnuauau}
x(\nu) &=& 0.015 + 0.08\left\{1 + \tanh[(\nu - 3)/0.8]\right\}/2
\eea
describes a ``sharp transition'' in jet-related data features~\cite{anomalous} near $\nu = 3$ -- from a \pp\ trend $x = \alpha \bar \rho_{spp} = 0.006 \times 2.5 = 0.015$ to $x \approx 0.095$ that describes yield data for more-central 200 GeV \auau\ collisions consistent with results reported in Ref.~\cite{kn}.%
\footnote{In Ref.~\cite{kn} the right-hand side of Eq.~(\ref{nchppb}) (first line) above is expressed as $\bar \rho_{0pp}[(1-x)(N_{part}/2) + xN_{bin}]$ with $x$ for central 130 GeV \auau\ collisions estimated as $0.09 \pm 0.03$. That value is interpreted in Ref.~\cite{kn} to indicate that more than one-third of hadrons in central 130 GeV \auau\ collisions are jet-related.} 

\subsection{Pion spectra}

For pion spectra from 5 TeV \ppb\ collisions PID parameters are $z_0 = 0.8$ and $z_{h}/z_s = 0.8$. Those parameters included in  Eq.~(\ref{rhosi}) cause pion $\bar \rho_{sNNi}$ to {\em increase} with increasing centrality which worsens the normalization problem for \auau\ pion data, whereas the \ppb\ data {\em require} increasing $\bar \rho_{sNNi}$.  Thus, just as for the \auau\ analysis in Ref.~\cite{hardspec} an additional normalization factor $1 + 0.08(\nu - 1)$ is introduced relative to the \ppb\ pion TCM to ensure coincident normalized spectra for $p_t \rightarrow 0$.

\begin{figure}[h]
	\includegraphics[width=3.3in]{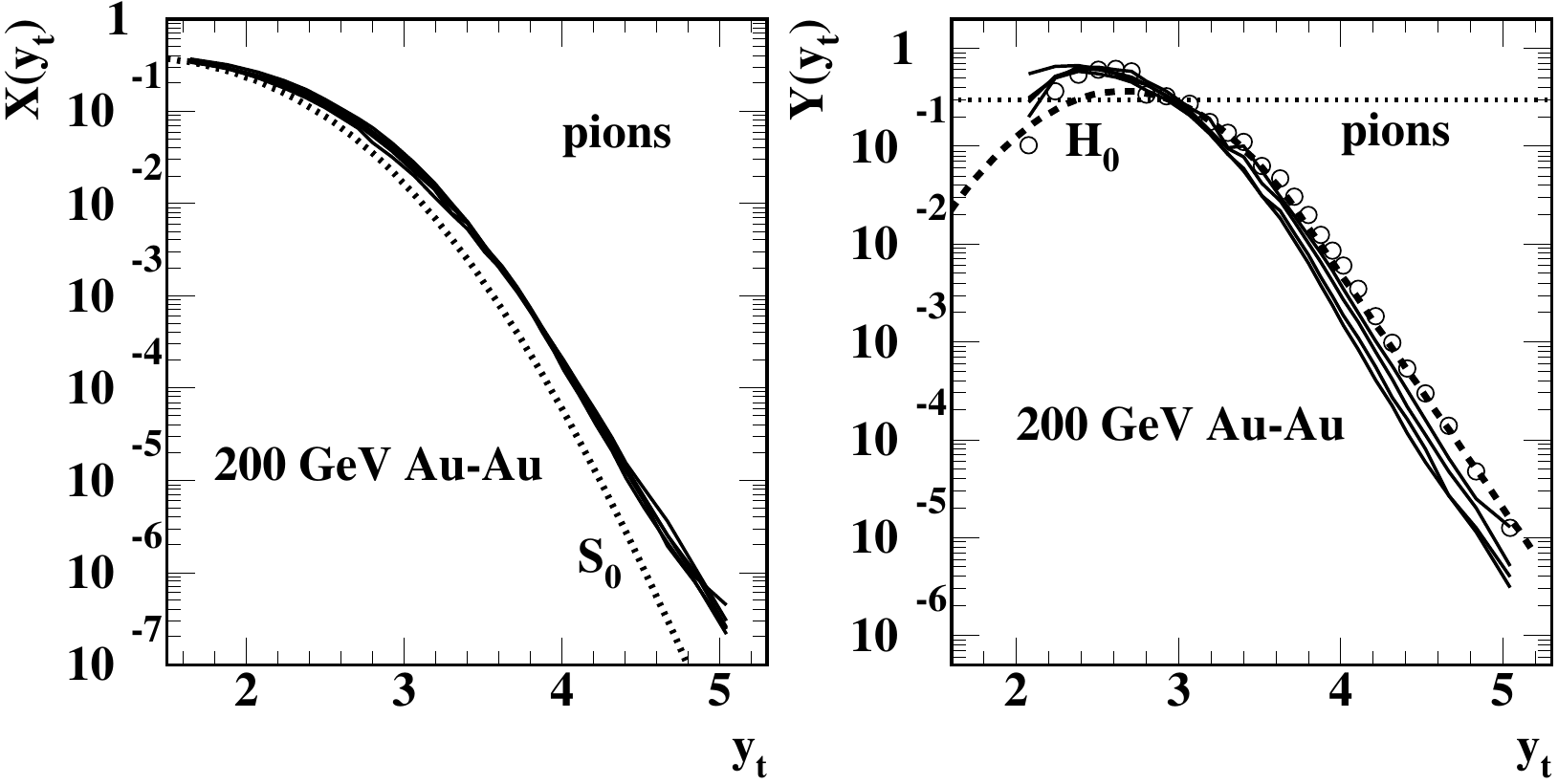}
	\caption{\label{pionauau}
		Left: pion transverse rapidity \yt\ spectra for five centrality classes of 200 GeV \auau\ collisions in the form $X(y_t)$ defined by the third line of Eq.~(\ref{pidspectcm})~\cite{hardspec}.
		Right: Pion spectrum hard components in the form $Y(y_t)$ defined by Eq.~(\ref{pidhard}). The open circles are peripheral 60-80\% data. $\hat S_0(y_t)$ (dotted) and $\hat H_0(y_t)$ (dashed) are TCM model functions. The horizontal dotted line at right is a normalization reference.
	} 
\end{figure}

Figure~\ref{pionauau} (left) shows pion spectra from five centrality classes of 200 GeV \auau\ collisions: 0-12\%, 10-20\%, 20-40\%, 40-60\%, 60-80\% normalized to the form $X_i(y_t)$ defined in  Eq.~(\ref{pidspectcm}). The bold dotted curve is unit-integral soft model function $\hat S_0(y_t)$ with L\'evy-distribution parameters $T = 145$ MeV and $n = 12.8$ as inferred from 200 GeV \pp\ unidentified-hadron spectra in Ref.~\cite{ppprd}.

Figure~\ref{pionauau} (right) shows pion hard components in the form $Y(y_t)$ as defined by Eq.~(\ref{pidhard}). The unit-normal hard-component reference is $\hat H_0(y_t)$ (dashed) with parameters $\bar y_t = 2.65$, $\sigma_{y_t} = 0.45$ and $q = 5.5$, also  as inferred from 200 GeV \pp\ unidentified-hadron spectra in Ref.~\cite{ppprd}. Significant differences from the 5 TeV \ppb\ hard-component parameter values in Table~\ref{pidparams} are expected given the substantial difference between underlying jet spectra at 200 GeV and 5 TeV~\cite{alicetomspec,jetspec2,fragevo}. In Eq.~(\ref{pidhard}) the preferred reference is the hard component for unmodified jets from \pp\ collisions, with $x(\nu) \rightarrow \alpha \bar \rho_{spp} = 0.015$ and $z_{h}/z_s = 0.8$ as for the \ppb\ analysis. If there were no jet modification \auau\ $Y(y_t)$ should then coincide with $\hat H_0(y_t)$. The horizontal dotted line in this and comparable plots is a reference at 0.3 to check the normalization procedure. If $\hat H_0(y_t)$ has a narrower width the amplitude at the mode should be slightly above the reference line as in this case for \auau\ pions. For LHC data the converse is true.

As observed in Ref.~\cite{hardspec} spectra displayed in the form $Y(y_t)$ exhibit high-\pt\ suppression above 4 GeV/c ($y_t \approx 4$) similar to conventional spectrum ratio $R_{AA}$ but also exhibit a corresponding {\em large enhancement} below 4 GeV/c that is concealed by $R_{AA}$. The centrality sequence near $y_t = 5$ is generally opposite the sequence at $y_t = 2$ supporting a common origin for balanced suppression and enhancement~\cite{fragevo}. The most peripheral data (open circles) match the reference closely above the mode near $y_t = 2.7$, consistent with no jet modification and also  with results from 2D angular correlation analysis~\cite{anomalous}.

\subsection{Proton spectra}

\begin{figure}[h]
	\includegraphics[width=3.3in]{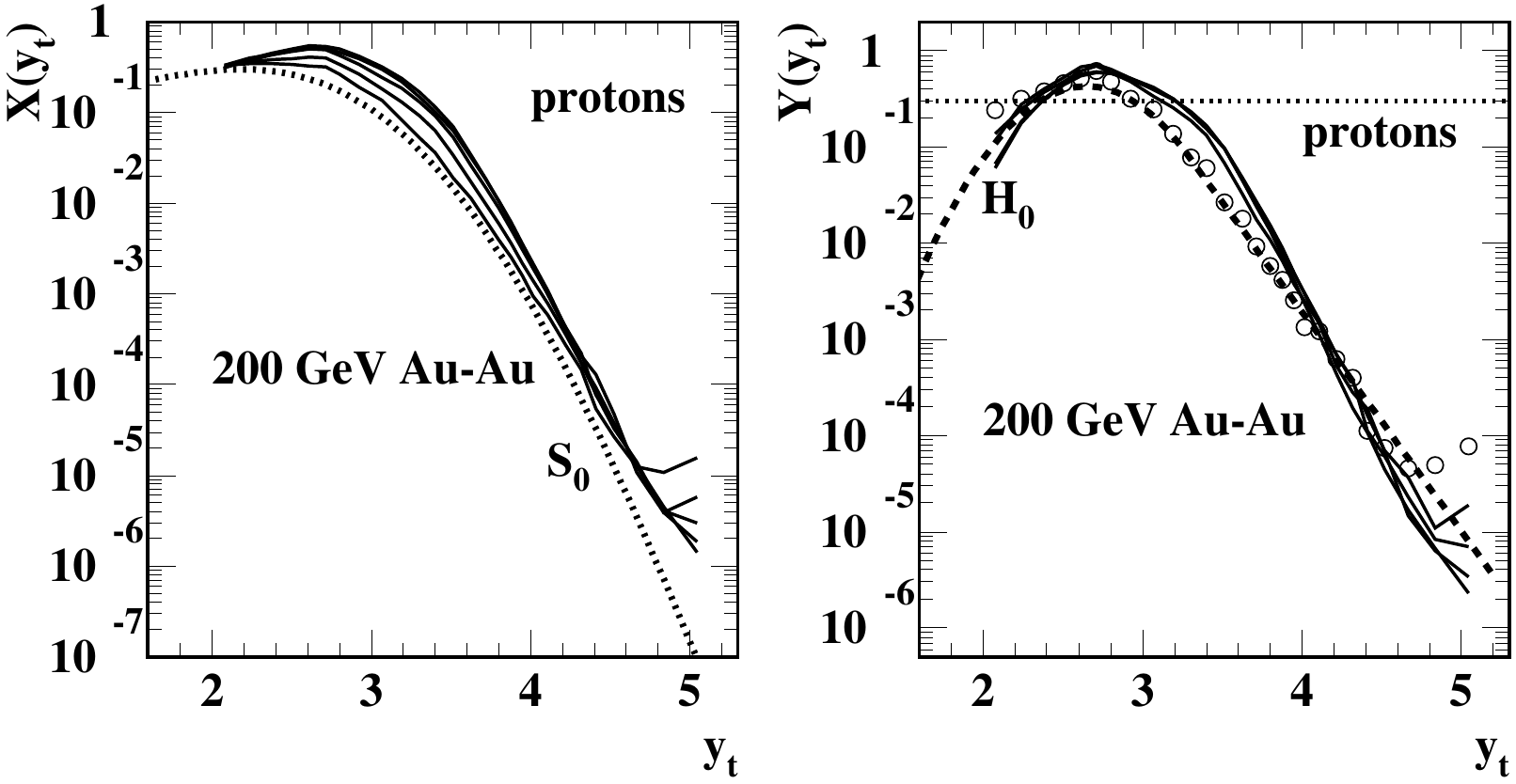}
	\caption{\label{protonauau}
		Left: Proton \yt\ spectra for five centrality classes of 200 GeV \auau\ collisions in the form $X(y_t)$ defined by the third line of Eq.~(\ref{pidspectcm})~\cite{hardspec}.
Right: Proton spectrum hard components in the form $Y(y_t)$ defined by Eq.~(\ref{pidhard}). The open circles are peripheral 60-80\% data. $\hat S_0(y_t)$ (dotted) and $\hat H_0(y_t)$ (dashed) are TCM model functions. The horizontal dotted line at right is a normalization reference.
	}  
\end{figure}

Figure~\ref{protonauau} (left) shows proton spectra from 200 GeV \auau\ collisions in the same format as for pion spectra. The $\hat S_0(y_t)$ L\'evy distribution (dotted) parameters are $T = 210$ MeV and $n = 14$ as for 5 TeV \ppb\ proton spectra in Table~\ref{pidparams}. Those numbers agree within uncertainties with 224 MeV and 17 from Ref.~\cite{hardspec}. The proton data are relatively insensitive to parameter $n$ because the hard component (jet fragments) dominates the spectra above $y_t = 2.3$ ($p_t \approx 0.7$ GeV/c). 
For $\bar \rho_{si}$ as defined in  Eq.~(\ref{rhosi}) the geometry parameters and $x(\nu)$ remain the same as for pion spectra. $z_0 = 0.065$ corresponds within uncertainties to 0.07 for \ppb\ data, but for proton spectra normalized via Eq.~(\ref{rhosi}) $z_h / z_s \rightarrow 3.5$ rather than 7 as for \ppb\ collisions, a major difference. 

Figure~\ref{protonauau} (right) shows proton spectrum hard components in the form $Y(y_t)$ compared to unit-normal TCM hard-component model $\hat H_0(y_t)$ (dashed). As for pion data the \pp\ value $x \rightarrow \alpha \bar \rho_{spp} = 0.015$ is used for Eq.~(\ref{pidhard}), and $z_h / z_s = 7$ is in this case consistent with the \ppb\ analysis but {\em not with the spectrum normalization noted above}.  Model parameters for $\hat H_0(y_t)$, adopted from the \auau\ analysis of Ref.~\cite{hardspec}, are $\bar y_t = 2.65$, $\sigma_{y_t} = 0.35$ and $q = 5.0$. The smaller proton width (relative to pion $\sigma_{y_t} = 0.45$) results in the amplitude at the mode being significantly above the dotted reference line. Just as for pion spectra substantial differences from 5 TeV \ppb\ parameters are expected due to the different underlying jet spectra. It is notable that in Ref.~\cite{hardspec} the reference for proton $H_{AA}(y_t)$ in Eq.~(\ref{auaunorm}) is $0.118 \hat H_0(y_t)$. Based on experience with the \ppb\ analysis reported in Ref.~\cite{ppbpid} $(z_h / z_s) x \rightarrow 7 \times 0.015 = 0.105$ in Eq.~(\ref{pidhard}) suggests the origin of that empirical value.

Again, the most-peripheral data (open circles) agree with the model within data uncertainties signaling no jet modification and consistent with evolution of jet-related angular correlations~\cite{anomalous}. However, the evolution of suppression at higher \pt\ and enhancement at lower \pt\ is notably different for protons as reported in Ref.~\cite{hardspec}. Rather than being located well below the hard-component mode as in the case of pions enhancement for protons, centered near $y_t = 3.4$ ($p_t \approx 2$ GeV/c), lies above the mode. The hard component below the mode is consistent with no jet modification for any \auau\ centrality. The two results suggest the possibility that jet modification for a given species has a lower bound consistent with hadron mass.

\subsection{Hard-component ratios $\bf r_{AA}(y_t)$}

\begin{figure}[h]
	\includegraphics[width=3.3in]{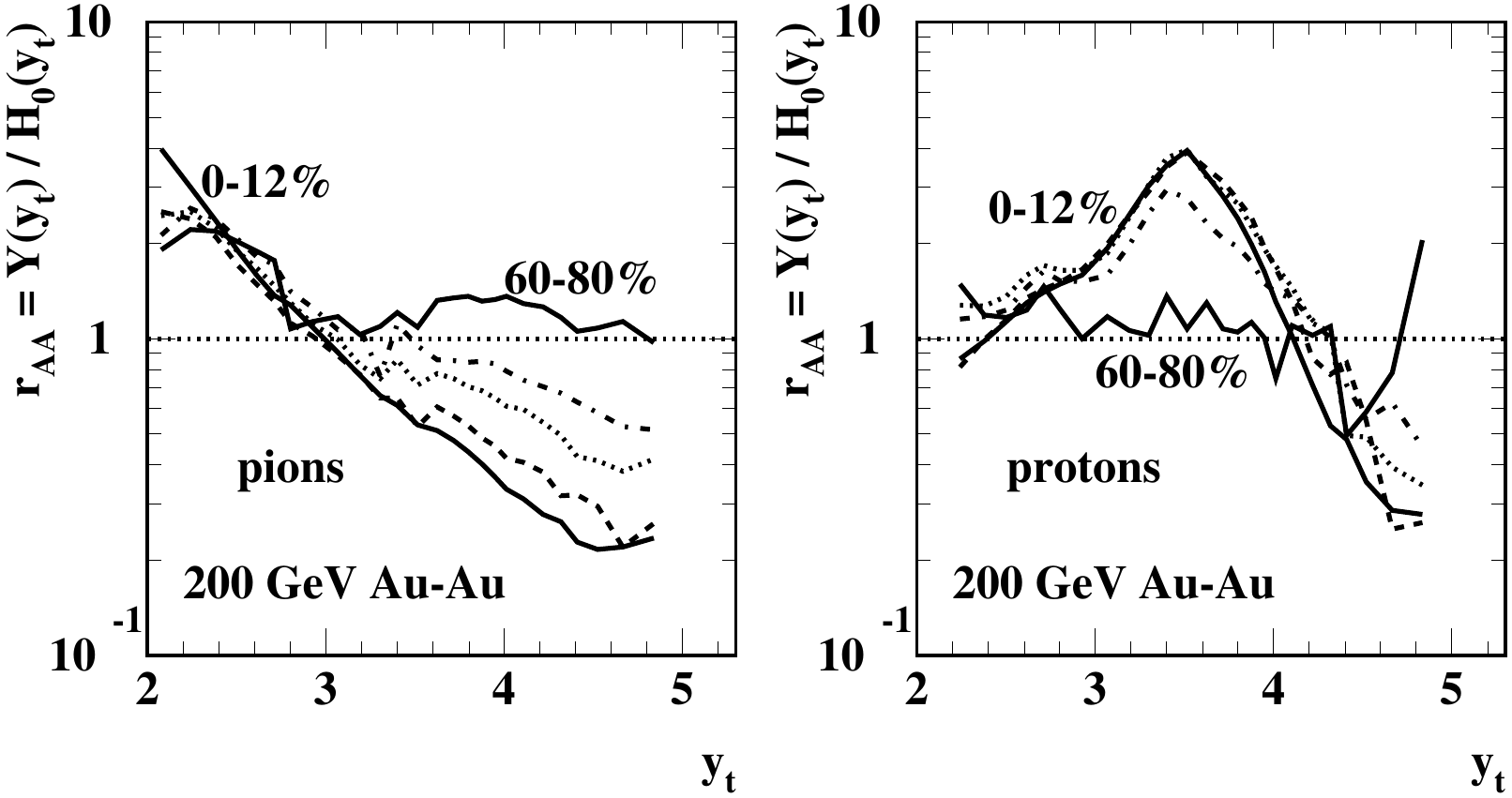}
	\caption{\label{raaauau}
		Spectrum hard-component ratios $r_{AA}(y_t)$ defined by Eq.~(\ref{raax}) for pions (left) and protons (right) from five centrality classes of 200 GeV \auau\ collisions. Deviations from unity indicate jet modification vs a \nn\ linear superposition model.
	}   
\end{figure}

Figure~\ref{raaauau} shows ratios $r_{AA}(y_t)$ for pions and protons from 200 GeV \auau\ collisions.  The results are generally quantitatively consistent with Ref.~\cite{hardspec}. The main results of the \auau\ TCM analysis are (a) the large pion enhancement below 4 GeV/c ($y_t \approx 4$) that is concealed by conventional spectrum ratio $R_{AA}(p_t)$, and (b) similar enhancement for protons but {\em above} the hard-component mode (explaining the baryon/meson ``puzzle''). Jet modification in spectra is negligible for the 60-80\% centrality class consistent with jet-related angular correlations~\cite{anomalous}.

 As noted, close correlation of suppression at higher \yt\ with enhancement at lower \yt\ strongly suggest that there is a common mechanism for both: modification of jet formation in more-central collisions. Thus, jet modification cannot be fully understood without examining jet fragment distributions down to at least 0.5 GeV/c ($y_t \approx 2$).

\section{2.76 $\bf TeV$ $\bf Pb$-$\bf Pb$ PID spectrum $\bf TCM$} \label{pidspec}

2.76 TeV \pbpb\ and \pp\ PID spectrum data from Sec.~\ref{alicedata} are analyzed via a TCM developed for \ppb\ PID spectra in Ref.~\cite{ppbpid}. \pbpb\ spectrum hard components for three hadron species are isolated and compared to hard components for \ppb\ and \pp\ collisions as references. The spectrum data are first described, then \pbpb\ geometry and TCM model parameters are discussed. 80-90\% central 2.76 TeV \pbpb\ data from  Ref.~\cite{alicepbpbspecx} are included as a reference, although the \pt\ acceptance is limited.

\subsection{Pb-Pb collision-geometry parameters} \label{geomparams}

To implement the PID spectrum TCM in Sec.~\ref{pidspecc} for 2.76 TeV \pbpb\ collisions, models for $x(n_s)$ and $\nu(n_s)$ are required. For \pp\ collisions $x(n_s) = \alpha \bar \rho_s$ applies and $\nu \equiv 1$. For \ppb\ collisions $x(n_s)$ and $\nu(n_s)$ are derived from \mmpt\ data as reported in Ref.~\cite{tommpt}. For \aa\ collisions there is considerable uncertainty about $x(n_s)$ and $\nu(n_s)$ modeling.%
\footnote{The number of binary \nn\ collisions $N_{bin}$ derived from a Glauber MC applied to A-B collisions is suspect based on a comparison of TCM vs Glauber \ppb\ centrality trends reported in Ref.~\cite{tomglauber}.} 
However, the {\em product} $x(n_s) \nu(n_s)$ can be inferred from the measured centrality trend of midrapidity charge density $\bar \rho_0$ via Eq.~(\ref{nchppb}) (second line) in the form
\bea \label{npartrho}
(2/N_{part}) \bar \rho_0 &=& \bar \rho_{sNN} [1 + x(n_s) \nu(n_s)].
\eea
Given measured charge density $\bar \rho_0$, a Glauber MC estimate of $N_{part}$ and an estimate of $\bar \rho_{sNN}$ from \pp\ data the product $x(n_s) \nu(n_s)$ can be derived directly from data.

\begin{figure}[h]
	\includegraphics[width=1.65in]{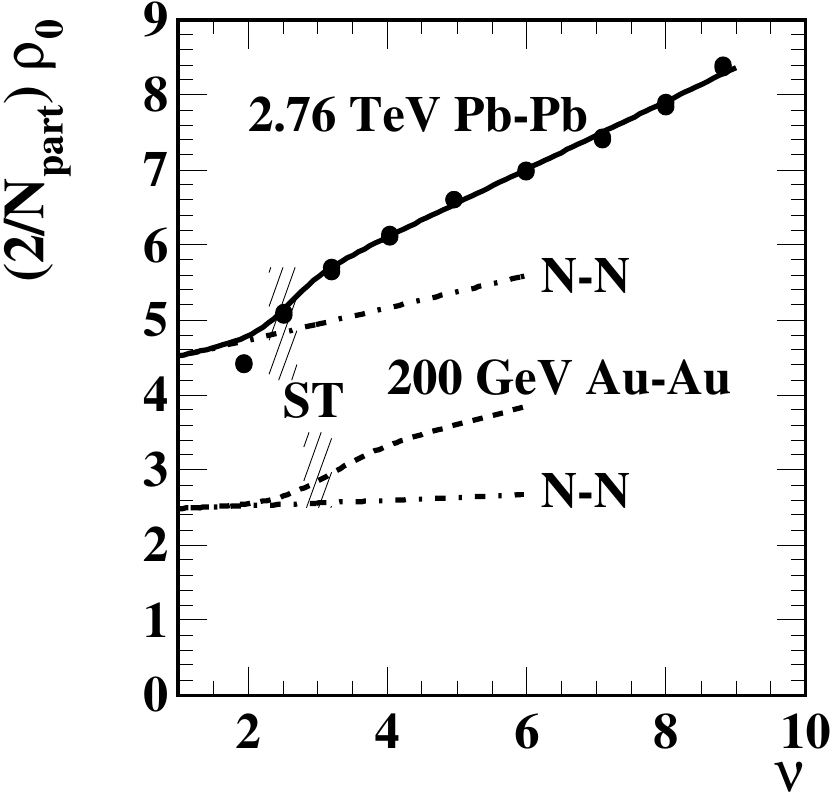}
	\includegraphics[width=1.65in]{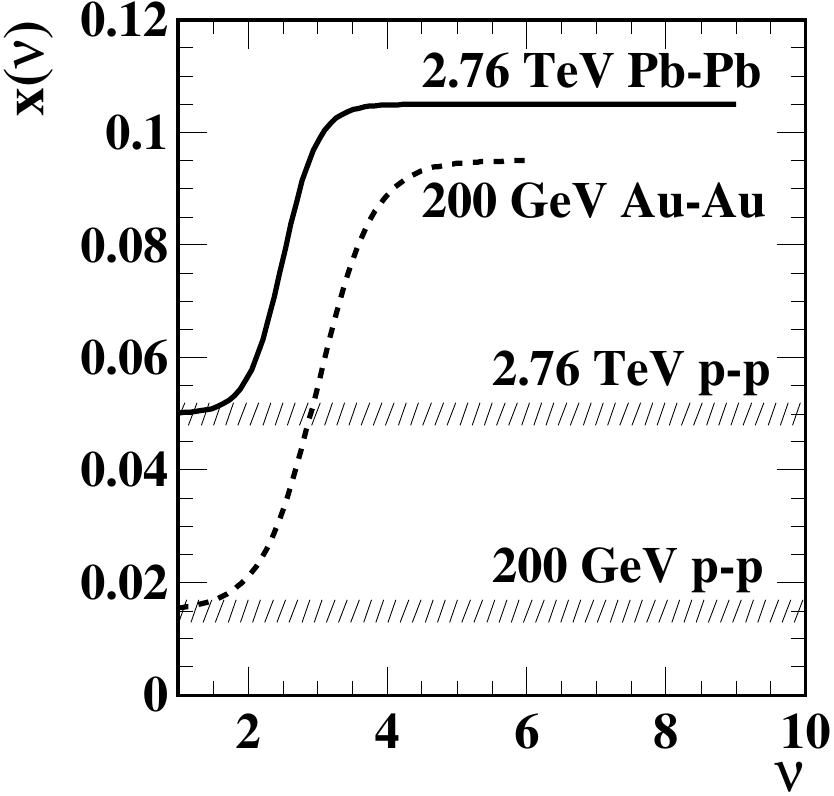}
	\caption{\label{900a}
		Left: Unidentified-hadron charge density $\bar \rho_0$ per participant-nucleon pair from Ref.~\cite{alicepbpbyields} (points) vs number of binary \nn\ collisions per participant pair $\nu$. The solid curve is defined by Eq.~(\ref{npartrho}). The dashed curve derived from Eq.~(\ref{xnuauau}) represents 200 GeV \auau\ collisions. The dash-dotted curves correspond to \nn\ linear superposition (no jet modification).
		Right: Hard/soft density ratio $x(\nu)$ for 2.76 TeV \pbpb\ collisions defined by Eq.~(\ref{16}) (solid). The dashed curve represents 200 GeV \auau\ collisions with $x(\nu)$ defined by Eq.~(\ref{xnuauau}).
	}  
\end{figure}

Figure~\ref{900a} (left) shows quantity $(2/N_{part}) \bar \rho_0$ for 2.76 TeV \pbpb\ collisions (solid points) as reported in Ref.~\cite{alicepbpbyields}. Values of $\nu$ are derived from Table~\ref{ppbparams1}. The solid curve is determined by $\bar \rho_{sNSD} = 4.3$ and an expression for $x(\nu)$ described in the right panel. The dashed curve representing 200 GeV \auau\ data is similarly constructed. The dash-dotted lines are extrapolations corresponding to \nn\ linear superposition within \aa\ collisions (no jet modification, $x \equiv x_{pp}$). The value $\bar \rho_{sNSD} = 4.3$ disagrees with the expectation $\bar \rho_{sNSD} \approx 0.81 \ln(\sqrt{s} / \text{10 GeV}) \approx 4.55$ for 2.76 TeV \pp~\cite{alicetomspec}. However, the $(2/N_{part}) \bar \rho_0$ data from Ref.~\cite{alicepbpbyields} require the value 4.3. The predicted value 4.55  is used for the PID spectrum analysis below. The hatched bands identify {\em sharp transitions} in jet characteristics (i.e.\ onset of jet modification, see below)~\cite{anomalous}.

Figure~\ref{900a} (right) shows $x(\nu)$ for 2.76 TeV \pbpb\ collisions (solid) given by the expression
\bea \label{16}
x(\nu) &=& 0.05 + 0.055 \left\{1+ \tanh[2(\nu - 2.5)] \right\}/2,
\eea
where the constant term is $x_{pp} \equiv \alpha \bar \rho_{spp} = 0.0113 \times 4.55 \approx 0.05$ (upper hatched band) and the coefficient of the second term 0.055 is adjusted to describe $\bar \rho_0$ data, as is the sharp-transition (ST) point 2.5. That expression differs from Eq.~(15) of Ref.~\cite{tommpt} because in the latter case 200 GeV $\nu$ values were used to describe LHC $\bar \rho_0$ data.  The dashed curve representing comparable 200 GeV \auau\ data is defined by Eq.~(\ref{xnuauau}).

\begin{figure}[h]
	\includegraphics[width=1.69in]{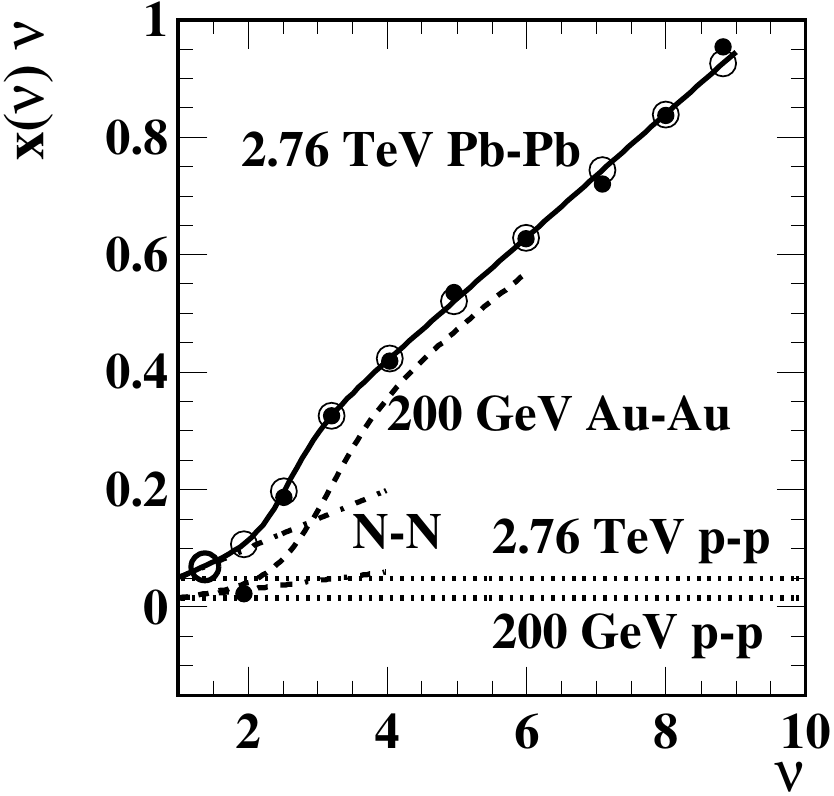}
	\includegraphics[width=1.61in]{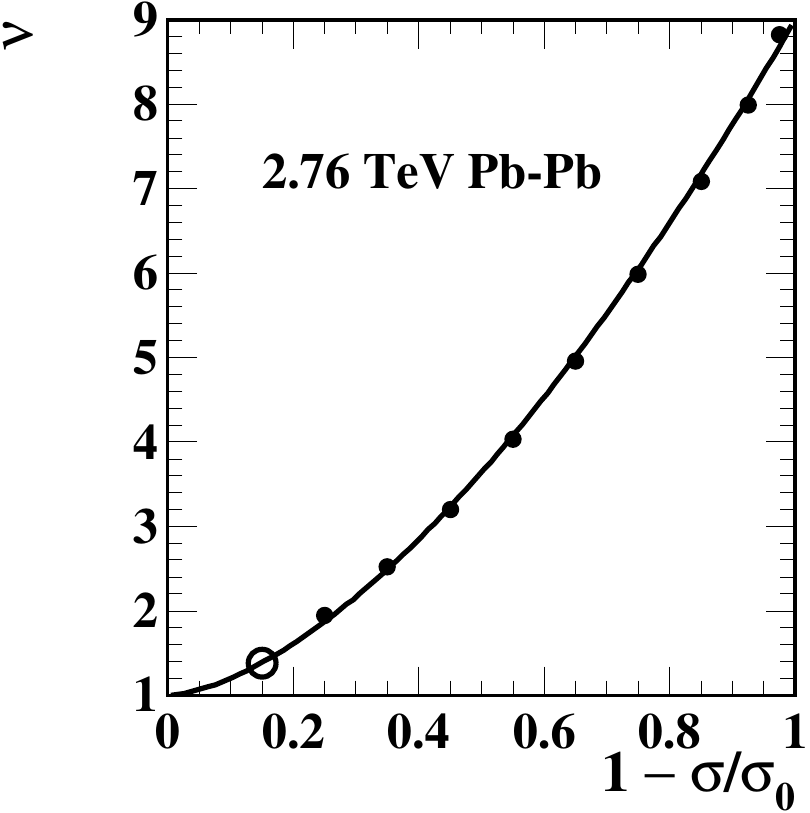}
	\caption{\label{900d}
Left: The product $x(n_s) \nu(n_s)$ vs $\nu$ required for the spectrum TCM as defined in Sec.~\ref{pidspecc} for 2.76 TeV \pbpb\ collisions (solid) and 200 GeV \auau\ collisions (dashed). The solid points and solid curve are transformed from Fig.~\ref{900a} via Eq.~(\ref{npartrho}). The open circles represent $x(n_s) \nu(n_s)$  values adopted for the present study. The dashed curve represents 200 GeV \auau\ collisions.
The dash-dotted lines represent \nn\ linear superposition (no jet modification).
		Right: Centrality parameter $\nu$ vs fractional cross section $1 - \sigma/\sigma_0$ (points) derived from  Ref.~\cite{alicenonpidspec}. The curve is used to extrapolate $\nu$ to 80-90\% central \pbpb\ collisions (open circle, see text).
	}  
\end{figure}

Figure~\ref{900d} (left) shows the product $x(n_s)\nu(n_s)$ (solid points) derived from solid points in Fig.~\ref{900a} (left) and Eq.~(\ref{npartrho}) assuming $\bar \rho_{sNN} = 4.3$. The solid curve is the solid curve from Fig.~\ref{900a} (left) also transformed via Eq.~(\ref{npartrho}). The dashed curve for 200 GeV \auau\ is similarly transformed. The dotted lines represent values $x_{pp} = \alpha \bar \rho_{spp}$ for RHIC and LHC \pp\ collisions. The advantage of such $x(n_s)\nu(n_s)$ product data, as noted, is that only $N_{part}$ and $\bar \rho_{sNN}$ are required for their derivation. The more-uncertain $N_{bin}$ is not utilized. The product values contributing to the PID spectrum analysis below are denoted by the open circles. The final values for product $x(\nu)\nu$ from six centralities of \pbpb\ collisions are  0.152  0.375 0.575 0.744 0.839 0.926. Note that the 70-80\% data point lies well below the inferred trend.

Figure~\ref{900d} (right) shows parameter $\nu$ vs centrality measured by fractional cross section in the form $1 - \sigma / \sigma_0$, where $\sigma_0$ is the integrated total cross section for 2.76 TeV \pbpb\ collisions. The nine solid points correspond to the nine centrality classes defined in Ref.~\cite{alicenonpidspec}. The solid curve $\nu = 1 + 8(1 - \sigma / \sigma_0)^{1.6}$ describes the $\nu$ centrality trend (consistent with Table~\ref{ppbparams1}) well.  The purpose of that panel is to evaluate $\nu$ for peripheral 80-90\% data reported in Ref.~\cite{alicepbpbspecx}. The corresponding value $\nu = 1.38$ is shown as the open circle. Given that value the corresponding product value is  $x \nu \approx 0.05 \times 1.38 = 0.069$ appearing as the peripheral open circle in the left panel. Similar means were used to extrapolate the trends from Table~\ref{ppbparams1} for $N_{part}/2$ and $N_{bin}$ to 80-90\% centrality, albeit with substantially greater uncertainty than for $\nu$. The minimum value for $N_{part}/2$ is 2.5, but the actual value could be in the range 3-4. The minimum value for $N_{bin}$ is 3.5 but the actual value could be in the range 4-5.5. For this analysis the values 3 and 4 respectively are utilized.

PID spectra from two event classes (mid-central class 4 and peripheral class 7) of 5 TeV \ppb\ collisions as reported in Ref.~\cite{ppbpid} and presented in Table~\ref{rppbdata} (unprimed values) are incorporated in this analysis as references. For class 4 $x = 0.16$, $\nu = 1.26$, $ \rho_{sNN} = 14.1$ and $N_{part}/2 = 1.35$. For  class 7 $x = 0.048$, $\nu \equiv 1$, $ \rho_{sNN} = 4.2$ and $N_{part}/2 \equiv 1$. Note that $\rho_{sNN} = 4.2$ is significantly below the NSD \pp\ value 5.0~\cite{alicetomspec} suggesting that jet production is significantly suppressed for that event class.
  
\subsection{$\bf Pb$-$\bf Pb$  PID spectrum plotting formats} \label{piddiff}

In the figures below, 2.76 TeV \pbpb\ and \pp\ PID \pt\ spectra introduced in Sec.~\ref{alicedata} are replotted in panels (a) multiplied by $2\pi$ (to be consistent with $\eta$ densities from other TCM analyses) and transformed to pion transverse rapidity $y_{t\pi}$. The \pbpb\ data are plotted as solid curves and the \pp\ data are plotted as open circles. As a reference 5 TeV \ppb\ PID spectra for multiplicity classes 4 (mid-central) and 7 (peripheral), as analyzed in Ref.~\cite{ppbpid}, are plotted as bold dashed (or  dash-dotted) curves. The advantage of spectrum plots on transverse rapidity should be evident: At lower \yt\ spectra transformed to \yt\ are nearly constant, permitting precise differential normalization comparisons. At higher \yt\ spectra are well approximated by a linear trend corresponding to a power-law dependence on \pt\ as presented in a log-log format which also allows precise differential comparisons.

Panels (b) show normalized \pbpb\ spectra in the form $X_i(y_t)$ defined by Eq.~(\ref{pidspectcm}) (thin solid) compared to TCM soft components $\hat S_{0i}(y_t)$ (bold dotted). In that format relative spectrum normalizations can be examined at the percent level. Also plotted are normalized \pp\ spectra (open circles) and normalized 5 TeV \ppb\ spectra for centrality classes 4 and 7 (bold dashed or dash-dotted).

Panels (c) show \pbpb\ spectrum hard components in the form $Y_i(y_t)$ defined by Eq.~(\ref{pidhard}) (thin solid) compared to TCM hard components $\hat H_{0i}(y_t)$ (bold dotted).  Also plotted are hard components from \pp\ spectra (open circles) and 5 TeV \ppb\ spectra for centrality class 4 (bold dashed or dash-dotted). Hard components for most-peripheral \ppb\ centrality class 7 are biased by the low \nch\ condition just as observed for \pp\ in Ref.~\cite{ppprd}.

Panels (d) show hard-component ratios $r_{\text{AA}i}(y_t) \equiv Y_i(y_t) / \hat H_{0i,pp}(y_t)$ that can be related to conventional spectrum ratios $R_{\text{AA}i}(p_t)$ defined by Eq.~(\ref{raa}) and appearing in Fig.~\ref{piddata} (right panels). Whereas conventional $R_{\text{AA}i}(p_t)$ for lower \pt\ approaches a reference trend increasingly insensitive to jet structure below $p_t \approx 4$ GeV/c (solid reference curves in Fig.~~\ref{piddata}) ratio $r_{\text{AA}i}(y_t)$ is sensitive to jet structure to well below 0.5 GeV/c ($y_t \approx 2$), thereby including the mode of the hard component (representing {\em all} jet fragments) near $y_t \approx 2.7$ ($p_t \approx 1$ GeV/c).

Hard-component ratio $r_{\text{AA}i}(y_t)$ can be related to full spectrum ratio $R_\text{AAi}(y_t)$ as follows. The latter ratio approaches the following limits at low and high \yt\
\bea \label{raaa}
R_{\text{AAi}}(y_t) &=& \frac{1}{N_{bin}'} \frac{(N_{part}/2) S_{AAi} + N_{bin} H_{AAi}}{S_{ppi}(y_t) + H_{ppi}(y_t)}~~~~~
\\ \nonumber
&\rightarrow& \frac{N_{bin}}{N_{bin}'} \frac{\bar \rho_{hNNi,AA}}{\bar \rho_{hppi}} \frac{ \hat H_{0i,AA}(y_t)}{ \hat H_{0i,pp}(y_t)}~~~\text{for higher $y_t$}
\\ \nonumber
&\rightarrow& \frac{N_{part}/2}{N_{bin}'} \frac{\bar \rho_{sNNi,AA}}{\bar \rho_{sppi}} \frac{ \hat S_{0i,AA}(y_t)}{ \hat S_{0i,pp}(y_t)}~~~\text{for $y_t \rightarrow 0$}
\eea
per Eq.~(\ref{pidspectcm}). Both limits implicitly include the ratio $q_{iAA}(\nu) / q_{ipp}$ through ratios $\bar \rho_{xNNi,AA} / \bar \rho_{xppi}$ per Eq.~(\ref{rhosi}). In contrast, the  expression for $r_{\text{AA}i}(y_t)$ over {\em all} \yt\ is
\bea \label{raax}
r_{\text{AA}i}(y_t) &\equiv& \frac{Y_i(y_t)}{\hat H_{0i,pp}'(y_t)} = \frac{\nu}{\nu'} \frac{x_{AA}}{x_{pp}'} \frac{ \hat H_{0i,AA}(y_t) }{\hat H_{0i,pp}'(y_t)}
\eea
per Eq.~(\ref{pidhard}), where for both ratios primed quantities are (possibly biased) model estimates and unprimed quantities are physical data values.  The ratio $\bar \rho_{hNN,AA}/\bar \rho_{hpp}' \approx x_{AA}/x_{pp}'$ is typically $\gg 1$ for more-central \aa\ collisions.
In both expressions $\hat H_{0i,AA}(y_t)$ is a unit-normal representation of the \aa\ data hard component including whatever jet modification occurs in the \aa\ collision system. In ratio $r_{\text{AA}i}(y_t)$, $\hat H_{0i,pp}'(y_t)$ is a unit-normal TCM model function: a Gaussian + exponential tail as described in Sec.~\ref{tcmmodel}. However, in ratio $R_{\text{AA}i}(y_t)$ $\hat H_{0i,pp}(y_t)$ represents a physically measurable \pp\ data feature that may be subject to new physics or experimental bias. That difference can be assessed by replacing \aa\ spectra in $r_{\text{AA}i}(y_t)$ by a measured \pp\ spectrum to form $r_{ppi}(y_t)$ [e.g.\ the open circles in panels (d) below].  

Finally, a ``ratio of ratios'' may be defined as
\bea \label{ratrat}
\frac{r_{AAi}(y_t)}{r_{ppi}(y_t)} &\equiv&  \frac{\nu}{\nu'} \frac{x_{AA} \hat H_{0i,AA}(y_t)}{x_{pp} \hat H_{0i,pp}(y_t)} 
\eea
that can be compared with $R_{\text{AAi}}(y_t)$ at higher \yt\ but remains a strictly hard-component ratio down to low \yt.  To make a direct, quantitative comparison of that ratio with spectrum ratio $R_{AAi}(y_t)$ as defined in Eq.~(\ref{raa}) requires further processing: (a)  Factors $q_i(n_s)$ defined in Eq.~(\ref{rhosi}) must be {\em added} to numerator (\pbpb) and denominator (\pp). (b) The extra factor 1.15 or 1.25 introduced to \pp\ spectra (to be consistent with \pbpb\ data, see Sec.~\ref{pidparamsx}) must be removed from the denominator. That combination is referred to below as ``modified'' $r_{AA}(y_t)$.

For the ideal or reference case that \aa\ collisions can be described as {\em linear superpositions} of \nn\ $\approx$ \pp\ collisions the relation in Eq.~(\ref{raaa}) can be modified to obtain
\bea  \label{raann}
R_{\text{AA}i}(y_t) &\rightarrow& \frac{1}{N_{bin}} \frac{(N_{part}/2)\bar \rho_{sNNi}\hat S_{0i} + N_{bin} \bar \rho_{hNNi}\hat H_{0i}}{\bar \rho_{sppi}\hat S_{0i}(y_t) +  \bar \rho_{hppi}\hat H_{0i}(y_t)}
\nonumber \\
&& \hskip -.3in =~\frac{1}{\nu} \frac{1 + (z_{hi}/z_{si}) x_{NN} \nu \, \hat T_{0i}(y_t)}{1 + (z_{hi}/z_{si})x_{pp}\hat T_{0i}(y_t)} \times \frac{q_{iAA}(\nu)}{q_{ipp}},
\eea
where the model functions are unmodified, as observed in \pp\ and \ppb\ collisions, $\hat T_{0i}(y_t) \equiv \hat H_{0i}(y_t) / \hat S_{0i}(y_t)$ and $x_{NN} \equiv \bar \rho_{hNN} / \bar \rho_{sNN}\approx x_{pp}$, with $(z_{hi}/z_{si})x(n_s)\hat T_0(y_t) \ll 1$ for low \yt\ and $(z_{hi}/z_{si})x(n_s)\hat T_0(y_t) \gg 1$ for high \yt~\cite{alicetomspec}. The ratio of $q_i$ factors arises from Eq.~(\ref{rhosi}). Equation~(\ref{raann}) generates the thin solid TCM reference curves in Fig.~\ref{piddata} (right) and comparable panels in Sec.~\ref{ratcompare}.

\subsection{Differential $\bf Pb$-$\bf Pb$  PID spectrum data} \label{piddiff2}

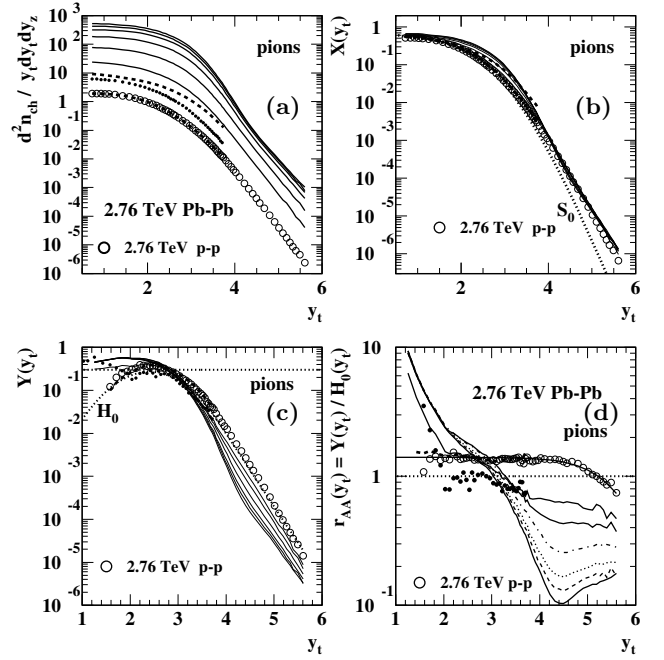
\begin{figure}[h]
	\includegraphics[width=3.3in]{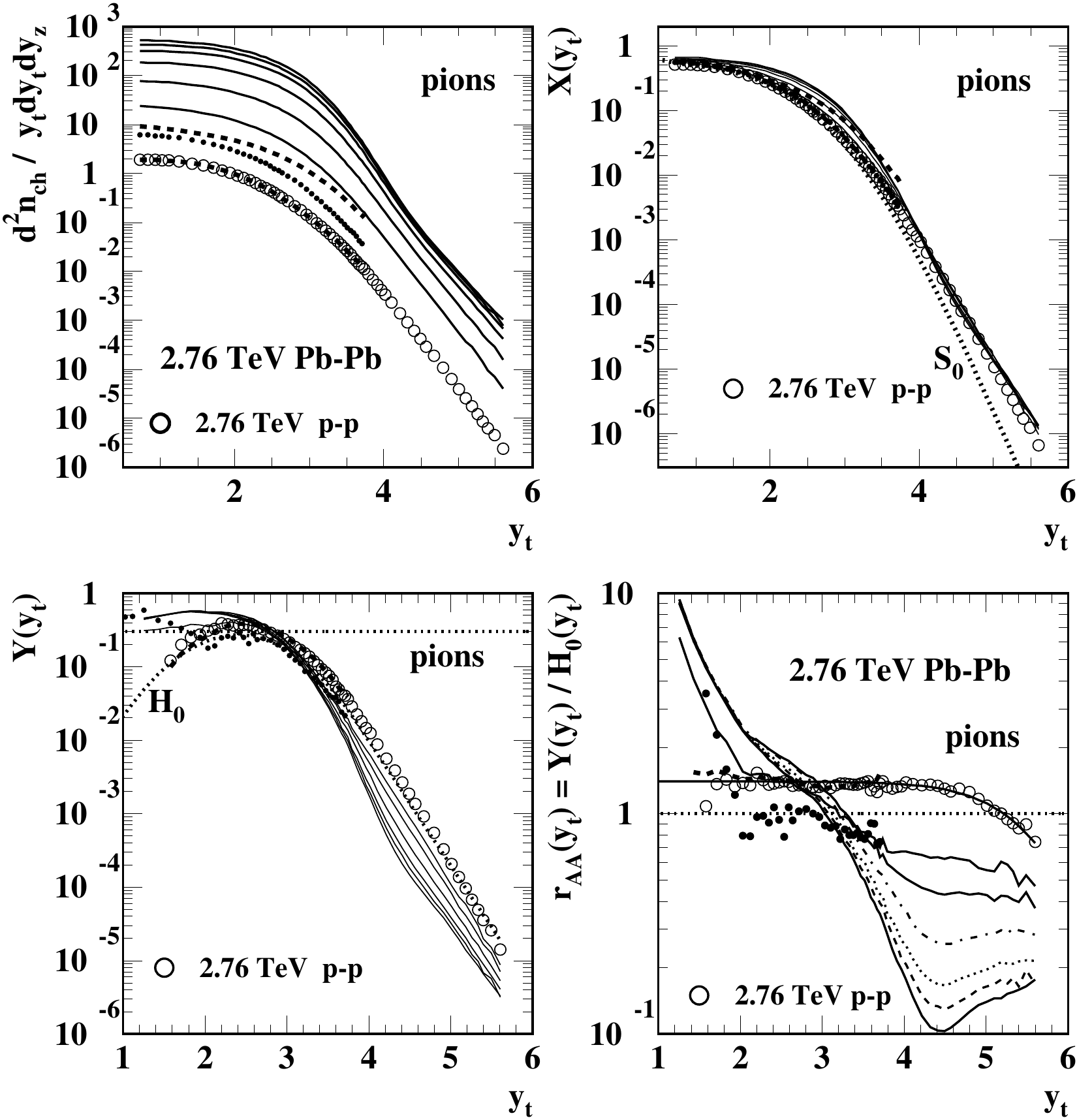}
\put(-142,205) {\bf (a)}
\put(-25,205) {\bf (b)}
\put(-142,89) {\bf (c)}
\put(-22,89) {\bf (d)}
	\caption{\label{pions}
		(a) Pion spectra from six centrality classes of 2.76 TeV \pbpb\ collisions (solid)~\cite{alicepbpbpidspec}  plotted on pion transverse rapidity $y_{t\pi}$. Spectra from midcentral and peripheral 5 TeV \ppb\ collisions (dashed) are included as a reference. Also included are peripheral 80-90\% central \pbpb\ data (solid points) and \pp\ data (open circles).
		(b) Data from panel (a) plotted in format $X(y_t)$ defined by the third line of Eq.~(\ref{pidspectcm}). 
		(c)  Data from panel (b) plotted in format $Y(y_t)$ defined by Eq.~(\ref{pidhard}).
		(d)   Data from panel (c) plotted as hard-component ratios $r_{AA}(y_t)$ defined by Eq.~(\ref{raax}).
	}  
\end{figure}

Figure~\ref{pions} shows data for identified pions from 2.76 TeV \pbpb\ (solid curves), \pp\ (open circles) and 5 TeV \ppb\ (dashed) for centrality classes 4 (mid central) and 7 (peripheral). The pion data extend down to $p_t = 0.1$ GeV/c  ($y_t \approx 0.7$). Also shown are 80-90\% central \pbpb\ pion data (solid dots) from Ref.~\cite{alicepbpbspecx} over a more-limited \pt\ range. Although panel (b) presents full spectra as differentially as possible the systematic trends for the pion spectra are difficult to discern albeit still visible. Such details are inaccessible in the plot format of Fig.~\ref{piddata} (a). 

Figure~\ref{pions} (c) shows finer details. The dotted line at 0.3 in this and other panels (c) is a reference to check for correct normalization of model $\hat H_0(y_t)$; the maxima vary slightly relative to the reference depending on peak widths. Both the \pp\ and \ppb\ hard components lie significantly above TCM reference $\hat H_0(y_t)$ (dotted). Since TCM PID parameters $z_{hi}/z_{si}$ and $z_{0i}$ are determined by \ppb\ spectra in a \pt\ interval below 0.5 GeV/c (\yt\ = 2) there is no possibility to adjust the TCM to accommodate data hard components. In contrast, data for all \pbpb\ centralities fall well below the model except for the excess below $y_t = 3$ (1.4 GeV/c) that is closely correlated with the rest of the hard-component evolution with centrality.

Figure~\ref{pions} (d) shows hard-component ratios $r_{\text{AA}i}(y_t)$. The percent-level correspondence between \ppb\ mid-central (class 4) (dashed) and \pp\ data (open circles) is notable. The 40\% average excess relative to the TCM for both collision systems is also apparent. The hard-component data trend should fall below the usual $\hat H_0(y_t)$ model at higher \pt\ (e.g.\ above 10 GeV/c) because the underlying jet spectrum decreases more rapidly than a power-law approximation at the higher jet energies~\cite{jetspec2}. The solid curve through \pp\ points is a tanh model representing that effect used below for efficiency corrections.

Concerning \pbpb\ data, in contrast to conventional ratio $R_\text{AA}(y_t)$ TCM ratio $r_{\text{AA}}(y_t)$ reveals a dramatic increase in jet-related fragments at lower \yt\ coordinated with suppression at higher \yt. The close correspondence suggests that jet modification in \pbpb\ proceeds such  that parton energy is retained within a jet but is transported to lower hadron \pt\ within the splitting cascade, as noted in Ref.~\cite{fragevo}. In each panel (d) the most-central data manifest the greatest suppression above \yt\ = 4. The 80-90\% data for pions (solid points) suggest no jet modification for that centrality interval but are not definitive.

\begin{figure}[h]
	\includegraphics[width=3.3in]{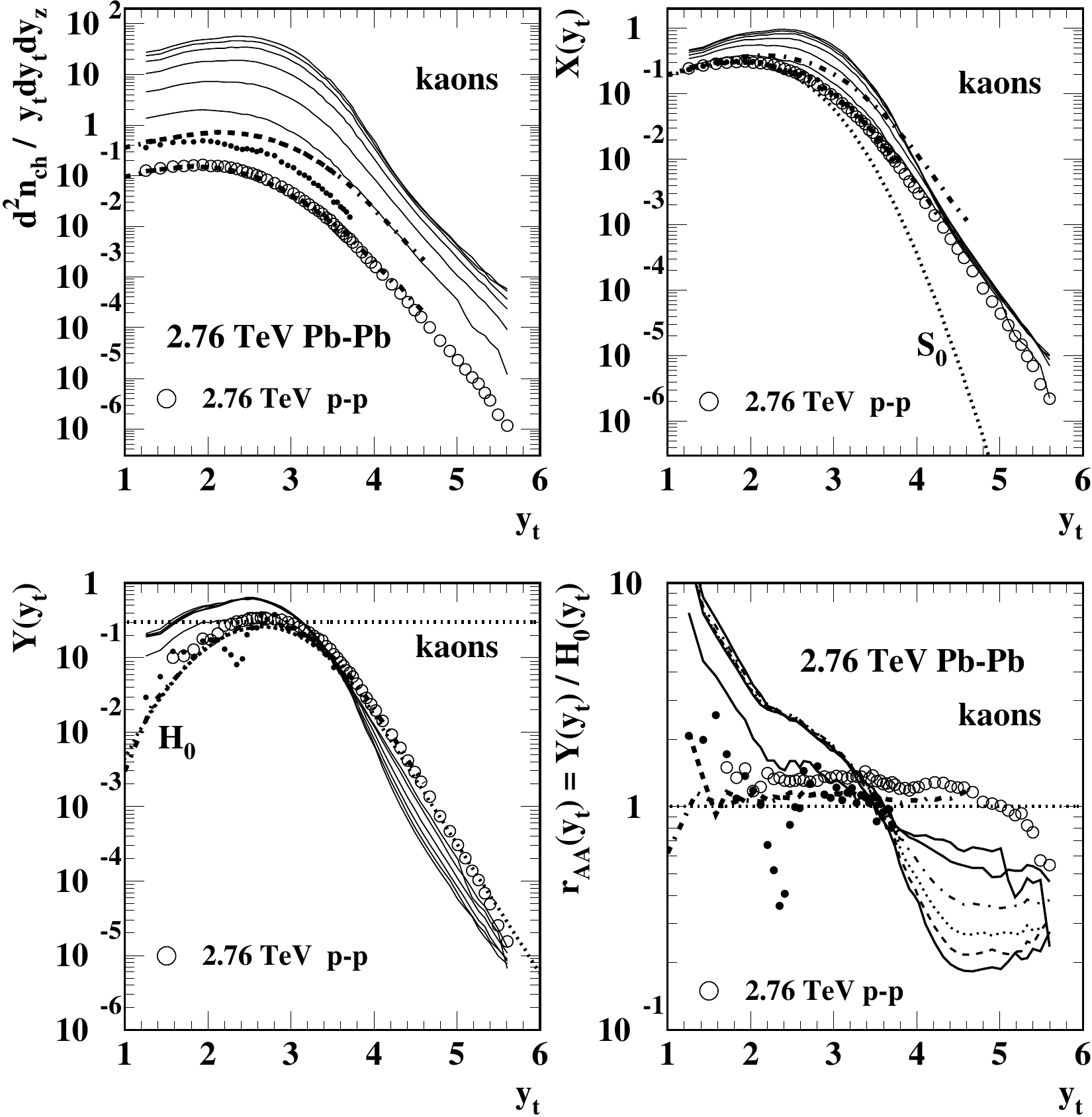}
\put(-142,205) {\bf (a)}
\put(-25,205) {\bf (b)}
\put(-142,85) {\bf (c)}
\put(-22,75) {\bf (d)}
	\caption{\label{kaons}
Kaon spectra from 2.76 TeV \pbpb\ and \pp\ collisions ($K^\pm$) and 5 TeV \ppb\ collisions ($K^\pm,~K^0_\text{S}$) plotted as in Fig.~\ref{pions}. In panels (b) and (c) only $K^0_\text{S}$ \ppb\ data are plotted.
	}  
\end{figure}
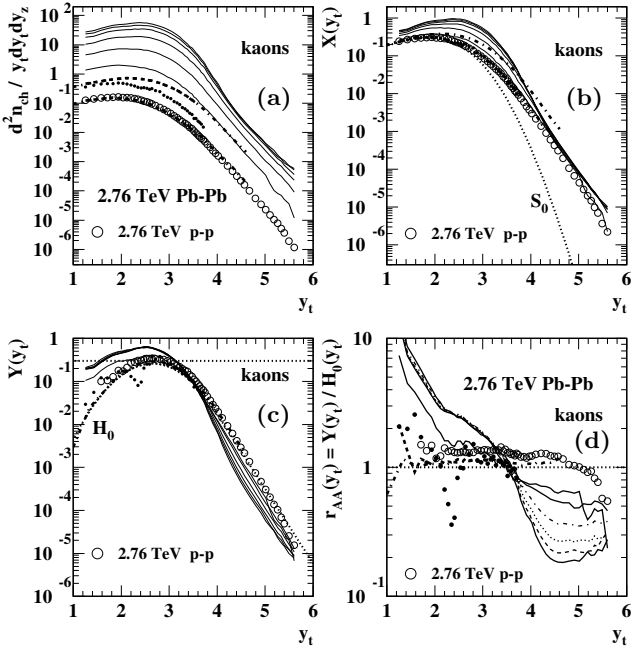

Figure~\ref{kaons} shows results for charged kaons similar to those for pions. The most obvious difference is the larger fractional contribution from jet fragments (hard component) relative to the kaon soft component, especially for more-central \pbpb\ collisions where the hard component remains dominant down to \yt\ = 1.2 ($p_t \approx 0.2$ GeV/c). Results for charged kaons (dashed) and $K^0_\text{S}$ (dash-dotted) from centrality classes 4 (mid-central) and 7 (peripheral) of 5 TeV \ppb\ collisions are included as a reference. Whereas Fig.~\ref{pions} (d) reveals a 40\% excess for \pp\ collisions compared to the TCM reference the \pp\ charged-kaon excess in panel (d) is about 30\%. $K^0_\text{S}$ (dash-dotted) and charged-kaon (dashed) \ppb\ data are consistent with the TCM  over the entire \yt\ acceptance. Only the $K_\text{S}^0$ data (dash-dotted) are shown in panels (b) and (c) for clarity.

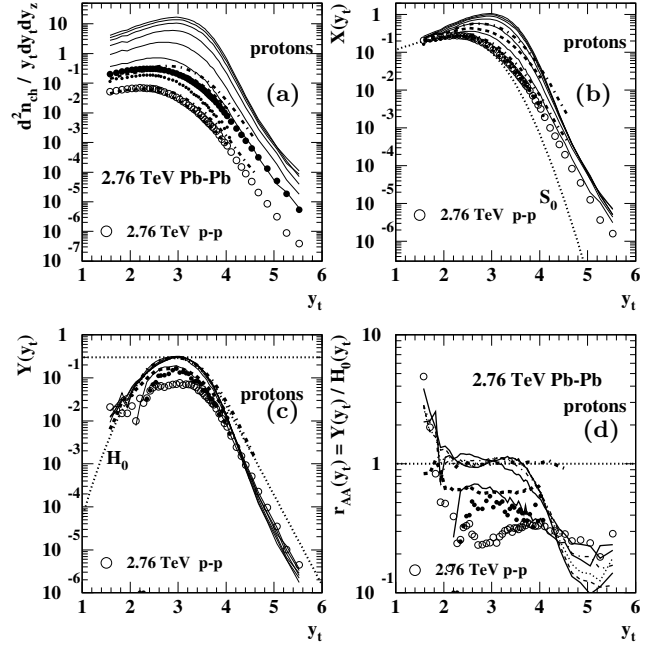
\begin{figure}[h]
	\includegraphics[width=3.3in]{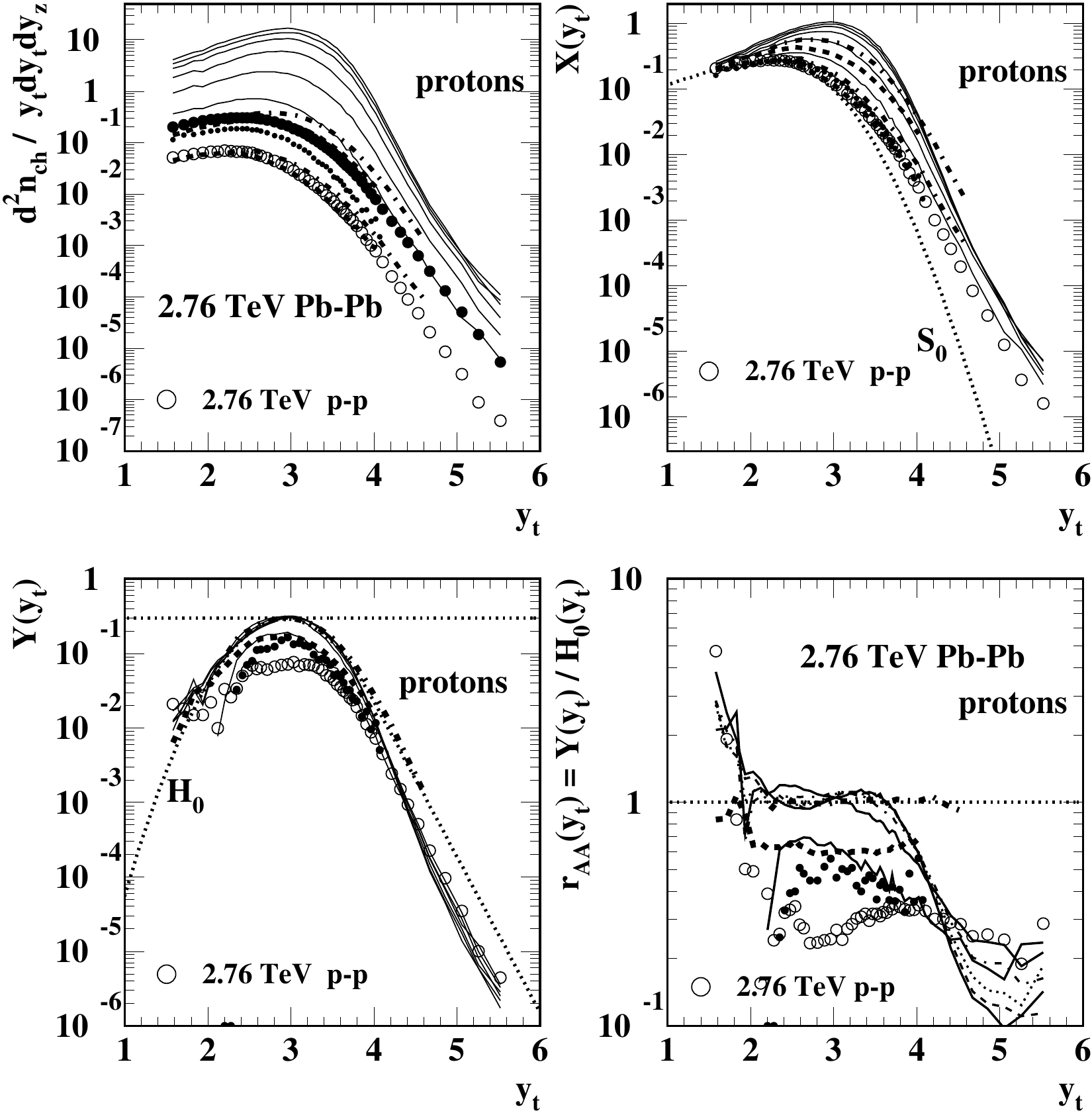}
	\put(-142,205) {\bf (a)}
	\put(-25,205) {\bf (b)}
	\put(-142,85) {\bf (c)}
	\put(-22,78) {\bf (d)}
	\caption{\label{protons}
Baryon spectra from 2.76 TeV \pbpb\ and \pp\ collisions ($p,\,\bar p$) and 5 TeV \ppb\ collisions ($p,\,\bar p,\, \Lambda,\, \bar \Lambda$)  as in Fig.~\ref{pions}.
	}  
\end{figure}

Figure~\ref{protons} shows results for protons from \pbpb\ (solid) and \pp\ (open circles) collisions. Also shown are \ppb\ results for protons (dashed) and Lambdas (dash-dotted) from centrality classes 4 (mid-central) and 7 (peripheral). Significant anomalies appear for charged-proton spectrum hard components compared to meson and neutral Lambda spectra, as first encountered in Ref.~\cite{ppbpid}. In panel (b) the \pp\ proton spectrum (open) is consistent with \ppb\ class 7 (most peripheral, lower dashed) but not with Lambdas from the same \ppb\ events (lower dash-dotted). The same difference appears for class 4 (mid-central, upper dashed and dash-dotted)

In panel (c) the Lambda data for \ppb\ class 4 (dash-dotted) accurately follow TCM model $\hat H_0(y_t)$ (dotted) as reported in Ref.~\cite{ppbpid}. \ppb\ class-4 protons (dashed) are suppressed by 40\% relative to Lambdas (as are \ppb\ class-7 data not shown). The 80-90\% peripheral \pbpb\ data (solid dots) approximately follow the \ppb\ class 4 data.  The \pp\ proton data (open) are further suppressed (and distorted) near the hard-component mode. 

Panel (d) shows those trends more clearly.  \pbpb\ data (several line types) deviate dramatically from the TCM and from \pp\ and \ppb\ data as expected, but further investigation reveals the likelihood of substantial inefficiencies in \pp\ and \pbpb\ proton detection. 
The proton results for $r_{AA}(y_t)$ are quite complex due to a combination of strong suppression of protons approximately uniform on \yt\ (40\%) compared to the \ppb\ TCM reference for \ppb, \pbpb\ and \pp\ data and increasing suppression of protons with increasing \yt\ for \pp\ and \pbpb\ spectra (see Sec.~\ref{fractions}). Systematic inefficiency for PID protons is investigated and a correction for \pp\ and \pbpb\ data devised in Sec.~\ref{correct}.

Some general trends emerge: (a) All \ppb\ neutral $K^0_\text{S}$ and Lambda data agree with the PID spectrum TCM within data uncertainties. In contrast, hard components for charged hadrons based on $dE/dx$ measurements deviate substantially from TCM predictions: strong excess for charged pions and strong suppression for protons. (b) TCM ratios $r_{\text{AA}}(y_t)$ for pions and kaons from \pp\ and \ppb\ collisions are approximately uniform on \yt\ (except for falloff at higher \yt\ following the underlying jet spectrum). There is no significant evidence for jet modification in \ppb\ (``jet quenching'' in \aa\ collisions) beyond few-percent peak shifts.  (c) In contrast, ratios $r_{\text{AA}}(y_t)$ for more-central \pbpb\ collisions vary dramatically with \yt. The trend with increasing centrality for given hadron species is transport from higher to lower \yt. Further comments on \pbpb\ proton trends appear in Sec.~\ref{correct} after efficiency corrections are applied to proton spectra.

\subsection{2.76 $\bf TeV$ $\bf p$-$\bf p$ and $\bf Pb$-$\bf Pb$ TCM parameters} \label{pidparamsx}

The PID spectrum TCM as expressed in Eqs.~(\ref{pidspectcm}) and (\ref{rhosi}) requires accurate parameter values for each collision system and hadron species. Previous results are presented in Tables~\ref{pidparams} and \ref{otherparams} from analysis of 5 TeV \ppb\ PID spectrum data in Ref.~\cite{ppbpid} and Glauber geometry parameters in Table~\ref{rppbdata} from analysis of 5 TeV \ppb\ mean-\pt\ data in Ref.~\cite{tommpt}. Geometry for 2.76 TeV \pbpb\ spectra is based on $N_{part}$ and $N_{bin}$ estimates from Table~\ref{ppbparams1} as reported in Ref.~\cite{pbpbcent} which also determine parameter $\nu$. $\bar \rho_{sNN} = 4.55$  for 2.76 TeV \pp\ and \pbpb\ is based on $\rho_{sNN} \approx 0.81 \ln(\sqrt{s} / \text{10 GeV})$ reported in Ref.~\cite{alicetomspec}. \pp\ spectra are treated as special cases of \pbpb\ spectra with $N_{part}/2 = N_{bin} = 1$. Collision geometry parameters are discussed in more detail in Sec.~\ref{geomparams}.

Parameters for TCM model functions $\hat S_{0i}(y_t)$ and $\hat H_{0i}(y_t)$ and PID parameters are reported below for each hadron species. Generally, TCM parameters are required to be precisely consistent across A-B systems and collision energies. TCM parameters are not adjusted to accommodate specific spectrum data.

{\bf Pions:} Soft-component parameters are as in Table~\ref{pidparams} except for \pp\ where $T \rightarrow 140$ MeV. Hard-component parameters differ somewhat from the \ppb\ values in Table~\ref{pidparams}, with $\bar y_t = 2.45$, $\sigma_{y_t} = 0.60$ and $q = 3.9$. Only the $\bar y_t$ change is significant relative to uncertainties in Table~\ref{pidparams}. Parameters $z_0$ and $z_h / z_s$ are as in Table~\ref{otherparams} except $z_0 \rightarrow 0.8$ for \pbpb. For pions $\hat S_{0i}(y_t)$ must be modified at lower \yt\ relative to the L\'evy distribution in Eq.~(\ref{s00}) by a factor gradually increasing from 1 below 0.5 GeV/c ($y_t \approx 2$) according to a $\tanh$ function. The same is true for \pp\ and \ppb\ data. Other hadron species do not require such modification. The same values are used for \pbpb\ and \pp\ models. The \pp\ spectrum is multiplied by factor 1.15 relative to standard normalization to agree with $\hat S_0(y_t)$ and \ppb\ spectra at low \yt.

{\bf Kaons:}  Soft-component parameters are as in Table~\ref{pidparams} except for \pp\ where $T \rightarrow 190$ MeV.  Hard-component parameters are as in Table~\ref{pidparams}.  Parameters $z_0$ and $z_h / z_s$ are as in Table~\ref{otherparams} except $z_h / z_s \rightarrow 2.8$ for $K^0_\text{S}$ data (within estimated uncertainties). The \pp\ spectrum is multiplied by factor 1.25 relative to standard normalization to agree with $\hat S_0(y_t)$ and \ppb\ spectra at low \yt.

{\bf Protons, Lambdas:}   Soft-component model parameters are as in Table~\ref{pidparams}.  Hard-component parameters are as in Table~\ref{pidparams}. Parameters $z_0$ and $z_h / z_s$ are as in Table~\ref{otherparams}. The \pp\ spectrum is multiplied by factor 1.25 relative to standard normalization to agree with $\hat S_0(y_t)$ and \ppb\ spectra at low \yt.

The TCM is a tightly-constrained model with few parameters, not an arbitrary many-parameter fit to individual data sets. The TCM is a stable reference against which any data may be compared. Data-model deviations may reflect experimental biases or novel physics relative to the basic assumptions of the TCM. Deviations are not addressed by arbitrary parameter variation, in part as a matter of principle and in part because the few model parameters cannot in practice accommodate some deviations, as in \pbpb. The TCM is therefore falsifiable.

\section{Comparing spectrum Ratios} \label{ratcompare}

Given detailed differential spectrum analysis from the previous section certain A-B comparisons may shed further light on the nature of jet contributions to spectra and jet modifications in more-central \aa\ collisions. In this section 2.76 TeV \pbpb\ $R_{AA}$ results are compared with TCM $r_{AA}$ data, $R_{CP}$ ratios for \ppb\ collisions are derived as a reference, and $r_{AA}$ data for 200 GeV \auau\ collisions are compared with those for \pbpb\ collisions.

\subsection{Comparison between $\bf R_{AA}(y_t)$ and $\bf r_{AA}(y_t)$}

In this subsection spectrum ratios $R_{AA}(y_t)$ (left panels) and the ``ratio of ratios'' $r_{AA}(y_t)/r_{pp}(y_t)$ (right panels) defined in Eq.~(\ref{ratrat}) are compared directly on transverse rapidity \yt. For $R_{AA}(y_t)$ plots the thin solid curves denote limiting case Eq.~(\ref{raann}) assuming \aa\ collisions are linear superpositions of \nn\ collisions (no jet modification). Major differences between $R_{AA}(y_t)$ and $r_{AA}(y_t)/r_{pp}(y_t)$ arise because ratios $R_{AA}(y_t)$ include spectrum soft components $\hat S_0(y_t)$ that dominate spectrum structure at lower \yt\ and block access to large jet contributions there. In addition, the definition of $R_{AA}(y_t)$ in Eq.~(\ref{raa}) includes implicitly a factor $q_{iAA}(\nu) / q_{ipp}$ as noted in Eq.~(\ref{raann}) that is not present in $r_{AA}(y_t)$ as defined by Eq.~(\ref{raax}). To make direct quantitative comparisons additional factors are required, as explained in text below Eq.~(\ref{ratrat}).

\begin{figure}[h]
	\includegraphics[width=1.65in]{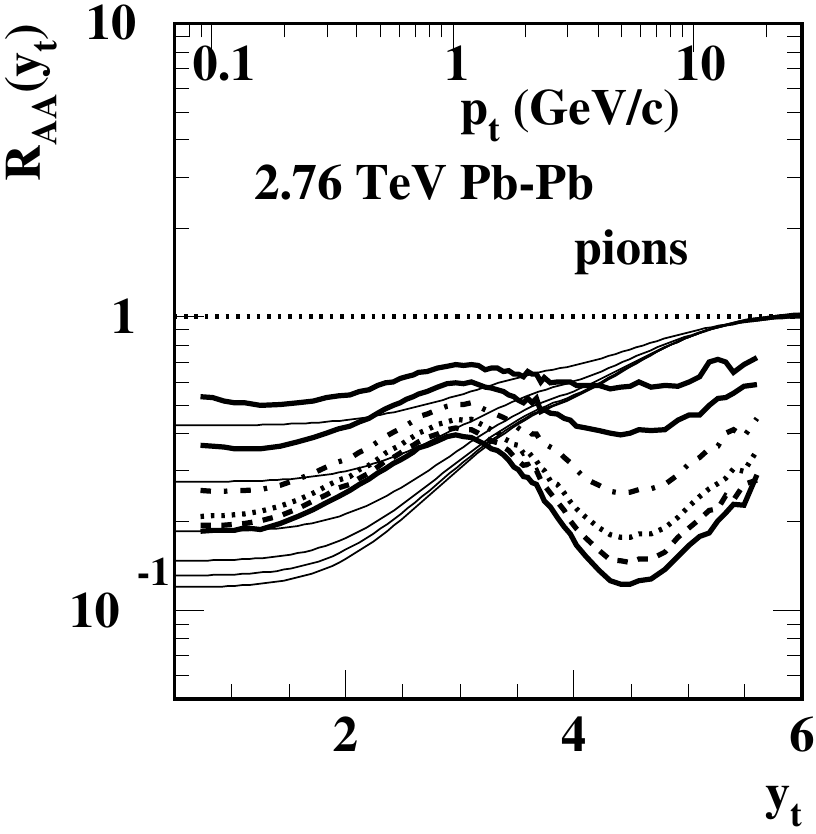}
	\includegraphics[width=1.65in]{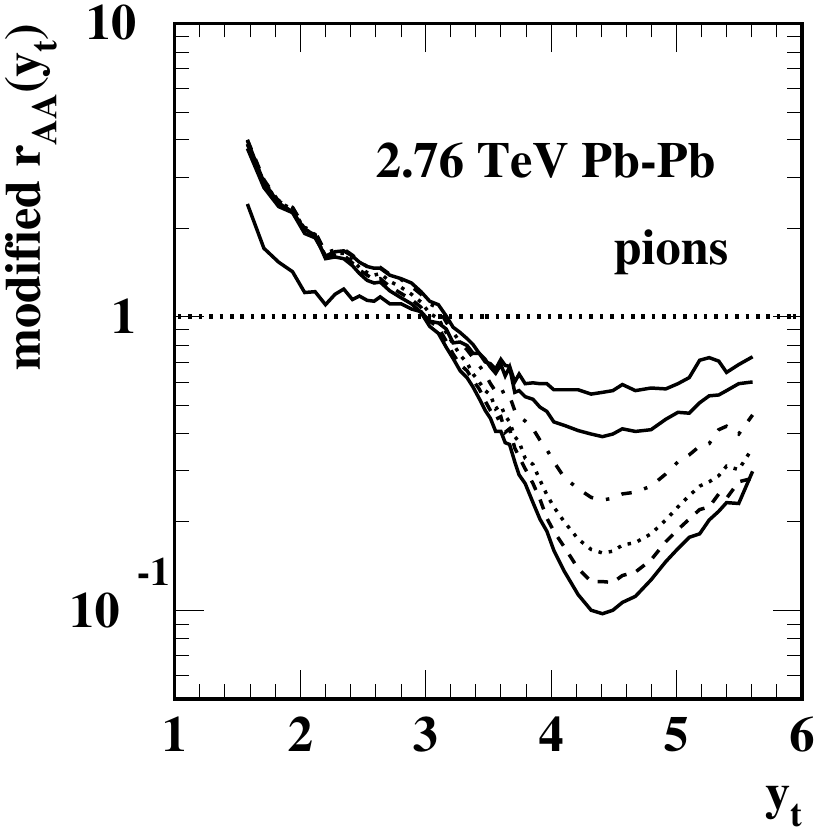}
	\put(-142,95) {\bf (a)}
	\put(-20,105) {\bf (d)}\\
	\includegraphics[width=1.65in]{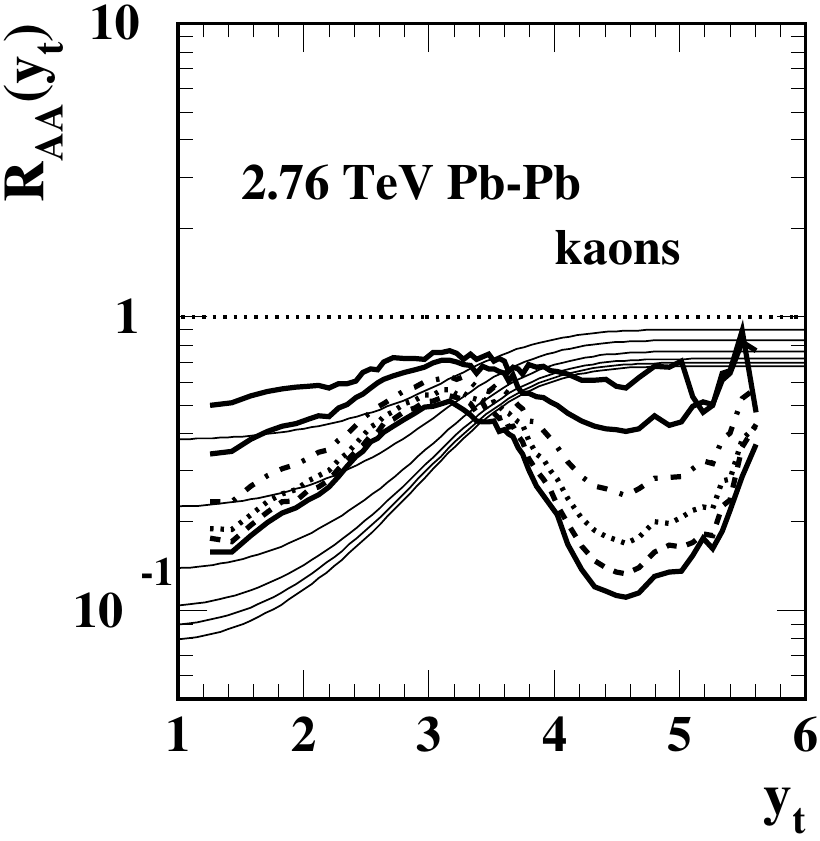}
	\includegraphics[width=1.65in]{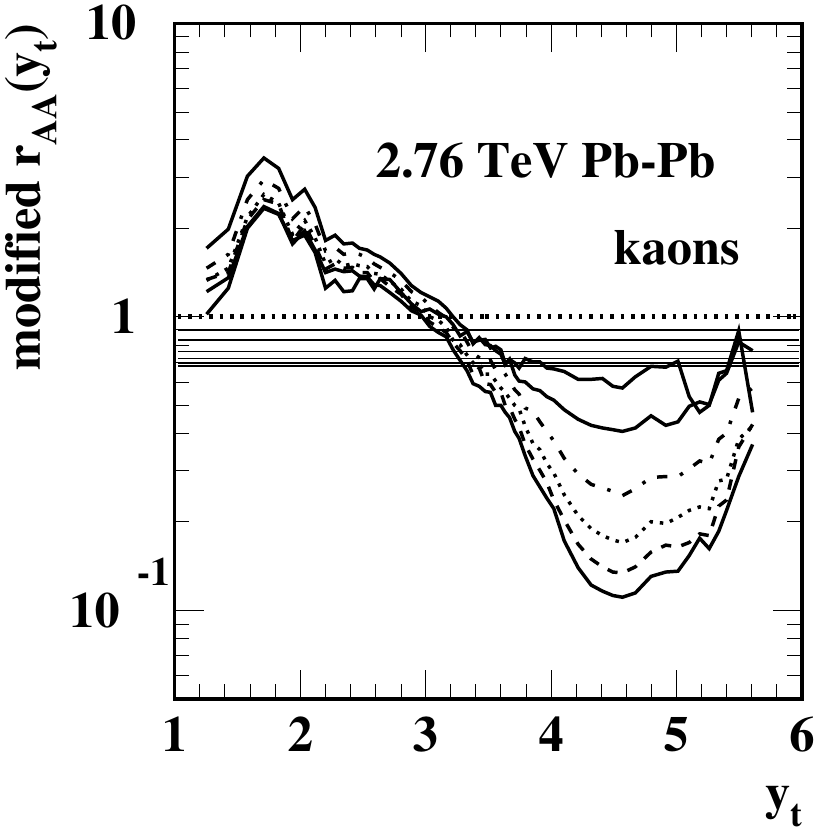}
	\put(-142,105) {\bf (b)}
	\put(-20,105) {\bf (e)}\\
	\includegraphics[width=1.65in]{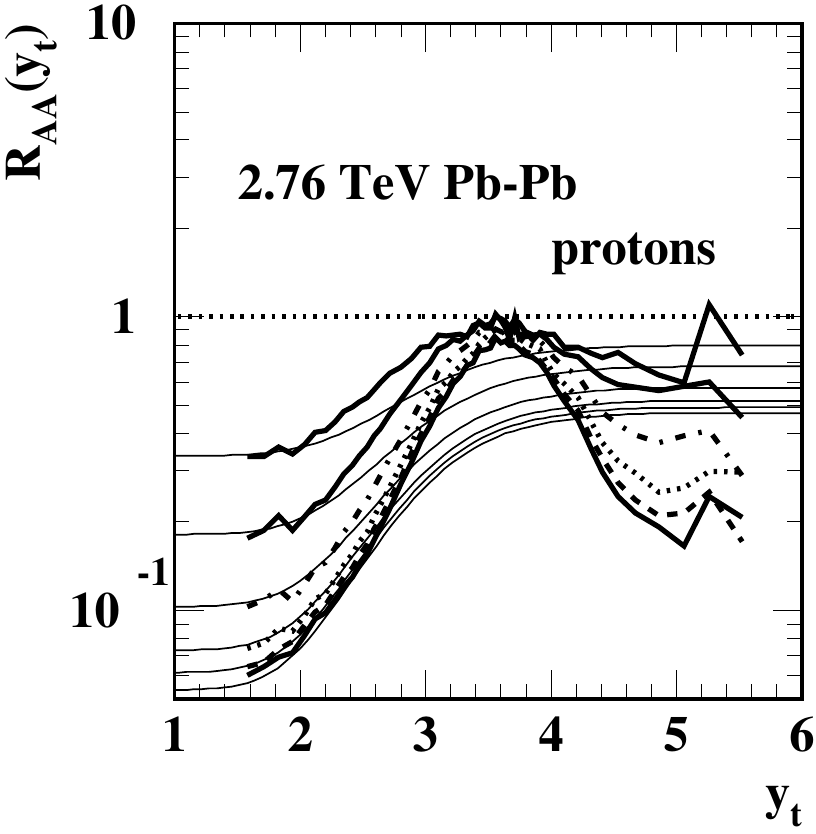}
	\includegraphics[width=1.65in]{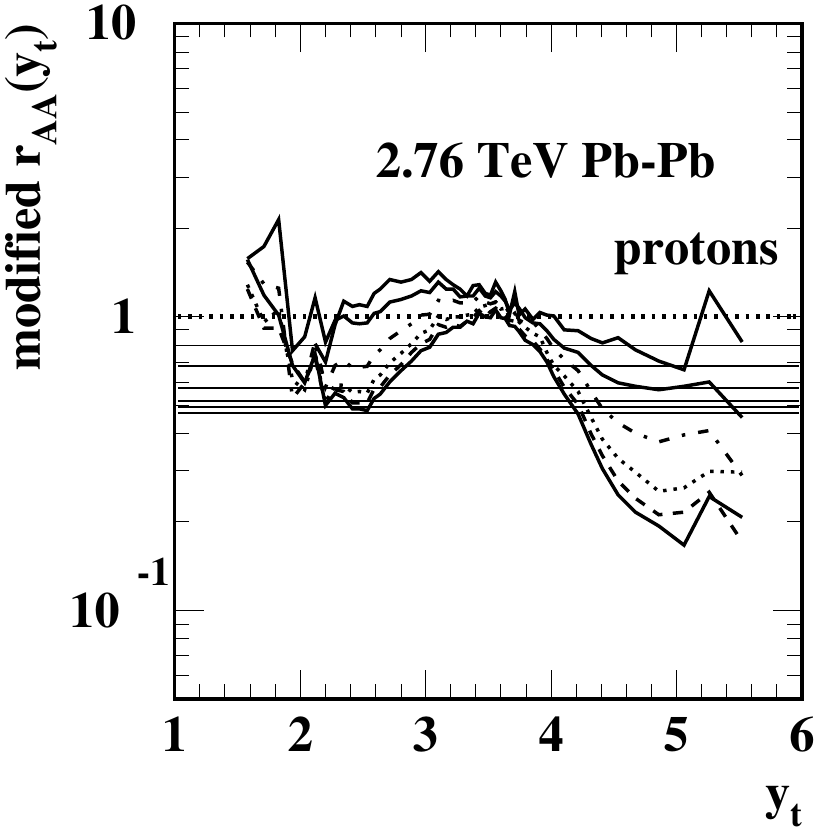}
	\put(-142,105) {\bf (c)}
	\put(-20,105) {\bf (f)}\\
	\caption{\label{pidrats}
	Left: Spectrum ratios $R_{AA}(y_t)$ defined by Eq.~(\ref{raa}) for three hadron species from six centralities of 2.76 TeV \pbpb\ collisions (curves of several line styles). The thin solid curves are references defined by Eq.~(\ref{raann}).
		Right: Spectrum hard-component ratios $r_{AA}(y_t)/r_{pp}(y_t)$, modified relative to Eq.~(\ref{ratrat}) as explained below that equation, for three hadron species from six centralities of 2.76 TeV \pbpb\ collisions. The solid lines are references representing linear superposition.
	}  
\end{figure}

Figure~\ref{pidrats} shows comparisons for three hadron species from 2.76 TeV \pbpb\ collisions. Given the structure of Eq.~(\ref{raann}) the $R_{AA}(y_t)$ thin reference curves in left panels go to limits $[q_{iAA}(\nu) / q_{ipp}] \times 1/\nu$ at low \yt\ that have no relationship to jet structure, and to limits $q_{iAA}(\nu) / q_{ipp}$ at high \yt. Above \yt\ = 4 $R_{AA}(y_t)$ and the {\em modified} version of $r_{AA}(y_t)/r_{pp}(y_t)$ should be numerically equivalent. For pions and kaons the shape of modified $r_{AA}(y_t)/r_{pp}(y_t)$ is fairly close to single ratio $r_{AA}(y_t)$ because $r_{pp}(y_t)$ is, within an $O(1)$ factor, approximately uniform on \yt. The large deviations of proton $r_{pp}(y_t)$ from uniformity may reflect instrumental bias or novel physics. However, the modified ratio of ratios appears to cancel part or all of that effect. It is notable  that despite apparent proton spectrum distortions the linear-superposition \ppb\ TCM in Eq.~(\ref{raann}) {\em predicts \pbpb\ $R_{AA}(y_t)$ proton data quantitatively} below \yt\ = 2.7 (1 GeV/c). That result is consistent with \auau\ proton data in Fig.~\ref{raaauau} (right)~\cite{hardspec}.

The greatest differences occur below $y_t = 4$ ($p_t \approx 3.8$ GeV/c) where $r_{AA}(y_t)$ may increase to large values $\gg 1$. As noted in Ref.~\cite{fragevo} the $r_{AA}(y_t)$ pattern of strong suppression at higher \yt\ coupled with strong enhancement at lower \yt\ (for pions and kaons) is consistent with modification of the splitting cascade for jet formation in \aa\ that transports parton energy to lower hadron \pt\ than for \pp\ but approximately conserves the parton energy within a jet. The strong low-\yt\ enhancement is effectively concealed by $R_{AA}(y_t)$, presenting major difficulties for theoretical interpretation. Comparing pions, kaons and protons the meson trends are similar, with a zero crossing near \yt\ = 3 ($p_t \approx 1.4$ GeV/c). In contrast, proton data show a pronounced peak near $y_t = 3.5$ discussed in Sec.~\ref{compare}, with no dramatic increase at lower \yt. 

\subsection{Ratio comparisons for p-Pb spectra}

It is informative to consider an $R_{AA}$-type spectrum ratio for \ppb\ collisions where a previous spectrum study has established that no significant jet modifications occur~\cite{ppbpid}.%
\footnote{``Modification'' refers here to changes in jet formation for a given jet energy, i.e.\ {\em significant} changes to FFs~\cite{fragevo}.
} For PID spectra an alternative to the simple $R_{AA}$ formula of Eq.~(\ref{raa}) must be derived that is based on the PID spectrum TCM of Eq.~(\ref{pidspectcm}) (first line). In that expression the quantities $z_{si}$ and $z_{hi}$ are dependent on A-B centrality, as noted in Ref.~\cite{ppbpid} Sec.~VI C and its  Eq.~(15). From that expression two factors from Eq.~(\ref{pidspectcm}) can in turn be defined in terms of non-PID charge densities:
\bea
\bar \rho_{sNNi} &\equiv& z_{si}  \bar \rho_{sNN}
~=~ q_i(n_s) \, z_{0i} \bar \rho_{sNN}
\eea
and
\bea
\bar \rho_{hNNi} &\equiv& z_{hi}  \bar \rho_{hNN}
 ~=~ q_i(n_s) \frac{z_{hi}}{z_{si}} \, z_{0i} \bar \rho_{hNN},
\eea
both also consistent with Eq.~(\ref{rhosi}). In what follows the fixed factors $z_{hi}/z_{si}$ and $z_{0i}$ cancel in the ratios.  

Spectrum ratio $R_{AA}(y_t)$ requires \pp\ spectra to serve as references. In the absence of reliable \pp\ PID data (to complement the 5 TeV \ppb\ PID spectra) a more-general central/peripheral or C/P ratio $R_{CP}$ is usually defined
\bea \label{rcp1}
R_{CP}(y_t) &\equiv& \frac{N_{bin,P}}{N_{bin,C}}  \times \frac{\bar \rho_{0i,C}}{\bar \rho_{0i,P}}.
\eea
However, in the context of the present analysis an alternative C/P ratio definition may be more instructive
\bea \label{rcp2}
\tilde R_{CP}(y_t) &\equiv& \frac{N_{bin,P}}{N_{bin,C}} \times \frac{q_{iP}}{q_{iC}} \times \frac{\bar \rho_{0i,C}}{\bar \rho_{0i,P}}.
\eea
That form is dictated by the condition that given no jet modification and linear superposition of \pn\ collisions, i.e.\ $\hat S_{0i}(y_t)$ and $\hat H_{0i}(y_t)$ are {\em universal and unchanging} with \ppb\ centrality, the following limits should arise: 
\bea
\tilde R_{CP}(y_t) & \rightarrow & 1~~~\text{for high \yt}
\\ \nonumber
&\rightarrow&\frac{\nu_P}{\nu_C} \frac{x_{i,P}}{x_{i,C}}\rightarrow \frac{\nu_P}{\nu_C} \frac{x_{P}}{x_{C}}~~~\text{for low \yt}.
\eea
where $x_i(n_s) = \bar \rho_{hNNi} / \bar \rho_{sNNi} = (z_{hi} / z_{si}) x(n_s)$. The {\em added tilde} indicates that this definition of $R_{CP}$ includes the extra ratio $q_{iP} / q_{iC}$ to achieve those limits. For \aa\ collisions with \pp\ as the reference ($\nu_P = 1$), and assuming all \nn\ collisions within \aa\ collisions are on average equivalent ($x_{P}/x_{C} = 1$), the low-\yt\  limit should be $1/\nu_C$.

\begin{figure}[h]
	\includegraphics[width=1.65in]{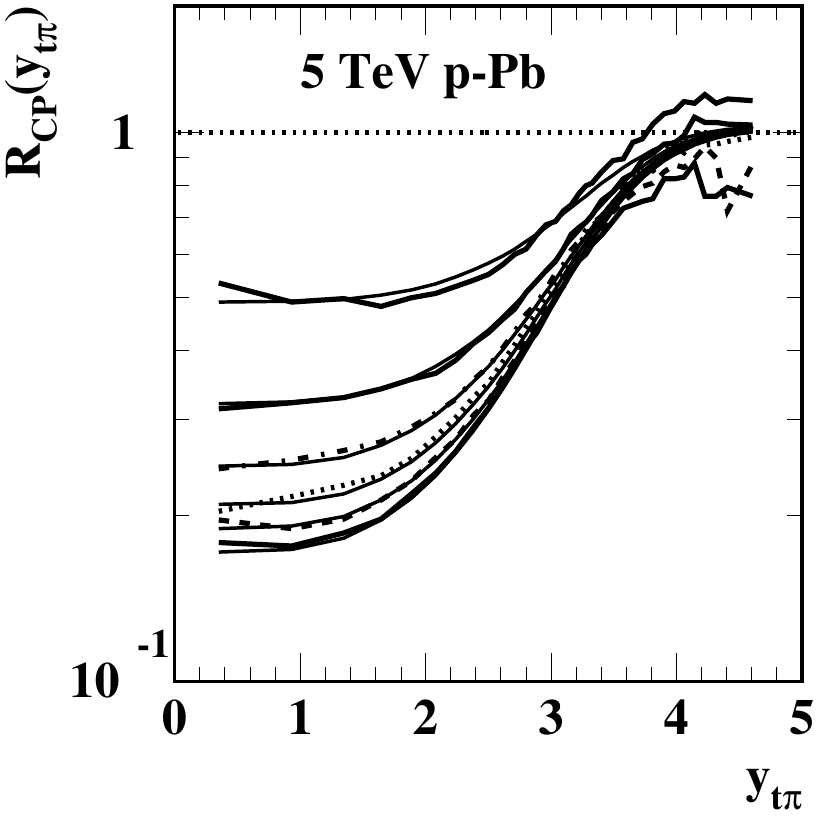}
	\includegraphics[width=1.65in]{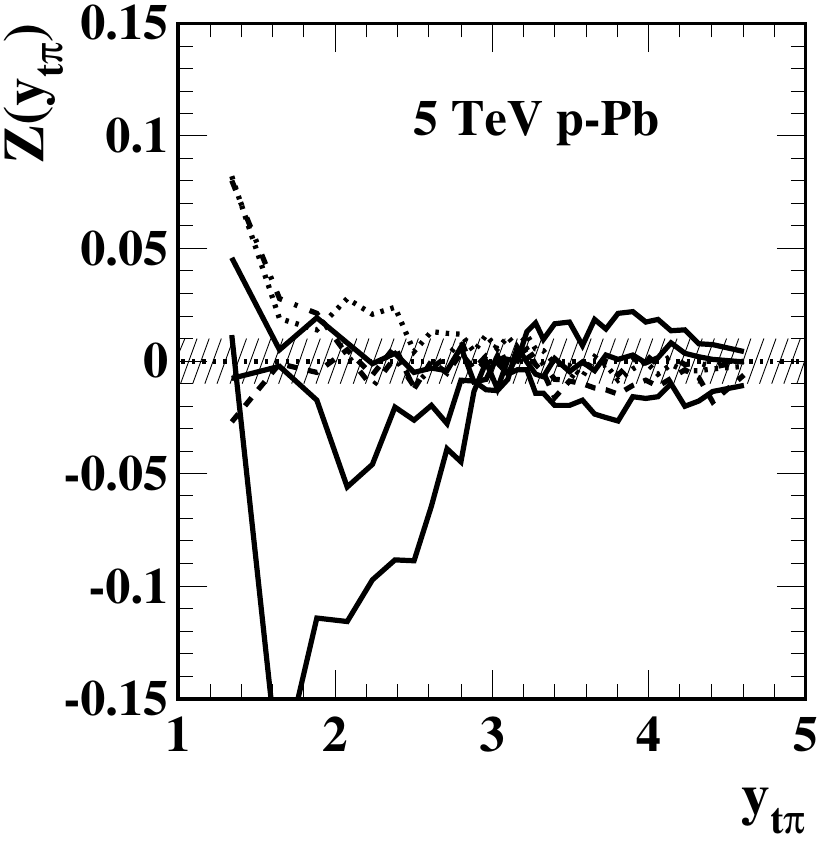}
	\caption{\label{rcp}
		Left: Spectrum ratios $\tilde R_{CP}(y_t)$ derived from 5 TeV \ppb\ PID spectra (several line styles) via Eq.~(\ref{rcp2}) compared to equivalent TCM ratios (solid), all relative to a peripheral TCM reference. TCM model functions are fixed independent of centrality.
		Right: PID spectrum hard-component ratios as in Eq.~(\ref{zee}) (curves) assuming fixed model functions.
	}  
\end{figure}

Figure~\ref{rcp} (left) shows \ppb\ $\tilde R_{CP}(y_t)$ (with tilde) data (curves of several line styles) for $K^0_\text{S}$ spectrum data from six centrality classes of 5 TeV \ppb\ collisions. Also included are TCM equivalents (thin solid curves). The P reference is in all cases the TCM for most-peripheral ($n = 7$) \ppb\ collisions. That \nch\ class has a smaller charge density (4.2) than 5 TeV NSD \pp\ collisions (5.0) per Table~\ref{rppbdata}. While this TCM version does approach unity at high \yt\ by construction there are significant deviations for the data there, although they are difficult to discern relative to statistical fluctuations at high \yt.

Figure~\ref{rcp} (right) shows a differential ratio comparison of TCM and data spectrum hard components in a form based on $r_{AA}(y_t)$ as it is defined in Eq.~(\ref{raax}):
\bea  \label{zee}
Z_i(y_t) &\equiv& \sqrt{\hat H_{0i}(y_t)} [r_{AAi}(y_t) - 1].
\eea
The square-root factor ensures that the statistical uncertainty of $Z_i(y_t)$ is approximately independent of \yt. Without it uncertainties would increase strongly for small and large \yt\ as discussed in Sec.~IV B of Ref.~\cite{alicetomspec}. Statistical uncertainty is estimated by the hatched band. The data for peripheral event classes 5 and 6 (lowest solid curves) show substantial systematic biases at low \yt\ typical of \pp\ and peripheral A-B collisions. The most-peripheral (sub-NSD) $n = 7$ ratio data are not shown.

In that data format systematic deviations from zero are apparent near $y_t = 3.8$ and are consistent with displacements of the spectrum hard component. Small centroid shifts were noted for protons and Lambdas in Sec.~VI B, Fig.~6 and Sec.~VIII, Fig.~9 of Ref.~\cite{ppbpid} according to $\bar y_{tn} = \bar y_{t0} - \delta y_t (n - 4)$. For protons $\delta y_t \approx +0.03$ and for Lambdas $\delta y_t \approx +0.015$, i.e.\ shifts to higher \yt\ with increasing \nch. No shift was reported there for pions and kaons. 
However, in Fig.~\ref{rcp} (right) ratio $Z(y_t)$ is more sensitive and significant shifts for kaons are observed, with $\bar y_{t0} \approx 2.66$ and $\delta y_t \approx -0.01 $, i.e. shifts to {\em lower} \yt. 
 
\begin{figure}[h]
	\includegraphics[width=1.65in]{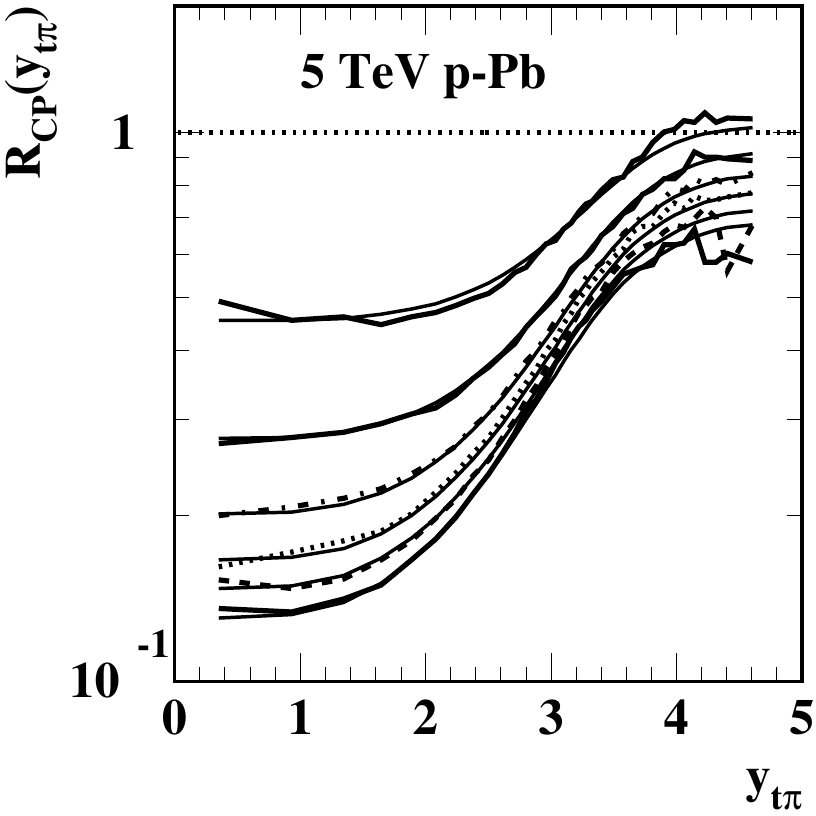}
	\includegraphics[width=1.65in]{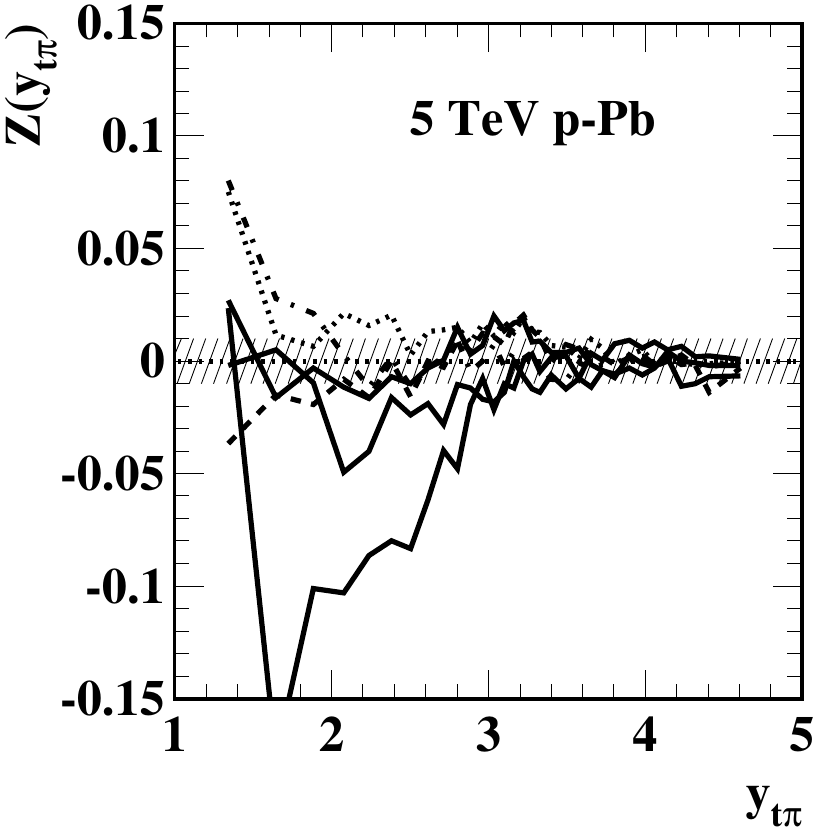}
	\caption{\label{rcpx}
		Left: Spectrum ratios $R_{CP}(y_t)$ derived from 5 TeV \ppb\ PID spectra (several line styles) via Eq.~(\ref{rcp1}) compared to equivalent TCM ratios (solid), all relative to a peripheral TCM reference. The TCM hard-component centroid shifts to lower \yt\ with increasing centrality as described in the text.
		Right: PID spectrum hard-component ratios as in Eq.~(\ref{zee}) (curves). The TCM hard-component centroid shifts on \yt\ with changing centrality as described for the left panel.
	}  
\end{figure}

Figure~\ref{rcpx} (left) shows  $R_{CP}(y_t)$ (no tilde). The TCM hard component is shifted with centrality as described above. Ratio $R_{CP}(y_t)$ defined via Eq.~(\ref{rcp1}) includes ratio $q_{i}(n_s) / q_{iP}$ that causes overall \yt-independent vertical shifts with changing centrality, the same trend as for $R_{AA}(y_t)$ in Fig.~\ref{pidrats} (left panels). This result demonstrates that conventional $R_{AA}(y_t)$ would indicate ``jet quenching'' at high \yt\ even in the absence of jet modification.

Figure~\ref{rcpx} (right) repeats Fig.~\ref{rcp} (right) but with the model $\hat H_0(y_t)$ centroid shifted with \nch\ as described above. The data description above \yt\ = 3 is now consistent with statistical uncertainties for all centralities.

The main conclusions are: 
(a) The TCM-based ratio measure in the right panels provides precise jet information across the large interval $p_t \in [0.25,7]$ GeV/c. 
(b) In contrast, the conventional $R_{CP}(y_t)$ spectrum ratio in the left panels provides significant jet-related information only above $y_t = 4$ ($p_t \approx 3.8$ GeV/c). Jet information below that point is concealed by the presence of the soft components in the spectrum ratio.
(c) The conventional spectrum ratios $R_{AA}(p_t)$ Eq.~(\ref{raa}) or $R_{CP}(p_t)$ Eq.~(\ref{rcp1}) include a centrality-dependent factor that further confuses the issue of jet modification at higher \pt.
(d) The highly differential presentation in the right panels does indicate detectable jet modification in \ppb\ collisions. For $K_\text{S}^0$ the data are consistent with slight softening of the spectrum hard component whereas for protons and Lambdas the data are consistent with hardening (shifts to higher \yt). In any case such ``modifications'' (shifts on \yt) are quite small and appear inconsistent with changes in jet formation {\em per se} (i.e.\ altered FFs) since the hard-component shapes do not change significantly.
(e) The excellent agreement between \ppb\ data and TCM below $y_t = 3.5$ down to zero momentum confirms no detectable evidence for radial flow in \ppb\ collisions.

\subsection{Correcting Pb-Pb and p-p proton inefficiencies} \label{correct}

The previous subsection establishes that an $R_{CP}$ model based on the TCM provides an accurate description of \ppb\ PID spectrum ratios. Figure~\ref{pidrats} (c) in turn demonstrates that \pbpb\ proton $R_{AA}$ data agree closely with a TCM {\em prediction} below \yt\ = 2.5 ($p_t \approx 0.85$ GeV/c), suggesting that jet modification plays no role there. Thus, whatever the anomalous behavior of \pbpb\ and \pp\ proton spectra separately, in ratio they may follow an informative trend. This result suggests that \pp\ proton data may be used to determine an efficiency curve applied to correct \pbpb\ proton data. The efficiency is determined by forming the ratio of the \pp\ proton spectrum (i.e.\ open circles in Fig.~\ref{protons}) to the TCM reference
\bea \label{ppprotontcm}
\bar \rho_{0ppi} &=& q_{ipp} z_{0i} \bar \rho_{spp} \left[ \hat S_0(y_t) + (z_{hi}/z_{si}) x_{pp} \hat H_0(y_t)   \right],~~
\eea
where for 2.76 TeV \pp\ protons $q_{ipp} = 1.05 / 1.35 = 0.78$, $z_{0i} = 0.07$, $\bar \rho_{spp} = 4.55$, $z_{hi}/z_{si} = 7$ and  $x_{pp} = 0.05$. Model functions are determined by \ppb\ parameters in Table~\ref{pidparams}. That model describes Lambda data within their uncertainties. The fall-off of \pp\ pion and kaon data at high \yt\ as in Figs.~\ref{pions} (d) and \ref{kaons} (d) (reflecting the underlying jet spectrum) is simply modeled with a tanh function [appearing in Fig.~\ref{pions} (d)] and included as an additional factor in Eq.~(\ref{ppprotontcm}). The resulting efficiency trend is shown as data points in Fig.~\ref{ratiox} (right). A tanh function representing that data trend (curve in the same panel) is used for efficiency corrections in what follows.

\begin{figure}[h]
	\includegraphics[width=3.3in]{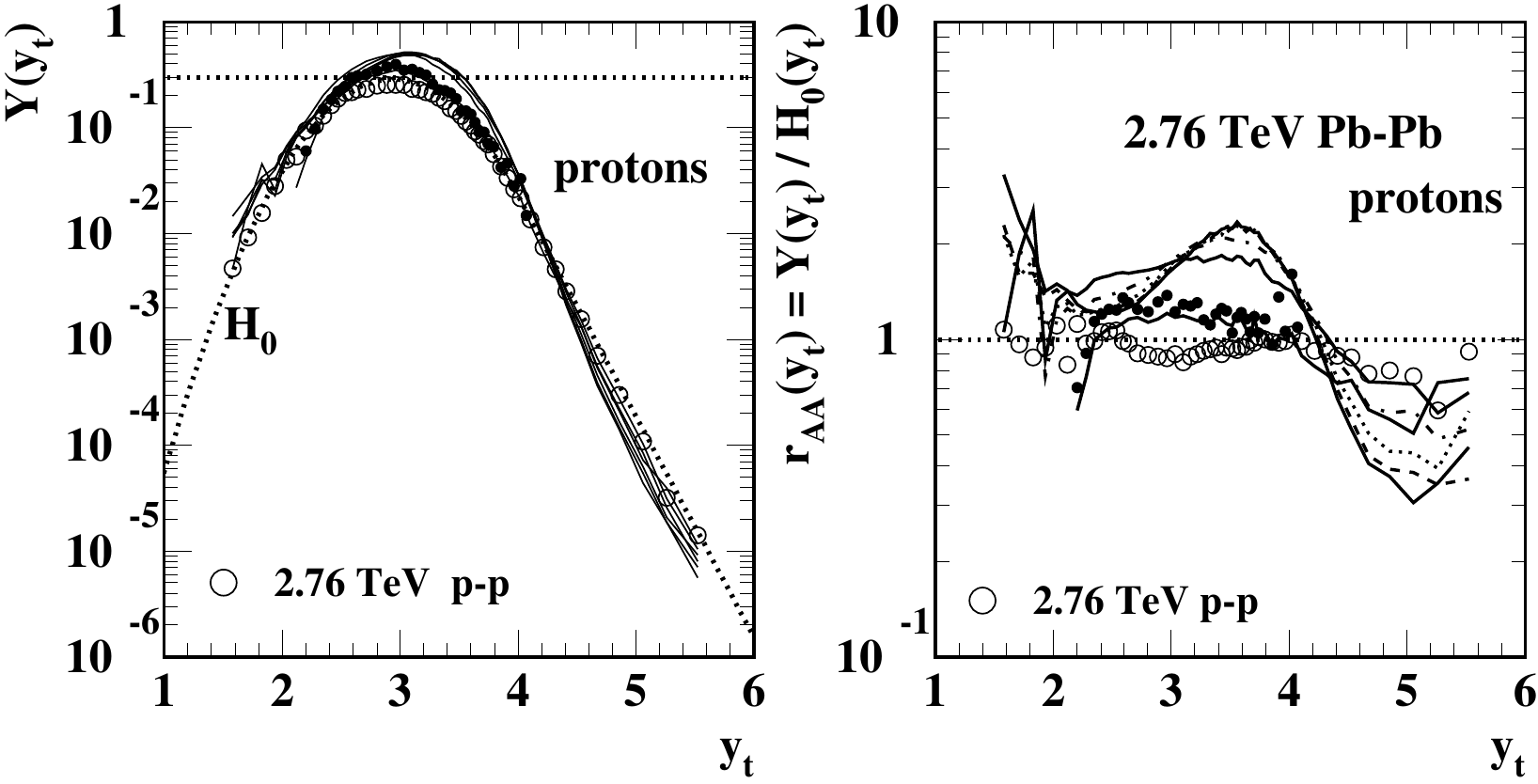}
	\caption{\label{rcpz}
		Left: Proton spectrum hard components $Y(y_t)$ as in Fig.~\ref{protons} (c) but with \pp\ and \pbpb\ data corrected for inefficiencies (see text).
		Right: 	 Proton hard-component ratios $r_{AA}(y_t)$ as in Fig.~\ref{protons} (d) but also corrected for inefficiencies.
	}  
\end{figure}

Figure~\ref{rcpz} (left) shows panel (c) of Fig.~\ref{protons} updated with \pp\ and \pbpb\ spectra corrected by the efficiency trend described above. The corrected \pp\ proton data (open circles) in the left panel now lie close to the TCM hard-component curve (dotted). Corrected \pbpb\ data (solid curves) can now be compared with \auau\ data in Fig.~\ref{protonauau} (right) -- the correspondence is striking.

Figure~\ref{rcpz} (right) shows panel (d) of Fig.~\ref{protons} updated with the efficiency correction. The \pp\ points lie near unity except for the expected (via jet spectrum) fall-off at high \yt. The \pbpb\ data are interpreted in the next subsection. It should be noted that this panel is exactly compatible with Fig.~\ref{pidrats} (f) per Eq.~(\ref{ratrat}) and modifications to $r_{AA}(y_t)$ in text below it.

\subsection{Comparison between two collision energies} \label{compare}

In this subsection spectrum ratios for pions (left panels) and protons (right panels) are compared for 200 GeV \auau\ collisions and 2.76 TeV \pbpb\ collisions. To aid comparison the plot formats are consistent for all panels.

\begin{figure}[h]
	\includegraphics[width=3.3in]{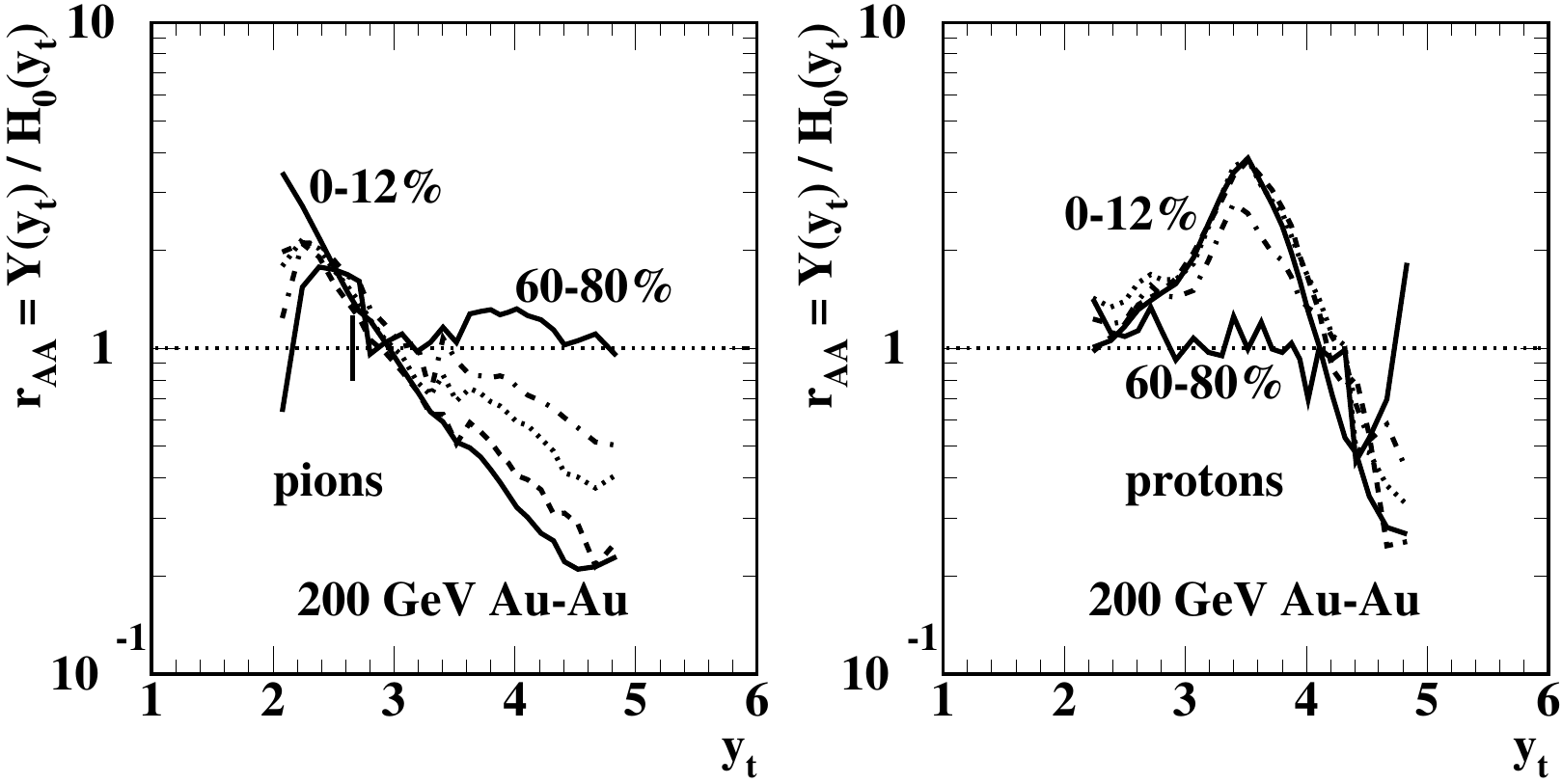}
\put(-142,105) {\bf (a)}
\put(-20,105) {\bf (b)} \\
	\includegraphics[width=1.65in]{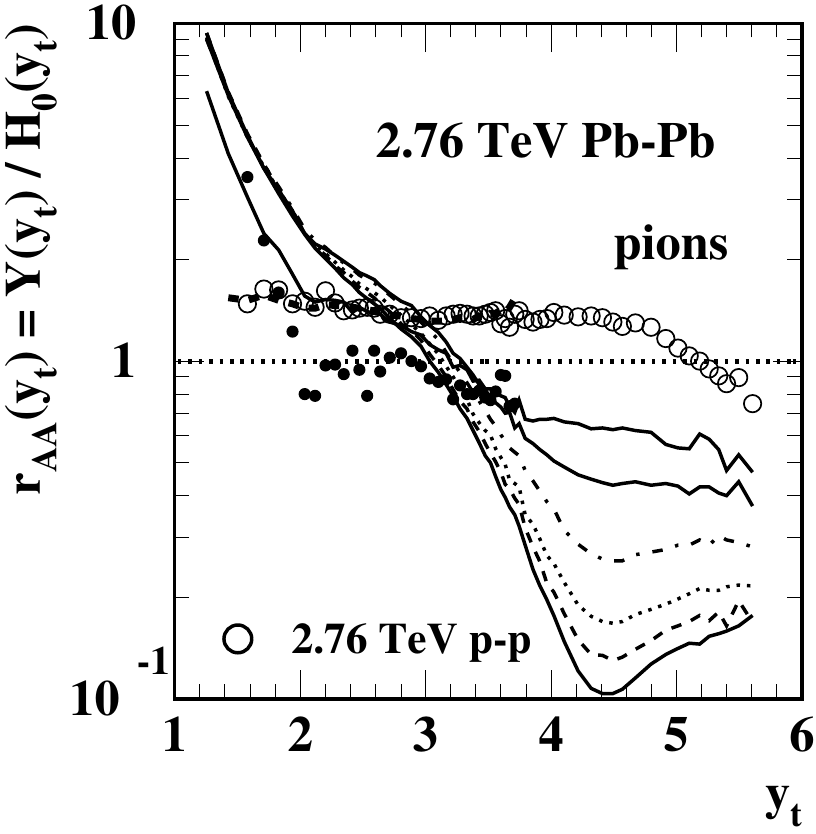}
\includegraphics[width=1.65in]{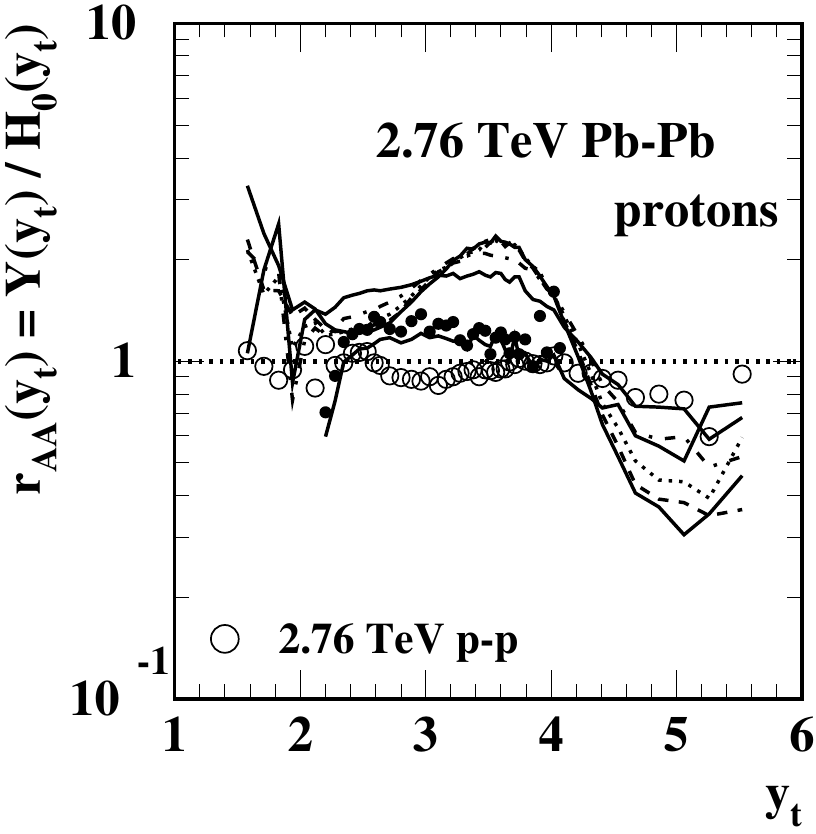}
 \put(-142,107) {\bf (c)}
\put(-20,107) {\bf (d)}
	\caption{\label{xxxx}
		Top: $r_{AA}(y_t)$ for pions (a) and protons (b) from 200 GeV \auau\ collisions.
		Bottom:  $r_{AA}(y_t)$ for pions (c) and protons (d) from  2.76 TeV \pbpb\ collisions. The proton data in panel (d) have been corrected for inefficiencies (see text).
	}  
\end{figure}

Figure~\ref{xxxx} shows spectrum ratio $r_{AA}(y_t)$ for pions (left) and protons (right) and for 200 GeV \auau\ data (upper) and 2.76 TeV \pbpb\ data (lower). The general features are similar for 200 GeV \auau\ and 2.76 TeV \pbpb. The 60-80\% centrality class for \auau\ collisions is  statistically consistent with unity independent of \yt\ indicating no jet modification whereas more-central classes are consistent with strong jet modification. The \auau\ $r_{AA}(y_t)$ data are thus consistent with an observed sharp transition in jet-related angular correlations near 50\% fractional cross section ($\nu \approx 3$, 40-60\% centrality class) as reported in Ref.~\cite{anomalous} and shown in Fig.~\ref{900a}. 

The lower panels show 2.76 TeV pion and proton data from \pp\ and \pbpb\ collisions. The proton spectra have been corrected for apparent detection inefficiency as described in the previous subsection based on comparing \pp\ data to a TCM prediction. The \pp\ proton data then lie along unity in panel (d) except for the expected fall-off at higher \yt. The \ppb\ proton (dashed) and Lambda (dash-dotted) trends are as described in Ref.~\cite{ppbpid}. 

The \pbpb\ 80-90\% (solid points) and 60-80\% proton data (lowest solid curve) near \yt\ = 3.5 are statistically equivalent and clearly separated from five more-central event classes. That result is not inconsistent with a \pbpb\ sharp transition positioned as indicated in Fig.~\ref{900a}. The seemingly larger variation with centrality of proton data at high \yt\ in Fig.~\ref{pidrats} (f) compared to Fig.~\ref{xxxx} (d) is a result of the centrality-dependent $q_{iAA}/q_{ipp}$ factor in {\em modified} $r_{AA}(y_t)$ and in $R_{AAi}(y_t)$ as shown in Eq.~(\ref{raann}).

As noted in Sec.~\ref{interpret} Ref.~\cite{alicepbpbpidspec} interprets $R_{AA}(p_t)$ data to indicate that for $p_t > 10$ GeV/c ($y_t > 5$) three hadron species follow {\em statistically similar trends}, suggesting that while ``jet quenching'' represents in some sense parton energy loss to a dense medium the hadrochemistry (``species composition'') of the jet ``core'' is not significantly influenced by the medium. That conclusion is based on $R_{AA}(p_t)$ data as presented in Fig.~\ref{piddata} (right panels). The hard-component trends in Fig.~\ref{xxxx} (c) and (d) strongly contradict that claim. There are actually dramatic differences between pion and proton hard components, with kaons exhibiting intermediate behavior. The misinterpretation arises because as noted $R_{AA}(p_t)$ (a) conceals most of the jet fragment distribution, (b) includes the hidden ratio $q_{iAA}(\nu) / q_{ipp}$ that is strongly centrality dependent and exaggerates the centrality variation of proton $R_{AA}$ compared to pion $R_{AA}$, and (c) when plotted on linear \pt\ further suppresses essential structure near 1 GeV/c.

\section{Comparing particle Fractions} \label{fractions}

The strategy for constructing PID spectra described in Ref.~\cite{alicepbpbpidspec} consists in first measuring {\em particle fractions} $f_{id}(p_t,\eta)$ via several mechanisms (e.g.\ Cerenkov radiation, time of flight, $dE/dx$ relativistic rise) and then combining those with previously measured non-PID hadron spectra from Ref.~\cite{alicenonpidspec}. The procedure is represented by Eq.~(5) from Ref.~\cite{alicepbpbpidspec}
\bea \label{badpid}
\frac{d^2N_{id}}{dp_Tdy} &=& \left[ J_{id}(y,\eta) \frac{\epsilon_{ch}}{\epsilon_{id}} f_{id}(p_t,\eta) \right] \times \frac{d^2N_{ch}}{dp_Td\eta},
\eea
where $J_{id}(y,\eta)$ is a Jacobian from $\eta$ to $y$, $\epsilon_{ch}/\epsilon_{id} \approx 0.95$ is an efficiency correction, $f_{id}(p_t,\eta)$ are the particle fractions resulting from PID $dE/dx$ measurements, and the quantity on the left represents PID spectra in Fig.~\ref{piddata}. 
In contrast, the PID spectrum TCM defined by Eq.~(\ref{pidspectcm}) is based on the assumption that separate fractions $z_{si}$ and $z_{hi}$ of TCM soft and hard components for each hadron species $i$ are {\em independent of \yt}. That assumption was confirmed via analysis of \ppb\ PID spectra as reported in Ref.~\cite{ppbpid}. The same assumption is applied to \pbpb\ PID spectra in the present study, and the method results in a precise and informative TCM \pbpb\ data reference.

It is interesting to compare the two procedures directly by (a) reconstructing the particle fractions $f_{id}(p_t)$ from published non-PID and PID spectra from Refs.~\cite{alicepbpbpidspec} and \cite{alicenonpidspec} and (b) generating the equivalent distributions from the PID TCM derived from \ppb\ data in Ref.~\cite{ppbpid}. To obtain item (a) each of three PID spectra must be converted to the point-by-point non-PID \pt\ coverage as a common point set by interpolation and extrapolation, accurate to within data statistical uncertainties. Item (b) is easily achieved given the functional form of the PID TCM.

\begin{figure}[h]
	\includegraphics[width=1.65in]{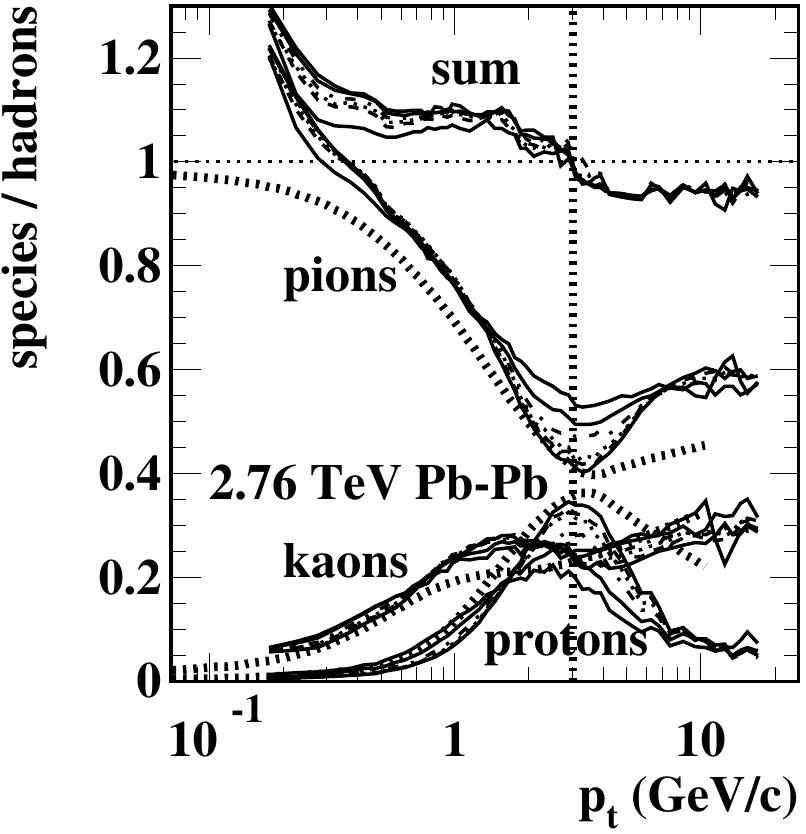}
	\includegraphics[width=1.65in]{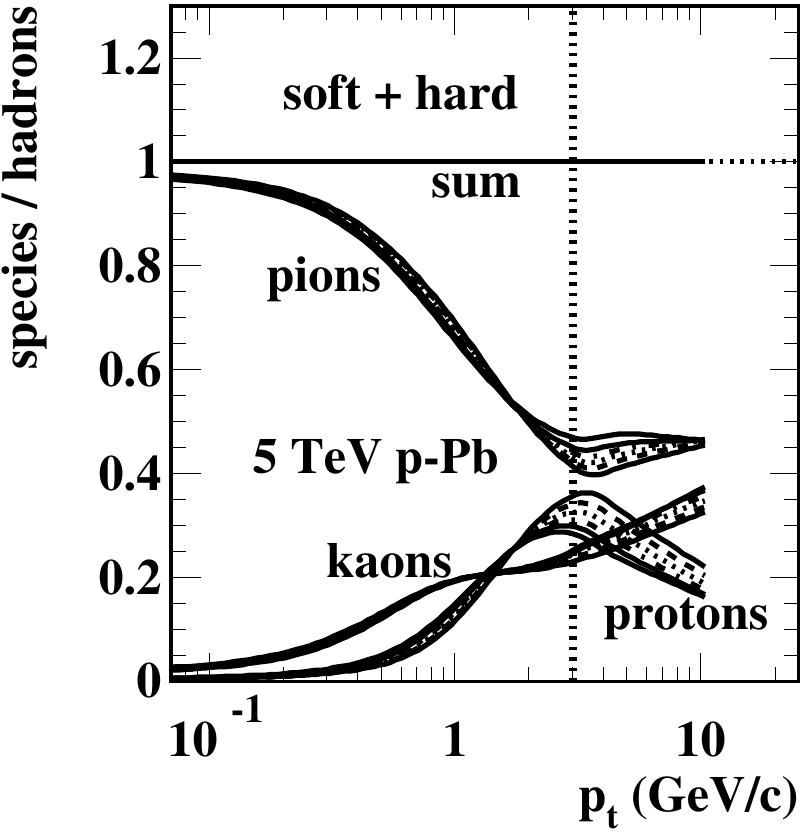}
	\caption{\label{ratio1}
	Left: Particle fractions  $f_{id}(p_t)$ for three hadron species from six centralities of 2.76 TeV \pbpb\ collisions reconstructed from PID~\cite{alicepbpbpidspec} and non-PID~\cite{alicenonpidspec} \pt\ spectra according to Eq.~(\ref{badpid}). Only the parts to the right of the vertical dotted line appear in Fig.~9 of Ref.~\cite{alicepbpbpidspec}. The sums of three distributions are shown at the top.
	Right: Results of the same procedure applied to PID and non-PID TCM spectrum models for spectra from 5 TeV \ppb\ collisions as reported in Ref.~\cite{ppbpid} with parameters appearing in Sec.~\ref{pidfracdata} of this article. TCM curves for the most-central collisions in the right panel appear as bold dotted curves in the left panel.
	} 
\end{figure}

Figure~\ref{ratio1} (left) shows  $f_{id}(p_t)$ [actually the quantity in square brackets appearing in Eq.~(\ref{badpid})] for three hadron species from 2.76 TeV \pbpb\ collisions reconstructed from the PID spectra shown in Fig.~\ref{piddata} (left panels) in ratio to non-PID spectra reported in Ref.~\cite{alicenonpidspec}. The sums of three hadron species are also included (uppermost curves). Those results are nominally equivalent to the contents of Fig.~9 of Ref.~\cite{alicepbpbpidspec}. However, they cover the entire \pt\ acceptance of the spectrum data whereas Fig.~9 of Ref.~\cite{alicepbpbpidspec} shows only the data above $p_t = 3$ GeV/c (see vertical dotted lines in Fig.~\ref{ratio1}). Also, a semilog plot format is used in Fig.~\ref{ratio1} to provide better access to the lower-\pt\ data. It is interesting that the sums at left deviate substantially from unity since the $f_{id}(p_t)$ data nominally result from model fits to $dE/dx$ curves ``normalized to have unit integrals'' according to Fig.~5 of Ref.~\cite{alicepbpbpidspec}. Note the strong suppression of \pbpb\ protons above $p_t = 3$ GeV/c relative  to the most-central \ppb\ result (bold dotted curve) that relates to Fig.~\ref{protons} (d).

Figure~\ref{ratio1} (right) shows comparable curves derived from the \ppb\ PID spectrum TCM described in Sec.~\ref{pidspecc}. As noted, those curves correspond to an assumption of hadron fractions $z_{si}$ and $z_{hi}$ independent of \yt\ along with TCM model functions describing individual \ppb\ hadron species accurately, and closely related to non-PID spectrum data and data from \pp\ collisions. The curves for three hadron species sum to unity by construction (upper solid line). Curves in the right panel for the most-central \ppb\ collisions are repeated in the left panel as bold dotted curves for direct comparison. Note that near 3 GeV/c in either panel the most-central curves are the lowest for pions and kaons and the highest for protons. The most-central \ppb\ data should be compared with more-peripheral \pbpb\ data.

Comparison of the two panels of Fig.~\ref{ratio1}) is interesting both for what corresponds and what does not. The general forms of the trends for three hadron species are similar as are their centrality dependences. However, there are major quantitative differences. As noted, the sums in the left panel deviate substantially from unity. The excess at low \pt\ arises from the pion fraction whereas near 1 GeV/c it also arises from the kaon fraction. The net deficiency at high \pt\ results from strong suppression of protons (relative to the \ppb\ dotted curve) partly offset by continuing pion excesses (relative to the \ppb\ dotted curves). Near 3 GeV/c in the left panel protons increase strongly with \pbpb\ centrality at the expense of pions, whereas that trend is much less strong for \ppb\ collisions in the right panel.

Based on a comparison of Tables~\ref{ppbparams1} and \ref{rppbdata} the most-central \ppb\ results should be roughly comparable to the most peripheral \pbpb\ results. However, some trends in Fig.~\ref{ratio1} do not agree with that expectation. Especially for protons the fraction for peripheral \pbpb\ collisions is strongly suppressed compared to any centrality of \ppb\ collisions. Examination of Fig.~5 of Ref.~\cite{alicepbpbpidspec} suggests that the suppression does not arise from misfitting the $dE/dx$ data; for all centralities the proton-kaon peak at high \pt\ (8-9 GeV/c, lower panels) is substantially lower than the pion peak for all centralities. Particle fractions derived from the TCM can be further separated into soft and hard components to provide additional information.

\begin{figure}[h]
	\includegraphics[width=1.65in]{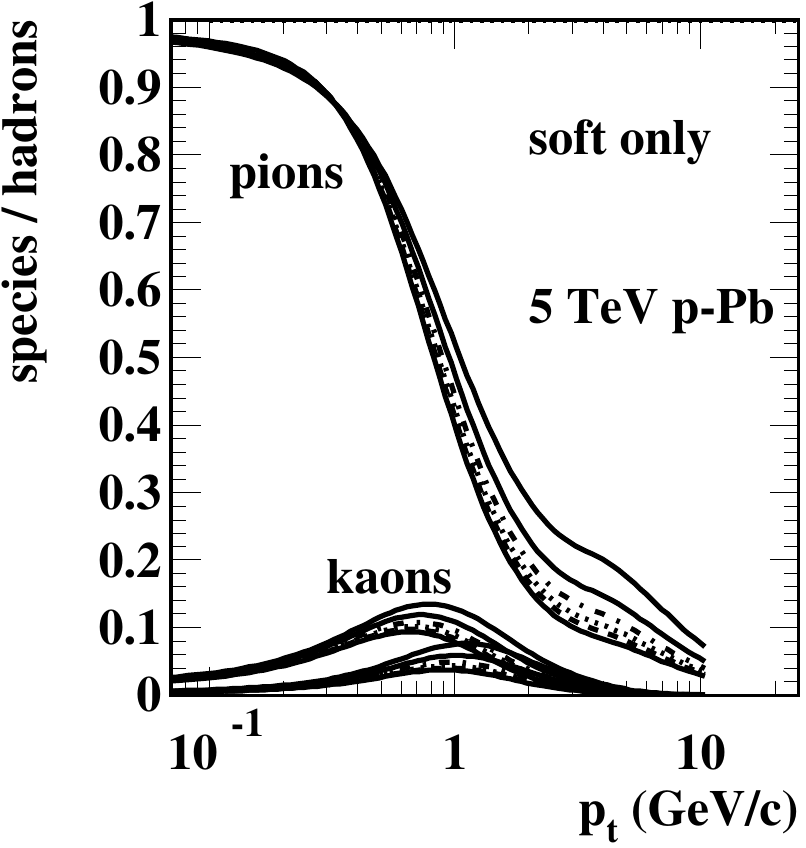}
	\includegraphics[width=1.65in]{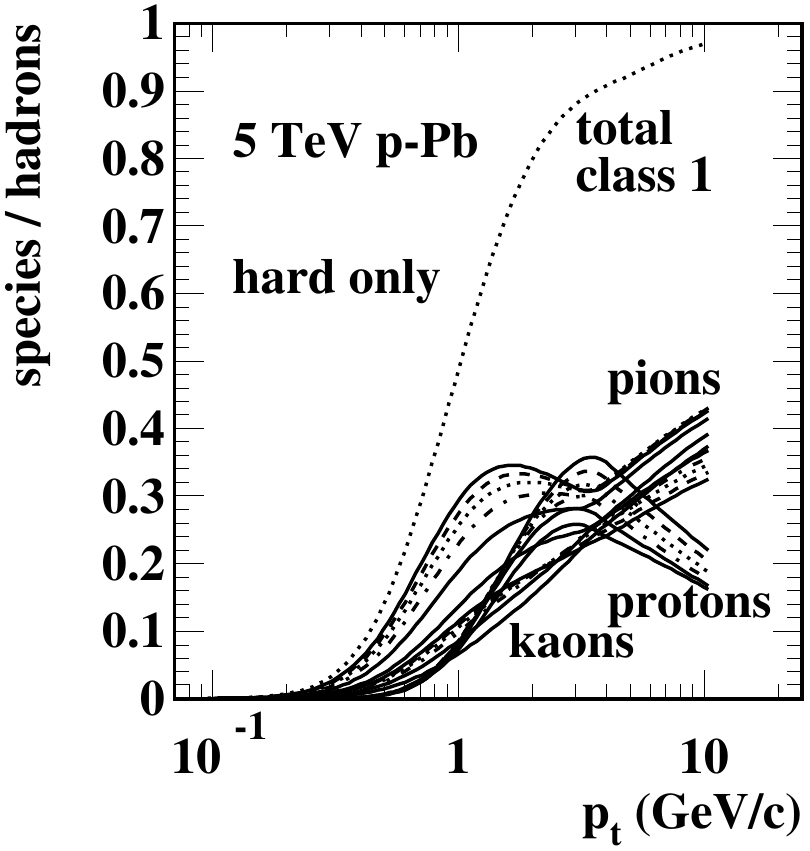}
	\caption{\label{ratio2}
		Left: Particle fractions from PID TCM spectrum soft components. The lowest curves correspond to most-central collisions.
		Right: Particle fractions from PID TCM hard components. The highest curves correspond to most-central collisions. The bold dotted curve is the sum of the three curves for most-central collisions (\ppb\ class 1). 
	} 
\end{figure}

Figure~\ref{ratio2} (left) shows particle-fraction soft components from 5 TeV \ppb\ collisions. The soft components are dominated by pions; protons comprise only a tiny fraction. All soft-component fractions decrease with increasing \ppb\ centrality. The maxima on \pt\ for kaons and protons arise from the dependence of \pt\ spectrum soft-component shapes on hadron mass.

Figure~\ref{ratio2} (right) shows particle-fraction hard components from \ppb\ collisions increasing strongly with \pt\ above 0.3 GeV/c consistent with spectrum hard-component \pt\ dependence. It is clear that jet fragments dominate spectra ($> 50\%$) above 1 GeV/c in more-central \ppb\ collisions. Hard-component fractions for pions and protons are greatest for central \ppb\ collisions whereas the kaon trend reverses near 3 GeV/c. In contrast to soft components the proton hard component is comparable in magnitude to the pion and kaon fractions. These results also make clear that jet formation dominates proton production in high-energy nuclear collisions.

It should be noted that while a PID spectrum TCM (as for \ppb\ collisions) makes possible the detailed comparisons in this section, the particle-fraction procedure described in Ref.~\cite{alicepbpbpidspec} is required to process the primary $dE/dx$ data obtained from particle detectors.

\begin{figure}[h]
	\includegraphics[width=1.63in]{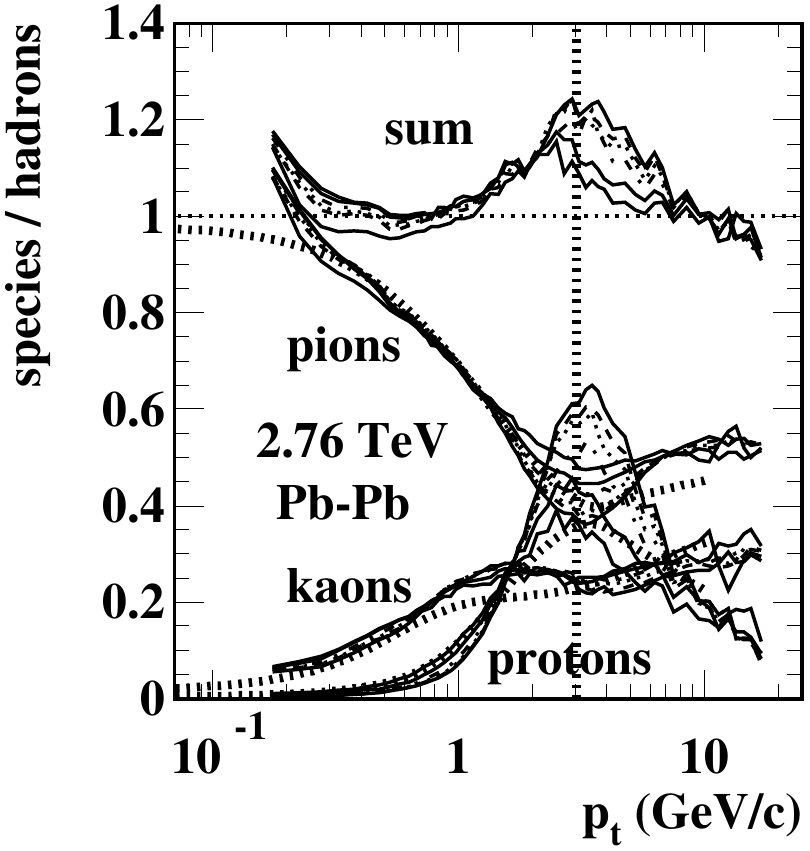}
	\includegraphics[width=1.67in]{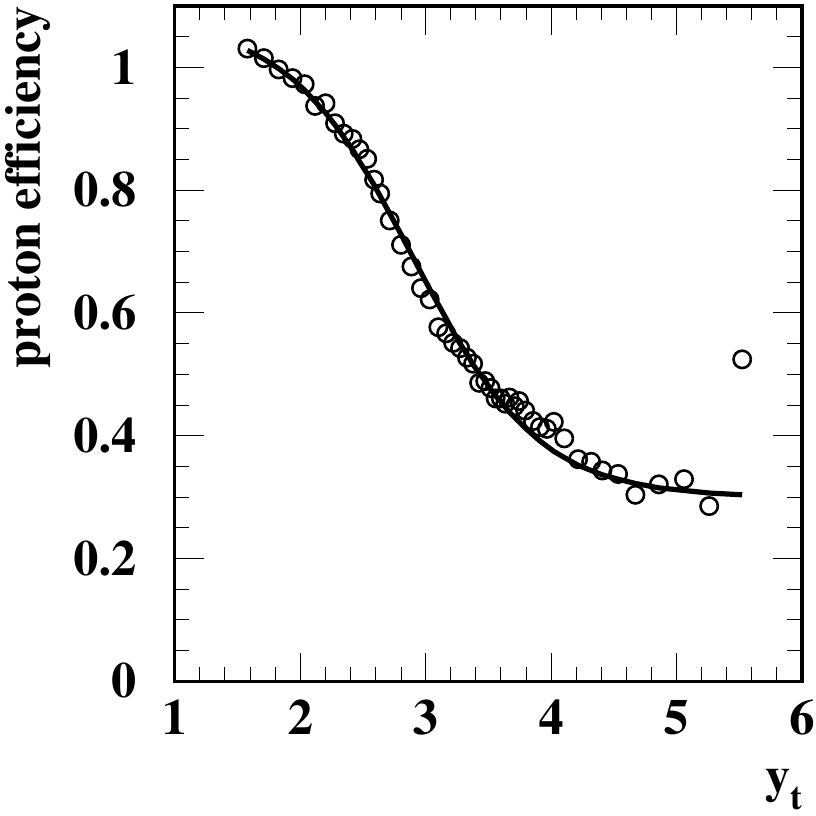}
	\caption{\label{ratiox}
		Left: Results from Fig.~\ref{ratio1} (left) replotted with proton spectra efficiency corrected. The pion data are multiplied by 0.9 to bring peripheral \pbpb\ curves into agreement with \ppb\ data. The proton data are multiplied by 0.8 to  accomplish the same.
		Right: Proton detection efficiency inferred by comparing \pp\ proton data with a TCM reference. The curve is a tanh model actually applied to correct \pbpb\ proton data.
		} 
\end{figure}

Figure~\ref{ratiox} (left) shows Fig.~\ref{ratio1} (left) updated with the proton efficiency correction described in Sec.~\ref{correct}. The pion fractions have been multiplied by factor 0.9 and the proton fractions by 0.8. Peripheral \pbpb\ proton fractions now correspond with central \ppb\ proton fractions as expected. The general result confirms that proton production in \aa\ collisions increases dramatically with centrality and is almost completely due to jet production.

Figure~\ref{ratiox} (right) shows an efficiency curve for proton detection within 2.76 TeV \pp\ and \pbpb\ collisions derived from 2.76 TeV \pp\ data. The efficiency is inferred from the ratio of \pp\ proton spectrum data from Ref.~\cite{alicepbpbpidspec} to the corresponding \pp\ proton TCM represented by Eq.~(\ref{ppprotontcm}). The TCM trend is modified to include the systematic falloff of \pp\ hard components at high \yt\ attributed to the underlying jet spectrum deviating from a simple power law~\cite{jetspec2}. The curve in the right panel is a simple tanh model that is actually used to correct the proton spectra for \pbpb\ and \pp\ collisions.

\section{Systematic uncertainties} \label{sys}

Uncertainties for \pbpb\ collision-geometry determination, TCM spectrum model functions and accuracy of spectrum models for PID data are discussed in the context of previous analysis of \pp\ \pt\ spectra~\cite{ppprd,ppquad,alicetomspec} and PID \ppb\ spectra~\cite{tommpt,tomglauber,tomexclude,ppbpid}. In particular, the 5 TeV \ppb\ PID spectrum TCM reported in Ref.~\cite{ppbpid} provides the basis for the present analysis. Uncertainties include those for primary spectrum data, for spectrum models and for physical interpretations of model results. Uncertainties in interpretation are discussed in Sec.~\ref{convention}

\subsection{5 TeV p-Pb PID spectrum uncertainties}

Uncertainties in \ppb\ geometry are discussed in Sec.~\ref{pbpbgeom} along with \pbpb\ geometry. Uncertainties in PID TCM model functions are provided in Table~\ref{pidparams}. Uncertainties in PID parameters are provided in Table~\ref{otherparams}. Details of uncertainty estimates are reported in Ref.~\cite{ppbpid}.

It can be questioned whether the pion excess and proton suppression reported in Ref.~\cite{ppbpid} are significant (i.e.\ exceed systematic uncertainties). That question relates to the structure of Eqs.~(\ref{pidspectcm}), (\ref{rhosi}) and (\ref{pidhard}). To extract data hard components accurately $X_i(y_t)$ must result from proper normalization of spectra $\bar \rho_{0i}(y_t)$ so as to match model function $\hat S_{0i}$ at low \yt. As described in the text the normalization factor is defined by Eq.~(\ref{rhosi}) and proper matching of data to $\hat S_{0i}$ defines parameters $z_{hi}/z_{si}$ and $z_{0i}$. There is then no freedom to adjust those parameters further. $z_{hi}/z_{si}$ in turn determines the scaling of $Y_i(y_t)$ that is compared directly with model function $\hat H_{0i}(y_t)$ to obtain hard-component ratios $r_{AAi}(y_t)$ as in Eq.~(\ref{raax}).

It is notable that the described TCM procedure applied to {\em neutral} hadrons -- $K_\text{S}^0$ and $\Lambda$ -- leads to an exact match between data and TCM within data uncertainties. The TCM parameters for $K_\text{S}^0$ vs $K^\pm$ and protons vs $\Lambda$s are equivalent within uncertainties. Those implicit A-B comparisons suggest that the pion excess and proton suppression are real effects. Whether they are due to novel physics or instrumental bias is an open question.
Further confirmation is provided by 2.76 TeV \pp\ PID spectra that, for pions and charged kaons, match the \ppb\ results. The proton spectra from \pp\ and \pbpb\ collisions appear to be problematic as discussed further below.
 
\subsection{$\bf p$-Pb and Pb-Pb collision geometry} \label{pbpbgeom}

A-B collision geometry in relation to observed charge density $\bar \rho_0$ as a control parameter depends on competition between increased \nn\ multiplicity and increased A-B centrality. For \pp\ collisions there is only one participant pair (centrality is not an issue) and $\bar \rho_0$ controls dijet production via $\bar \rho_h \approx \alpha \bar \rho_s^2$ with $\bar \rho_0 = \bar \rho_s + \bar \rho_h$~\cite{ppprd,ppquad,alicetomspec}. For \ppb\ collisions the competition between increasing \pn\ multiplicity and \pa\ centrality can be monitored by dijet production in \pn\ collisions, for instance as manifested by \mmpt\ systematics assuming no jet modification~\cite{tommpt,tomglauber} (assumption verified by subsequent spectrum analysis~\cite{ppbpid}).

Those considerations argue against assuming that physics conditions in different A-B collision systems with the same \nch\  may be similar. For a charge multiplicity equivalent to $\bar \rho_{sNN} \approx 30$ the ratio $x(n_s)$ of hard (jet) to soft charge densities is 0.36, 0.18 and 0.07 respectively for \pp, \ppb\ and \pbpb\ collisions. The smallest collision system contains the largest {\em relative} jet contribution, and that ordering is reflected in the systematics of ensemble \mmpt\ data~\cite{tommpt} following a similar trend for \pt\ spectra from which \mmpt\ data are derived~\cite{ppbpid}. However, the same data features interpreted within a flow context lead to very different conclusions: In Ref.~\cite{aliceppbpid} it is concluded (from BW fits to \ppb\ spectrum data) that results are ``consistent with the presence of radial flow in \ppb\ collisions.'' It is further noted that ``a larger [inferred] radial velocity in \ppb\  [vs \pbpb] collisions has been suggested as a consequence of stronger radial gradients.'' 

For \pbpb\ collisions the assumption of no jet modification is not valid and jet modification is an object of study. However, an alternative assumption is suggested by the fact that for central \ppb\ collisions $N_{bin} \approx 7$~\cite{tomglauber} whereas for central \pbpb\ collisions $N_{bin} \approx 1700$ (Table~\ref{ppbparams1}). The \ppb\ centrality trend in Table~\ref{rppbdata} suggests that initially, as the \nch\ condition is increased, the required increase is much more likely provided by isolated peripheral \pn\ interactions ($N_{part} = 2$). Only for greater \nch\ is \pa\ centrality driven to increase ($N_{part} > 2$). In contrast, increasing \nch\ for \pbpb\ collisions may be immediately more likely satisfied by an increasing number of \nn\ binary collisions via \pbpb\ centrality increase, and \nn\ multiplicity may then not increase significantly.

The procedure for \pbpb\ analysis in the present study is based on assuming a fixed value for $\bar \rho_{sNN}$ and accepting $N_{part}$ from Table~\ref{ppbparams1} as a reliable estimate. The value $\bar \rho_{sNN} \approx 4.55$ as for 2.76 TeV \pp\ collisions is obtained from Ref.~\cite{alicetomspec}.  $N_{part}/2$ has asymptotic limits 1 and A for \aa\ collisions. Changes with collision energy (due to change of inelastic cross section $\sigma_{NN}$) are greatest for peripheral 
\pbpb\ collisions but smaller for more-central collisions. Thus, $N_{part}/2$ from a Glauber Monte Carlo (Table~\ref{ppbparams1}) should be fairly reliable for most centralities.

Glauber estimates for $N_{bin}$ can be questioned  based on a comparison of TCM vs Glauber \ppb\ centrality trends reported in Ref.~\cite{tomglauber}. As described in Sec.~\ref{geomparams} a strategy is adopted to extract the product $x(\nu)\, \nu$ (all that is required for the spectrum TCM) from measured charge densities combined with $N_{part}/2$ from Table~\ref{ppbparams1}. Possible large uncertainties from Glauber estimates are then confined to the separate factors but not the product. The result is shown in Fig.~\ref{900d} (left). There is an overall scale uncertainty of 5\% because that plot was generated with $\bar \rho_{sNN} = 4.3$, not 4.55 as expected from \pp\ collisions~\cite{alicetomspec}.

\subsection{2.76 TeV $\bf p$-$\bf p$ and Pb-Pb PID uncertainties}

There are several distinct elements to the uncertainties for the \pbpb\ PID spectrum TCM and resulting inferred PID data hard components. Because this analysis is highly differential it is sensitive to three aspects: (a) biases or distortions in the primary PID and non-PID particle data, (b) uncertainties in the TCM model and (c) uncertainties in physical interpretations (Sec.~\ref{convention}).

(a)  Distortions or biases in primary PID particle data: Examples appear in Fig.~\ref{kaons} for 80-90\% central \pbpb\ kaon data (solid points), in Fig.~\ref{protons} for \pp\ proton data (open circles) and in the same figure for \pbpb\ proton data (thin curves of several line styles). Distortions in \pbpb\ proton data are also suggested by comparison of particle fractions in left and right panels of Fig.~\ref{ratio1} (see below).

(b) Uncertainties in PID spectrum TCM: The TCM arising from the 5 TeV \ppb\ study reported in Ref.~\cite{ppbpid} is adopted as a reference essentially unchanged for the 2.76 TeV \pbpb\ PID spectrum TCM.
Uncertainties are therefore essentially as described in that article. The main difference between \ppb\ TCM and data is enhancement of the \ppb\ data pion hard component (40\% increase) and suppression of the data proton hard component (40\% decrease) relative to the TCM. It is notable that the 2.76 TeV \pp\ pion data in Fig.~\ref{pions} (d) follow the same trend as the 5 TeV \ppb\ data. The same is true for kaon data in Fig.~\ref{kaons}. The 2.76 TeV \pp\ proton data in Fig.~\ref{protons} (d) are consistent with \ppb\ data at lower \pt\ (e.g.\ $\approx 1$ GeV/c) but fall much below the \ppb\ trend at higher \pt. 

The difference between \ppb\ and \pbpb\ proton trends is consistent with the particle fraction data in Fig.~\ref{ratio1} (left). Comparing peripheral \pbpb\ fraction data for protons (lowest proton curves) with highest-\nch\ curves for \ppb\ (bold dotted curve) the \pbpb\ trend falls to 25\% of the \ppb\ curve above 2 GeV/c (\yt\ $\approx$ 3.4), consistent with the peripheral \pbpb\ vs \ppb\ curves in Fig.~\ref{protons} (d). 

In contrast, the fraction data for pions (highest pion curves) lie substantially above  the \ppb\ trend (bold dotted curve) consistent with Fig.~\ref{pions} (d) above 1 GeV/c (hard component). However, the pion fraction curve below 0.5 GeV/c (not shown in Fig.~9 of Ref.~\cite{alicepbpbpidspec}) is not self-consistent, exceeding unity by a substantial amount. \ppb\ TCM fraction trends (right panel) correspond to the ideal case of an ideal spectrum TCM wherein there is no pion excess, proton suppression or jet modification.

\subsection{Conventional plot formats and analysis methods} \label{convention}

As noted in the previous subsection, uncertainty estimation should extend beyond uncertainty estimates for primary particle data. Uncertainties arising from plotting formats, uncertainties in applied data models and uncertainties in physical interpretations are also relevant.

The choice of plot format and plotted variables can strongly influence the accessibility of the physical information carried by particle data (what research funding pays for). For example, plotting spectrum data in a semilog format on linear \pt\ as in Fig.~\ref{piddata} exaggerates the high-\pt\ interval above 5 GeV/c while the interval below 3 GeV/c, where {\em almost all jet fragments reside}, is compressed and thereby visually obscured. In contrast, plots on transverse rapidity \yt\ as in Sec.~\ref{piddiff2} reveal essential details about jet structure below 3 GeV/c (\yt\ $\approx 3.8$) that are inaccessible via conventional plotting formats. 

By convention, comparisons between theoretical models and spectrum data are presented as theory/data {\em ratios}, again on linear \pt. That format choice greatly suppresses theory-data {\em differences} at lower \pt\ since the ratio of statistical-error values to spectrum values falls by orders of magnitude with decreasing \pt. Thus, theory-data differences at lower \pt\ may be tens of statistical error bars (thus falsifying the theory) but nevertheless remain invisible in the conventional ratio plotting format~\cite{alicetomspec}.

The BW spectrum model assumes a radially-expanding dense medium as the dominant hadron source for $p_t < 3$ GeV/c~\cite{blastwave,hydro}. One consequence should be a boosted hadron spectrum reflecting a broad boost distribution (Hubble expansion) and manifesting as spectrum {\em suppression} at lower \pt\ to complement enhancement at higher \pt\ (spectrum ``flattening'' or ``hardening'') as illustrated by Fig.~3 of Ref.~\cite{hydro}. The BW spectrum model is applied to restricted \pt\ intervals typically determined by \pt\ acceptance cutoff as a lower limit and relying on fit quality to determine an arbitrary upper limit~\cite{aliceppbpid}.

However, even if spectra seemed to evolve according to a flow scenario (e.g.\ spectrum ``hardening'' increasing with \nch) there is no guarantee that hydro expansion is the mechanism. Simple application of a hydro-based spectrum model cannot ``prove'' the existence of flow.  \pt\ intervals for model-data comparison are determined so as to accommodate the model, which is then not falsifiable. And alternative hadron production mechanisms, most notably MB dijet production but also including simple longitudinal projectile dissociation, have not been ruled out. Thus, whether uncertainties are quoted for BW model fit parameters or not the relevance of those parameters to nuclear collisions is highly questionable.

Spectrum ratio $R_{AA}(p_t)$ is conventionally used to assess the extent of jet modification (``quenching'') in nuclear collisions, specifically by the degree of suppression {\em relative to unity} at higher \pt\ (e.g. above 4 GeV/c). However, $R_{AA}(p_t)$ fails dramatically as a diagnostic for jet formation below 4 GeV/c, the interval within which {\em almost all jet fragments appear} as illustrated in Sec.~\ref{ratcompare}. Whether systematic uncertainties have been assessed within its region of validity or not, the overall uncertainty of $R_{AA}(p_t)$ results is overwhelming because a misleading picture of jet formation and modification is presented. Theoretical models of jet modification that address only $R_{AA}(p_t)$ data above 4 GeV/c and do not acknowledge the large enhancements at lower \pt\ revealed by hard-component ratio $r_{AA}(y_t)$ should be summarily rejected.

The combination of BW fits to selected lower \pt\ intervals and spectrum ratio $R_{AA}(p_t)$ in effect conspire to misrepresent the great majority of jet fragments as a flow manifestation.  $R_{AA}(p_t)$ transitions from a jet-dominated measure above 4 GeV/c ($y_t > 4$) to a trend dominated by the {\em nonjet} spectrum soft component below that point: contrast left and right panels of Fig.~\ref{pidrats}. Propagation of particle-data uncertainties to $R_{AA}(p_t)$ uncertainties via standard statistical methods must greatly underestimate the true uncertainty relative to jet-related information carried by particle data. With jet contributions at lower \pt\ effectively concealed by $R_{AA}(p_t)$ BW model fits to spectra in lower-\pt\ intervals  dominated by jet contributions may be misinterpreted as measuring radial flow.

\section{Discussion} \label{disc}

 This section considers three questions that emerge from a TCM analysis of 2.76 TeV \pp\ and \pbpb\ PID spectrum data in the context of previous TCM analysis:
 (a) Based on extensive measurements of jet properties and TCM analysis of spectrum and correlation data how are MB jet fragments actually distributed on \pt\ or \yt?
 (b) Based on current experience with TCM analysis of PID hadron spectra what is ``jet quenching?''
 (c) Based on comparison of hydro theory with TCM analysis of PID spectra can radial flow be inferred from \pt\ spectra?

\subsection{Where do jet fragments reside in $\bf p_t$ spectra?} \label{where}

Interpretation of \pbpb\ PID \pt\ spectra as reported in Ref.~\cite{alicepbpbpidspec} is summarized in Sec.~\ref{interpret}. Specifically, two \pt\ intervals are  assumed to reflect distinct physical mechanisms: For $p_t < 3$ GeV/c spectrum features are assumed to reflect ``bulk production'' (including flow manifestations), and for $p_t > 10$ GeV/c jet formation and in-medium jet modification (``jet quenching'') should be the relevant issue for produced hadrons. If that description were valid jet fragments would comprise a tiny fraction of hadron production, and almost all hadrons (i.e.\ most of the nonjet fraction) would emerge from a flowing dense medium (QGP). A similar interpretation of \ppb\ PID \pt\ spectra is presented in Ref.~\cite{aliceppbpid}. Between separate low-\pt\ and high-\pt\ intervals lies an intermediate region where the hadron production mechanism is uncertain. Such descriptions are based on the assumption that distinct production mechanisms relate to separated \pt\ intervals.

In contrast to spectrum models based on such {\em a priori} assumptions the TCM is based on {\em observed} scaling of distinct soft and hard hadron fractions that happen to overlap strongly on \pt. Derivation of the \pp\ TCM from \nch\ systematics of \pt\ spectra is first described in Ref.~\cite{ppprd}. Accurate isolation of the two spectrum components then leads to direct comparison between the spectrum hard component and measured jet properties~\cite{fragevo} and to development of an accurate description of the \pp\ collision-energy evolution of jet energy spectra over  three orders of magnitude~\cite{jetspec2}. In essence, measured jet spectra and measured jet FFs predict the shape and absolute magnitude of the spectrum hard component for any \pp\ collision energy. In particular, any predicted \pp\ hard component has a mode near 1 GeV/c, and most jet fragments appear {\em below} 3 GeV/c. The spectrum TCM soft component does not change with A-B centrality and changes only slowly $\sim \log(\sqrt{s_{NN}})$ with collision energy.

With the spectrum hard component predicted quantitatively via measured jet properties~\cite{fragevo}, the spectrum interpretations in Refs.~\cite{aliceppbpid} and \cite{alicepbpbpidspec} can be reexamined. Describing the lower-\pt\ interval  ($< 3$ GeV/c) is the statement  ``hardening of the spectra'' with increasing centrality ``...is mass dependent and is characteristic of hydrodynamic flow....'' That qualitative observation may be consistent as far as it goes but does not establish  that flow plays any role in nuclear collisions. In fact, the predicted trend for the MB jet contribution to spectra has exactly those characteristics: (a) With increasing A-B centrality the jet contribution increases much faster than the soft component, thus making the spectrum ``harder,'' and (b) as demonstrated for \ppb\ collisions in Ref.~\cite{ppbpid} the hard/soft ratio for baryons is much greater than that for kaons which is greater than that for pions, as presented in Table~\ref{otherparams}. 

Describing a higher-\pt\ interval ($> 10$ GeV/c) is the statement ``the spectra follow a power-law shape as expected from perturbative QCD (pQCD) calculations.'' But pQCD theory is not required to predict the high-\pt\ trend that is derived quantitatively from the convolution of a measured jet spectrum~\cite{jetspec2} with measured FFs~\cite{eeprd}. The power-law trend is only an approximation, over a limited \pt\ interval, to the actual jet spectrum. The same convolution predicts the jet fragment distribution down to 0.5 GeV/c and confirms that the majority of fragments appear near 1 GeV/c~\cite{hardspec,fragevo}.

Applied to an intermediate-\pt\ region (e.g.\ 2-10 GeV/c) is the statement ``...it is an open question if additional physics processes [e.g.\ the  ``baryon/meson puzzle'' \cite{rudy,duke,tamu}] occur in the intermediate $p_T$ region...''~\cite{alicepbpbpidspec}. In Ref.~\cite{alicespec2} is the statement ``Below the [proton/pion] peak [on \pt], $p_t < 3$ GeV/c, both ratios [$p/\pi$, $K/\pi$] are in good agreement with hydrodynamical calculations, suggesting that the peak itself is dominantly the result of radial flow....'' But that region is actually dominated by jet fragments, and results in Sec.~\ref{piddiff2} above confirm that spectrum structures relevant to the peak in the baryon/meson {\em ratio} scale {\em individually} with the number of binary \nn\ collisions as expected for jet production. The \ppb\ PID spectrum study of Ref.~\cite{ppbpid} reveals that baryons are copiously produced by MB dijets, and the present \pbpb\ PID study reveals that centrality evolution of the baryon/meson ratio is a direct consequence of PID jet modification.

\subsection{What is ``jet quenching?''} \label{quench}

Spectrum ratio $R_{AA}(p_t)$ as defined in Eq.~(\ref{raa}) was formulated in anticipation of RHIC startup in 2000 to search for evidence of parton energy loss in a dense QCD medium (``jet quenching'')~\cite{daufinalstate,starraacu}. Suppression of  $R_{AA}(p_t)$ at ``high \pt'' (e.g.\ $p_t > 6$ GeV/c or $y_t > 4.5$), an interval attributed exclusively to jet production, was expected to reveal jet quenching, and the degree of suppression might indicate quantitatively the extent of parton energy loss~\cite{jetquenching}. This study and previous PID spectrum analyses in Refs.~\cite{hardspec,fragevo} cast doubt on that conjecture.

In order to develop a realistic theoretical picture of jet modification in A-B collisions complete jet fragment distributions over the largest possible \pt\ interval must be determined accurately. In particular, the interval from 0.15 to 4 GeV/c ($y_t \in [1,4]$), where almost all jet fragments reside, should be a central focus of study. Reliance on biased $R_{AA}(p_t)$ measurements, with theoretical analysis applied only at high \pt, cannot achieve a correct understanding. Those criteria are consistent with results presented in Secs.~\ref{pidspec}, \ref{ratcompare} and \ref{where} of the present study.

Jet-related trends as revealed by hard-component ratio $r_{AA}(y_t)$, comparing data fragment distributions to a TCM reference, reveal basic elements of jet modification. Modification in \aa\ collisions, for instance as in Fig.~\ref{pidrats} (d) and (e), is consistent with large overall shifts $\Delta y_t$ of the spectrum hard component to {\em lower \yt}. Hard-component model $\hat H_0(y_t)$ is a Gaussian with exponential tail. The log of the ratio of shifted to unshifted such functions, expressed as a Taylor series, has the leading term $\Delta y_t \, d\log[\hat H_0(y_t)] / dy_t$. The derivative then has the form of a sloped line passing through zero and transitioning to a negative constant above $y_t \approx 3.6$ ($p_t \approx 2.5$ GeV/c). That scenario is described in Sec.~VIII B of Ref.~\cite{hardspec}.

\begin{figure}[h]
	\includegraphics[width=1.62in]{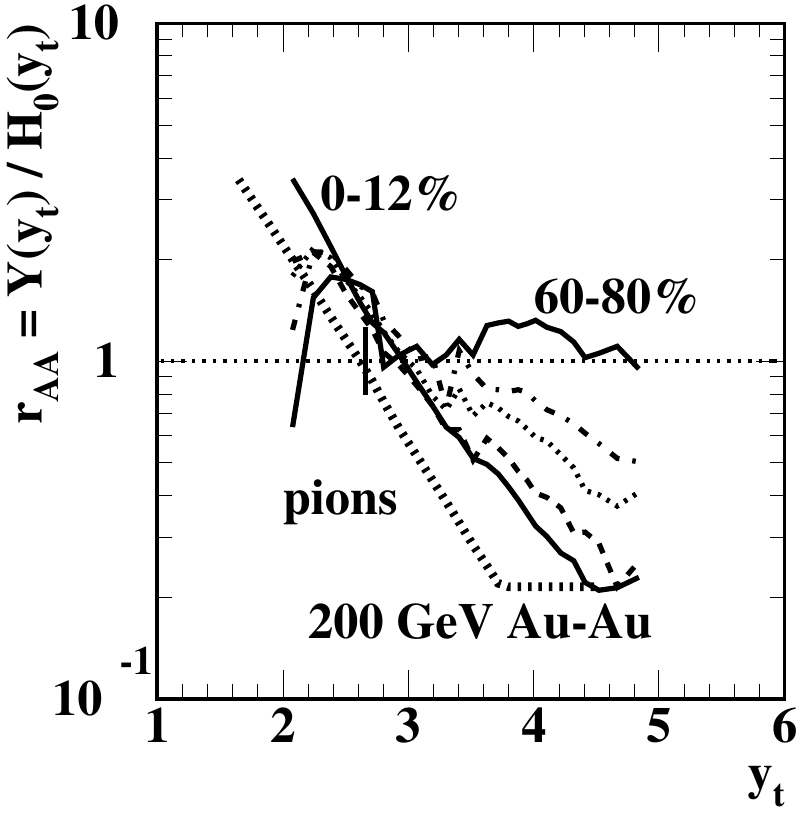}
	\includegraphics[width=1.69in]{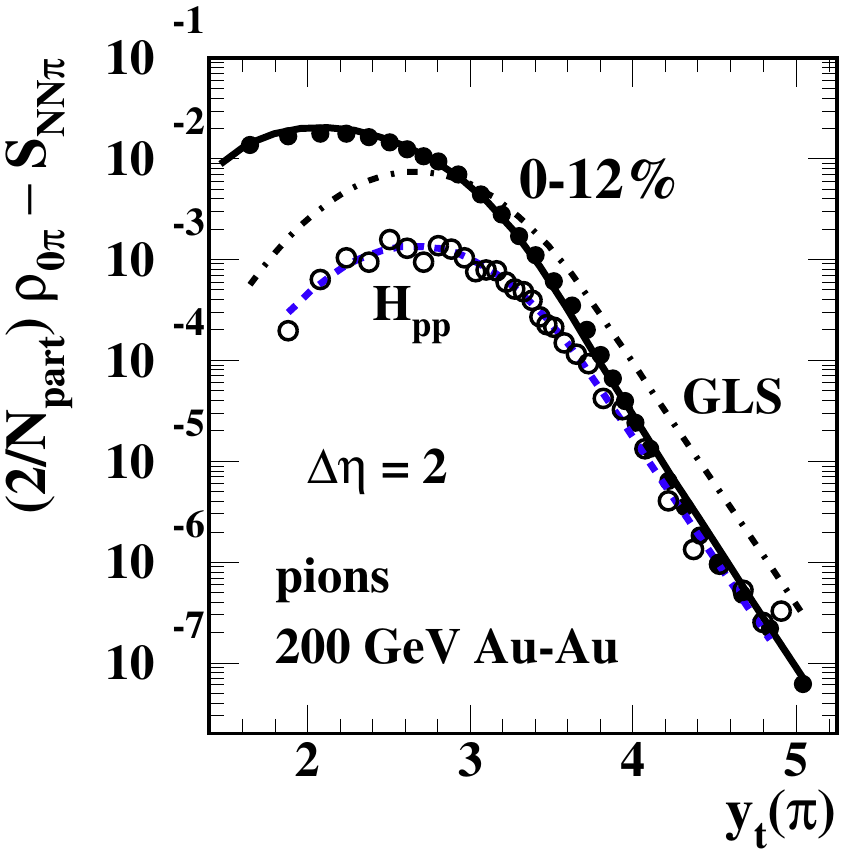}
	\caption{\label{quenchx}
		Left: Figure~\ref{raaauau} (left) with added jet-modification model based on shifted $\hat H_0(y_t)$ (bold dotted). The vertical solid line marks the $\hat H_0(y_t)$ mode at $y_t = 2.66$.
		Right: Pion spectrum hard components from 200 GeV \pp\ collisions (open circles) and 0-12\% central \auau\ collisions (solid dots). TCM models are shown for \pp\ collisions (dashed) and central \auau\ collisions (dash-dotted). The solid curve is a pQCD prediction for central \auau\ based on convoluting a measured jet spectrum with measured FFs modified according to a prescription in Ref.~\cite{borghini}.
	}  
\end{figure}

Figure~\ref{quenchx} (left) repeats Fig.~\ref{xxxx} (a). The scenario above is illustrated by the added bold dotted curve, with unmodified $\hat H_0(y_t)$ for pions and $\Delta y_t \approx 0.3$. The general shape of $r_{AA}(y_t)$ for central \auau\ data is reproduced, but with a zero intercept located at the mode of $\hat H_0(y_t)$ ($y_t = 2.66$) per the above formula, inconsistent with data.

Figure~\ref{quenchx} (right) shows the actual 0-12\% central \auau\ hard-component data (solid points) compared to a pQCD {\em prediction} (solid curve) based on modified FFs (see below)~\cite{fragevo}. \pp\ data (open circles) are accurately described by the TCM  $H_{pp}$ (dashed). The TCM reference for central \auau\ data is the dash-dotted curve $\nu H_{pp}$. Those results explain the discrepancy in the left panel. Jet modification does not only consist in a simple shift $\Delta y_t$; the fragment distribution also increases dramatically at lower \yt\ in a manner consistent with conservation of parton energy {\em within} the modified jet, as implied by modification of FFs consistent with the DGLAP equations. As a result, the crossover between \auau\ data and TCM reference ($\rightarrow$ unity crossing at left) occurs near $y_t = 2.9$, not 2.66.

Large shifts of the fragment distribution or spectrum hard component to lower \yt\ are consistent with transport of the splitting cascade within a jet to lower \pt\ that may be modeled by increasing the coefficient of the gluon splitting function within the DGLAP equations~\cite{borghini}. That picture is confirmed by comparing \auau\ $r_{AA}(y_t)$ data to the pQCD convolution of a measured jet spectrum with modified \pp\ FFs. The solid curve in Fig.~\ref{quenchx} (right) is obtained with \pp\ FFs corresponding to a $\approx$ 10\% increase in the coefficient of the gluon splitting function in the DGLAP equations (per Ref.~\cite{borghini})~\cite{fragevo}. The resulting agreement with the spectrum hard component from central 200 GeV \auau\ collisions (solid dots) is within data uncertainties. 

Such agreement excludes the possibility of any loss (by absorption in a dense medium) of jets as low as 3 GeV energy. Spectrum hard components are typically consistent with geometry parameter $N_{bin}$ as expected for jet production and as confirmed again in the present study. Whereas \ppb\ data are consistent with small shifts up or down on \yt\ \pbpb\ data for pions and kaons are consistent with much larger shifts always to lower \yt. Proton data are not as simple, suggesting  that jet modification in \aa\ is restricted, for baryon fragments at least, to \pt\ values above the hadron mass (e.g.\ $\approx$ 1 GeV/c for protons)~\cite{hardspec}.

There is a further issue for interpretation of $R_{AA}$ or $r_{AA}$ spectrum ratios based on \pp\ spectra as the reference. Fragmentation functions for \pp\ collisions are already modified relative to those for $q$-$\bar q$ or $g$-$g$ (inferred from three-jet events) dijets from \ee\ collisions~\cite{eeprd}. \pp\ FFs are strongly suppressed below 1 GeV/c compared to \ee\ FFs~\cite{fragevo,mbdijets}. For high-energy jets the suppression corresponds to a small fraction of the total jet energy but a substantial fraction of the fragment number. For lower-energy jets (i.e.\ most jets in a MB sample relevant to A-B spectra) the fractional suppression is correspondingly larger. Thus, the low-\pt\ enhancement observed for $r_{AA}$ in more-central \aa\ collisions {\em relative to \pp\ spectra}, apparently relating to ``jet quenching,'' could be seen instead as {\em restoration} of the fragment distribution for in-vacuum jets from \ee\ collisions. The correct reference for and physical interpretation of spectrum responses to more-central \aa\ collisions is then an open question.

\subsection{Can radial flow be inferred from $\bf p_t$ spectra?} \label{noflow}

The conventional interpretation of PID \pt\ spectra from high-energy nuclear collisions is expressed in terms of radial flow and jet quenching. The principal conclusions from Ref.~\cite{alicepbpbpidspec} are summarized in Sec.~\ref{alicedata}, among which are:  For $p_t < 3$ GeV/c spectrum features are assumed to ``provide information on bulk production....''  With increasing centrality ``...hardening of the spectra'' at lower \pt\ is observed. The effect ``...is mass dependent and is characteristic of hydrodynamic flow....'' Reference~\cite{alicepbpbspecx} reports details from lower-\pt\ intervals of the same PID spectra (below 3 or 4.5 GeV/c) and asserts that ``hydrodynamics has been very successful in describing [\pt\ spectra] up to a few GeV/c. In its Table~5 the reference reports parameters $\bar \beta_t$ and $T_{\text{kin}}$ from a BW model  fitted to limited \pt\ intervals. Reference~\cite{alicepbpbspecx} concludes that ``These features are compatible with the development of a strong collective flow...which dominates the spectral shapes up to relatively high $p_T$ in central collisions.'' 

While that interpretation might be seen as a plausible conjecture in more-central \aa\ collisions difficulties arise from recent \ppb\ PID spectra as reported in Ref.~\cite{aliceppbpid}, where it is observed that BW model parameters $\bar \beta_t$ and $T_{kin}$ have similar values for \ppb\ and \pbpb\ collisions. Those results are then interpreted as ``consistent with the presence of radial flow in \ppb\ collisions.'' Alternatively, one could interpret the \ppb\ spectrum results as casting doubt on hydro interpretations for any A-B spectra~\cite{ppbpid}. 

As an exercise consider the example of a thin cylindrical shell expanding with some velocity $\beta_t$ that corresponds to a boost on transverse rapidity denoted by $\Delta y_{t0}$ with the relation $\beta_t = \tanh(\Delta y_{t0})$. An example is given for kaons by Fig.~3 of Ref.~\cite{hydro} where the spectrum shape without transverse flow is assumed to be exponential and parameters are defined in the context of RHIC energies. With increasing flow velocity the {\em apparent} slope parameter increases, referred to as ``flattening'' or ``hardening.'' But presenting such an example within the conventional semilog plot format on linear \pt\ or \mt\ as  in Ref.~\cite{hydro} obscures essential features relating to evidence for {\em or against} radial flow from PID spectra.  Below is a detailed analysis of $K^0_{\text{S}}$ spectra from 5 TeV \ppb\ collisions to buttress that conclusion. $K^0_\text{S}$ data are compared to TCM model functions as reported in Ref.~\cite{ppbpid} using kaon transverse rapidity $y_{t\text{K}}$ as the independent variable.

\begin{figure}[h]
	\includegraphics[width=1.63in]{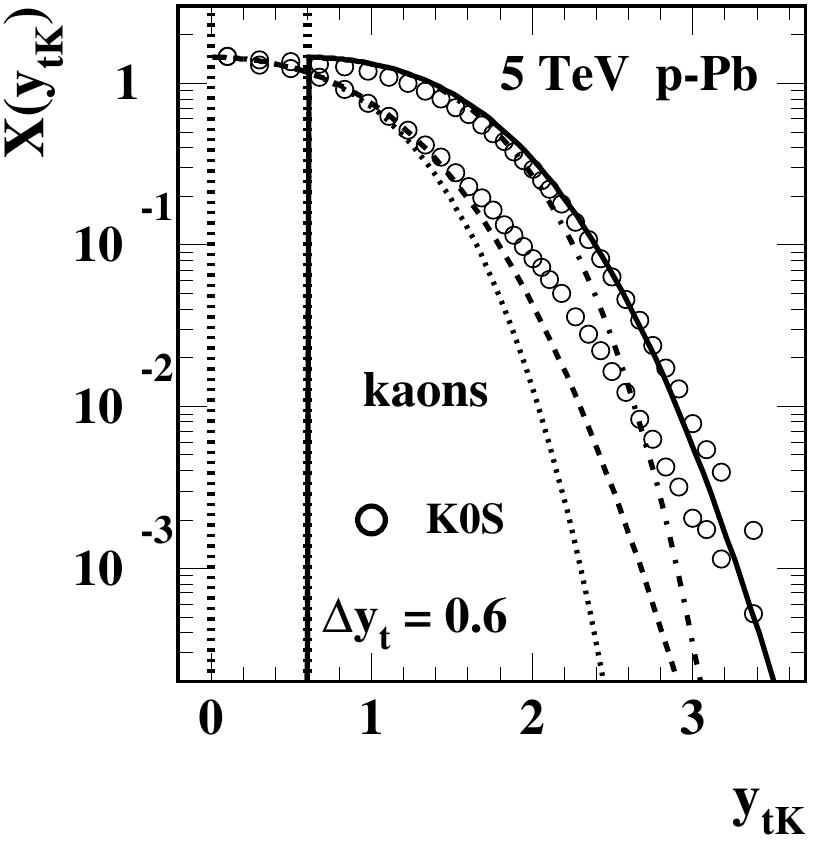}
	\includegraphics[width=1.67in]{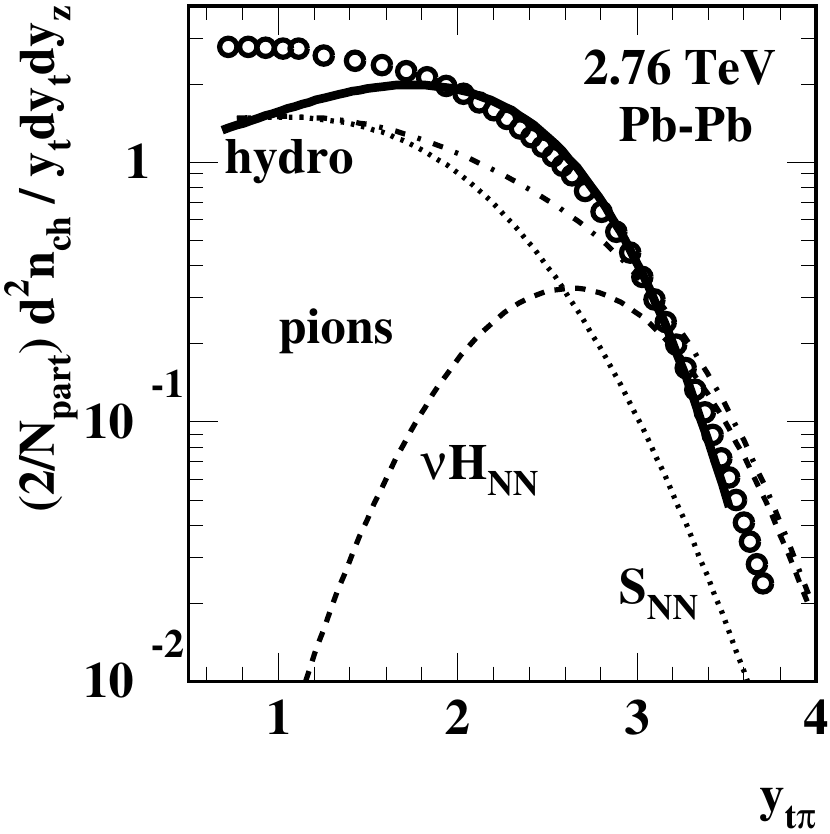}
	\caption{\label{950}
		Left: $K^0_{\text{S}}$ spectra for most-peripheral and most-central 5 TeV \ppb\ collisions (open circles) from Ref.~\cite{ppbpid} plotted on kaon rapidity $y_{t\text{K}}$. The dashed and dotted curves are respectively the \ppb\ TCM soft component model $\hat S_0$ and an exponential as a limiting case. The solid and dash-dotted curves are the dashed and dotted curves shifted to higher \yt\ by source boost $\Delta y_{t0} = 0.6$.
		Right:  Hydro theory curve for pions from 0-5\% central 2.76 TeV \pbpb\ collisions from Ref.~\cite{galehydro2} (solid) compared to the measured pion spectrum (open circles) plotted on pion transverse rapidity $y_{t\pi}$ as it appears in Fig.~\ref{pions} (b). The dashed and dotted curves are TCM model functions derived from 7 TeV \pp\ and 5 TeV \ppb\ spectrum data~\cite{alicetomspec,ppbpid} with TCM coefficients for 0-5\% 2.76 TeV \pbpb\ assuming \nn\ linear superposition (no jet modification), with no adjustment for this comparison.
	}  
\end{figure}

Figure~\ref{950} (left) shows $K^0_\text{S}$ data for most-peripheral and most-central 5 TeV \ppb\ collisions (open circles)~\cite{ppbpid}. The imported data are densities on pion rapidity $y_{t\pi}$ in the form $X(y_{t\pi})$ as defined in Eq.~(\ref{pidspectcm}). To convert those data to densities on kaon $y_{t\text{K}}$ requires the Jacobian $(y_{t\pi} m_{t\text{K}})/(y_{t\text{K}} m_{t\pi})$. The soft-component model function for that case, defined on kaon $m_{tK}$ by Eq.~(\ref{s00}), uses parameters from Table~\ref{pidparams}: $T = 200$ MeV and $n$ = 14. It is then converted, in this case, from kaon $m_{t\text{K}}$ to kaon rapidity $y_{t\text{K}}$ (dashed curve) via the Jacobian $m_{t\text{K}} p_{t} /y_{t\text{K}} $. The kaon rapidity is required because a source boost must be applied to the proper rapidity for a given hadron species. The dotted curve is the exponential limit with $n \rightarrow \infty$. The exponential is included in response to claims that data tend to follow an exponential trend below 3 GeV/c ($y_{tK} \approx 2.5$)~\cite{hydro}. With slope parameter determined by the great majority of (soft) hadrons below 0.5 GeV/c the exponential demonstrates that spectrum data are generally inconsistent with that model function. 

The solid curve is the dashed curve boosted by $\Delta y_{t0} = 0.6$, and the dash-dotted curve is the corresponding exponential. That specific boost value is relevant because it corresponds to the inferred value of $\bar \beta_t$ for more-central \pbpb\ collisions~\cite{alicepbpbspecx} and also corresponds to the transverse boost inferred from quadrupole spectra reconstructed from $v_2(p_t)$ data as in Refs.~\cite{quadspec,njquad}. It is intriguing that the boosted soft component coincides with central \ppb\ kaon data over some substantial range of \yt\ values, seemingly consistent with a flow manifestation. 

However, it is clearly evident that the $K^0_\text{S}$ data, extending down to $p_t = 0$ (\yt\ = 0), dramatically contradict the boosted model that corresponds to zero particle density below $\Delta y_{t0} = 0.6$ (for kaons corresponding to $p_t \approx$ 0.3 GeV/c). One could argue that real collisions correspond to a broad boost distribution with mean value $\bar \beta_t$, but that does not eliminate the required large reduction at lower \yt\ compared to an unboosted reference. It is notable that the unboosted TCM soft reference describes the $K^0_\text{S}$ data for {\em all} \ppb\ {\em centralities} down to zero \yt\ within the small data uncertainties. The analysis in Ref.~\cite{ppbpid} demonstrates that the difference between peripheral and central \ppb\ data is not due to a source boost. The difference is a {\em fixed} hard component for all \ppb\ centralities that varies {\em only in amplitude}, in a manner predicted by \ppb\ centrality and \nn\ binary-collision scaling as expected for dijet production. For spectra viewed in the conventional semilog plotting format with linear \pt\ or \mt\ (e.g.\ Fig.~\ref{piddata}) the essential details evident in Fig.~\ref{950} (left) (i.e.\ below 0.3 GeV/c) are invisible. Misinterpretation of data in favor of a flow narrative results.

Figure~\ref{950} (right) shows the pion spectrum from 0-5\% central 2.76 TeV \pbpb\ collisions (points) as it appears in Fig.~\ref{pions} (b). The published spectrum data have been multiplied by $2\pi$ to correspond with $\eta$ densities used in this study and divided by participant pair number $N_{part}/2 \approx 190$ corresponding to 0-5\% central \pbpb\ collisions. The solid curve is a hydro result for the same collision system from Ref.~\cite{galehydro2}. The other curves are TCM soft component $S_{NN}(y_t)$ (dotted), hard component $\nu H_{NN}(y_t)$ (dashed) and the sum $(2/N_{part})\bar \rho_0(y_t)$ of TCM soft and hard components (dash-dotted) as in Eq.~(\ref{auaunorm}) (second line). The \pbpb\ TCM model functions are adopted from the spectrum TCM for 7 TeV \pp\ and  5 TeV \ppb\  collisions~\cite{alicetomspec,ppbpid} and therefore constitute an absolute {\em prediction} for \pbpb\ assuming \nn\ linear superposition (no jet modification). TCM parameters are not varied to accommodate the \pbpb\ data. The measured pion spectrum is suppressed at larger \yt\ compared to the TCM reference (consistent with spectrum ratio $R_{AA}$) but {\em strongly enhanced} at smaller \yt\ suggesting approximate energy conservation within modified jets~\cite{hardspec}. 

The main feature of the right panel is comparison of the hydro result (solid) with data (points). The hydro curve falls well below the data at smaller \yt\ where the great majority of hadrons are located but is consistent with data at larger \yt\ where the spectrum is dominated by the jet-related hard component, {\em including the effect of jet modification}. The shape of the hydro curve is indeed consistent with a large-amplitude and broad source boost distribution (i.e.\ radial flow) as illustrated by Fig.~3 of Ref.~\cite{hydro}. But the strong decrease of the curve at lower \yt\ is generally not observed in spectrum data, and the increase at higher \yt\ produced by the source boost describes a data trend consistent with MB dijet production ($\propto N_{bin}$). In effect, the two panels of Fig.~\ref{950} support the same conclusion -- for a single boost value in the left panel or for a broad boost distribution in the right panel. Both are falsified by spectrum data below 0.5 GeV/c.

The logical form of argument in favor of a hydro mechanism based on qualitative observation of \pt\ spectrum ``hardening'' or ``flattening'' increasing with collision centrality and hadron mass (as expected for radial flow) is an example of the ``undistributed middle'' fallacy (referring to syllogism terminology): Cause A is assumed to produce effect B. I observe effect B (approximately) and  conclude that cause  A is responsible. But effect B may  be produced by some other cause C, and some effect D, observed as well, may exclude A as a cause. In the present case cause C is MB dijets as the signature manifestation of QCD in high-energy nuclear collisions, and effect D is the detailed shapes of \pt\ spectra at low momentum.

\section{Summary}\label{summ}

This article reports a differential analysis of  identified-hadron (PID) \pt\ spectra from six centrality or \nch\ classes of 2.76 TeV \pbpb\ collisions and from \pp\ collisions at the same energy. The PID spectra are described by a two-component (soft + hard) model (TCM)  of hadron production in high-energy nuclear collisions. The soft component is associated with longitudinal projectile-nucleon dissociation and the hard component is associated with large-angle scattering of low-$x$ gluons to form jets. 

The present \pbpb\ study relies on a recent TCM analysis of PID spectra from 5 TeV \ppb\ collisions to provide an essential reference in that the \ppb\ data are consistent with linear superposition of \pn\ collisions, no jet modification and no evidence for radial flow. The \ppb\ analysis then establishes TCM model functions and PID parameters for five hadron species (charged pions, charged kaons, neutral kaons, protons and Lambdas) that are adopted with almost no change for the present \pbpb\ analysis of charged pions, kaons and protons.

2.76 TeV \pp\ and \pbpb\ proton spectra appear to exhibit substantial inefficiencies above 1 GeV/c. A correction is estimated by comparing the $p$-$p$ spectrum with the corresponding TCM proton spectrum derived from \ppb\ collisions that also describes Lambdas within data uncertainties. After inefficiency correction Pb-Pb proton spectra are similar to 200 GeV \auau\ proton spectra. 

Jet-related Pb-Pb and Au-Au spectrum hard components exhibit strong suppression at higher $p_t$ in more-central collisions corresponding to results from spectrum ratio $R_{AA}$ but also (for pions and kaons) exhibit dramatic enhancements below $p_t = 1$ GeV/c that are concealed by $R_{AA}$. In contrast, enhancements of proton hard components appear only above 1 GeV/c suggesting  that the baryon/meson ``puzzle'' is a jet phenomenon. 

The principal results of the present analysis are as follows: 
(a) PID spectra from peripheral 2.76 TeV \pbpb\ and \pp\ collisions  are consistent with 5 TeV \ppb\ results including no significant evidence for radial flow or jet modification for any \ppb\ \nch\ class spanning a thirty-fold variation in dijet production. 
(b) For more-central \pbpb\ collisions jet-related spectrum hard components deviate strongly from \pp\ and \ppb\ references, including suppression at higher \pt\ and complementary {\em large enhancement} at lower \pt. 
(c) A {\em sharp transition} in the jet modification trend is similar to the transition observed in jet properties from 200 GeV \auau\ collisions. 
(d) For all \pbpb\ spectra soft components are consistent with a fixed shape as for \pp\ and \ppb\ reference spectra, providing no significant evidence for radial flow (i.e.\ no significant boost of the soft component on transverse rapidity \yt). 
(e) Overall spectrum shape evolution is dominated by hard-component effects that scale with  the number of \nn\ binary collisions as expected for dijet production.

This \pbpb\ analysis illustrates the precision achievable via the TCM applied as a common data reference across an ensemble of collision systems. The analysis also makes clear the large amount of information accessible within particle data given application of differential methods.


\end{document}